\documentclass[12pt]{iopart}

\usepackage{iopams}
\usepackage{cite}
\expandafter\let\csname equation*\endcsname\relax
\expandafter\let\csname endequation*\endcsname\relax

\usepackage{amsmath}
\usepackage{iopams}
\usepackage{graphicx}
\usepackage[breaklinks=true,colorlinks=true,linkcolor=blue,urlcolor=blue,citecolor=blue]{hyperref}
\usepackage{url}
\usepackage[T1]{fontenc}
\usepackage[british]{babel}
\usepackage[british]{babel}

\newcommand{\clr}{\color{black}}

\def\x{\mathbf{x}}
\def\pa{{\partial \Omega}}
\def\ve{\varepsilon}

\def\P{\mathbb{P}}
\def\C{\mathbb{C}}
\def\R{\mathbb{R}}

\def\T{\mathcal{T}}

\def\S{\mathcal{S}}
\def\L{\mathcal{L}}

\def\s{\mathbf{s}}

\def\nmax{n_{\rm max}}

\def\acos{\mathrm{acos}}
\def\Li{\mathrm{Li}}

\def\res{\mathop{\rm Res}}

\begin{document}

\title[Reduction-of-dimensionality scenario versus direct diffusive
search]{Search efficiency in the Adam-Delbr\"uck reduction-of-dimensionality scenario versus direct diffusive search}

\author{Denis S Grebenkov$^{\dagger}$, Ralf Metzler$^{\ddagger}$ \& Gleb
Oshanin$^\sharp$}
\address{$\dagger$ Laboratoire de Physique de la Mati\`{e}re Condens\'ee
(UMR 7643), CNRS -- Ecole Polytechnique, IP Paris, 91120 Palaiseau, France}
\address{$\ddagger$ Institute of Physics and Astronomy, University of Potsdam,
14476 Potsdam-Golm, Germany}
\address{$\sharp$ Sorbonne Universit\'e, CNRS, Laboratoire de Physique
Th\'eorique de la Mati\`ere Condens\'ee (UMR CNRS 7600), 4 Place Jussieu,
75252 Paris Cedex 05, France}

\begin{abstract}
The time instant---the first-passage time (FPT)---when a diffusive particle
(e.g., a ligand such as oxygen or a signalling protein) for the first
time reaches an immobile target located on the surface of a bounded
three-dimensional domain (e.g., a hemoglobin molecule or the cellular
nucleus) is a decisive characteristic time-scale in diverse biophysical and
biochemical processes, as well as in intermediate stages of various inter-
and intra-cellular signal transduction pathways. Adam and Delbr\"uck put
forth the reduction-of-dimensionality concept, according to which a ligand 
first binds non-specifically to any point of the surface on which the target is
placed and then diffuses along this surface until it locates the target. In
this work, we analyse the efficiency of such a scenario and confront it with
the efficiency of a direct search process, in which the target is approached
directly from the bulk and not aided by surface diffusion. We consider two
situations: (i) a single ligand is launched from a fixed or a random position and searches for the target,
and (ii) the case of "amplified" signals when $N$ ligands start either from
the same point or from random positions, and the search terminates when the
fastest of them arrives to the target. For such settings, we go beyond the
conventional analyses, which compare only the mean values of the corresponding
FPTs. Instead, we calculate the full probability density function of FPTs
for both scenarios and study its integral characteristic---the "survival"
probability of a target up to time $t$. On this basis, we examine how the
efficiencies of both scenarios are controlled by a variety of parameters
and single out realistic conditions in which the reduction-of-dimensionality
scenario outperforms the direct search.
\end{abstract}

\section{Introduction}
\label{sec:intro}

More than five decades ago Adam and Delbr\"uck put forth an idea how
organisms may handle some problems of efficiency and timing, limited
by molecular diffusion, by reducing the dimensionality in which the
diffusion takes place from the three-dimensional (bulk) space to
two-dimensional surface diffusion \cite{adam}. A similar claim was
previously made in \cite{trur}, suggesting that acetyl choline may get
faster via surface diffusion from its site of action to the site where
it is destroyed, and in \cite{buch}, arguing that surface diffusion
may result in higher turnover numbers for membrane-bound enzymes.
However, these earlier works did not present quantitative estimates
substantiating such a claim, while Adam and Delbr\"uck were the first to
provide theoretical arguments showing that the diffusive molecules may
indeed reduce the reaction times by subdividing the diffusion process
into successive stages of lower spatial dimensionality. Subsequently,
their analysis has been generalised in a number of directions (see, e.g.,
\cite{eigen,richter,kozak1,berg,ast,kozak2,szabo,zwan,axelrod,schmick}).
Its applicability has also been questioned \cite{mac}, indicating
situations in which the "reduction-of-dimensionality" scenario can be
advantageous and therefore plausible, and situations in which it is not
beneficial for the search process and is therefore not likely to occur.  

Concurrently, the concept of a dimensionality reduction together with
the notion of "facilitated" diffusion provide an explanation why
experimentally observed rates for binding of proteins to special sites
on DNA molecules are much larger than predictions based on the
Smoluchowski approach \cite{richter,
berg2,berg3,coppey,marco,hu,mirny,mirny2,lomholt,lomholt1,klenin,otto,slutsky,
kolesov,max,sheinman}. The Adam-Delbr\"uck scenario (ADS) was also
invoked to explain a fast translocation through the nuclear pore
complexes \cite{peters}.  This scenario also prompted further
investigations giving rise to the idea of so-called intermittent
search strategies \cite{benichou1,Benichou11a} in which the fine
tuning of systems parameters may further reduce the mean first-passage
time
\cite{Benichou10,Benichou11,Rojo11,Rupprecht12a,Rupprecht12b,Rojo13,
Benichou15,lomholt2} (see however \cite{tamm}) or the "survival"
probability---the probability that the target is not found up to time
$t$ \cite{katja1,katja2,katja3} (in transient processes it is also of
interest to consider the search reliability, the probability that the
target has not been found up to $t\to\infty$
\cite{vlad,vlad2}).

\begin{figure}
\begin{center}
\includegraphics[width=6cm]{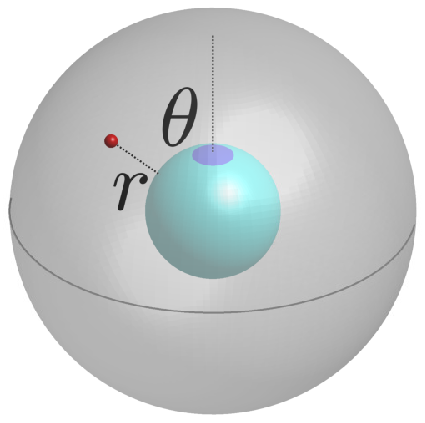}
\includegraphics[width=6cm]{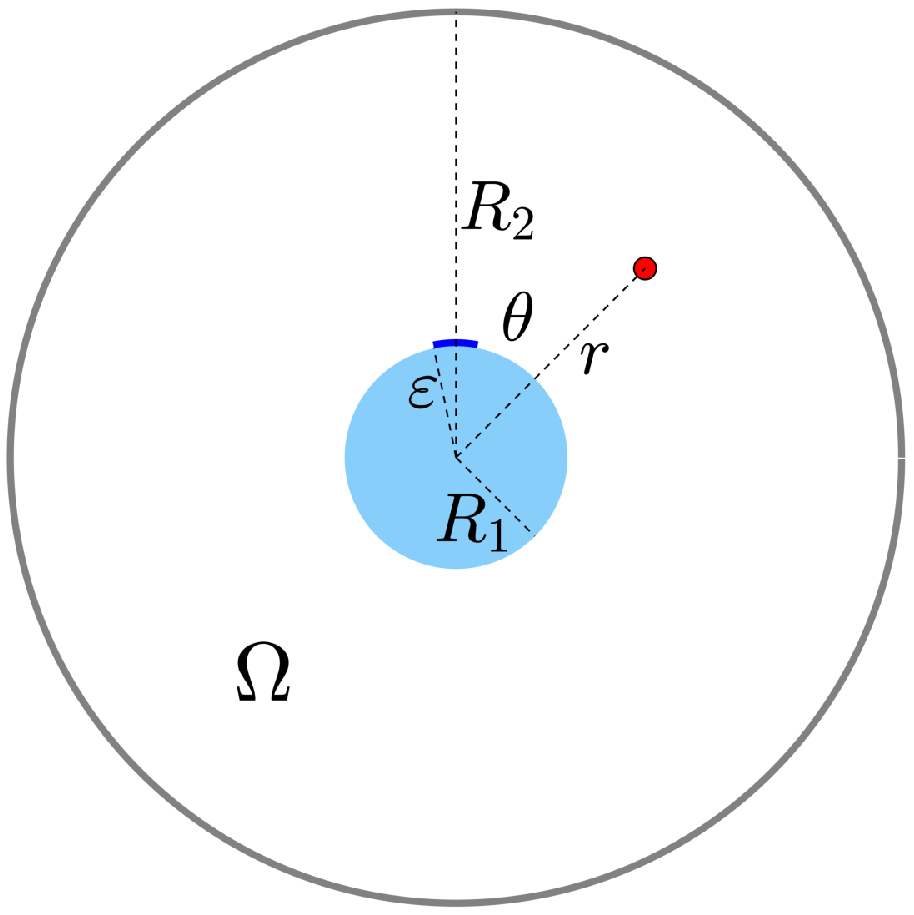}
\end{center}
\caption{Sketch of the geometrical setup in the study of the ADS: a spherical
shell domain $\Omega$ is enclosed by impermeable boundaries consisting of two
concentric nested spheres with the radii $R_1$ (the inner sphere) and $R_2$
(the outer, fully reflective sphere). A target region is located at the North
pole of the inner sphere and has the form of a spherical cap of radius $\rho$
(and angular size $\ve=\arcsin(\rho/ R_1)$). A small red ball denotes the
starting position $\x=(r,\theta,\phi)$ of the diffusive ligand. Within the
ADS, the ligand diffuses (with diffusion coefficient $D_b$) within $\Omega$
until it encounters the surface of the inner sphere, which happens at an
arbitrary point $\s=(R_1,\theta', \phi')$. Then, it non-specifically binds to
the surface and diffuses along it with diffusion coefficient $D_s$, until it
finds the boundary of the target region. Within the direct search scenario,
the surface of the inner sphere is assumed to be perfectly reflecting such
that the ligand bounces back to $\Omega$ once it hits the inner sphere
and may reach the target region only directly from the bulk $\Omega$.}
\label{fig1}
\end{figure}

To better illustrate the possible advantage of the ADS, it is
instructive to dwell on a particular geometrical setting, which was
discussed by Adam and Delbr\"uck themselves \cite{adam} and will be
also used in the present paper.  Namely, we consider two nested
concentric spheres (of radii $R_1$ and $R_2$ ($R_1 < R_2$),
respectively, as shown in figure \ref{fig1}) with impermeable
boundaries, a small immobile circular target of radius $\rho$ placed
at some fixed position on the inner sphere, and a ligand that starts
from a fixed location and diffuses with the diffusion coefficient
$D_b$ within the spherical-shell domain $\Omega$ between two spheres
(here and in what follows we use the term "ligand" to denote any
diffusing entity, e.g., a signalling protein; similarly, the term
"target" will generally denote a binding site, an adsorbed chemically
active particle or an entrance to a nuclear pore).  In such a bounded
domain, a ligand is certain to eventually find the target (see,
e.g. \cite{olivier}), whichever motion scenario it undertakes---and
therefore the only question is how long the search process will last
for a given scenario. We therefore are interested in the first-passage
time (FPT) $\T$, i.e., a random time instant when a ligand reaches the
target for the first time (see, e.g.,
\cite{150,151}).

Conventional one-stage (or direct search) scenario presumes that the
inner sphere is reflecting for the ligand such that it bounces back to
the bulk domain $\Omega$ once it hits the inner sphere everywhere
except for the target. The binding event thus occurs only when the
target is approached by the ligand directly from the bulk. Assuming
that the starting position of the ligand is uniformly distributed
within the spherical-shell domain, Adam and Delbr\"uck \cite{adam}
(see also \cite{berg}) estimated the mean "diffusion" time $\tau_{\rm
dir}$ necessary for the ligand to arrive to the target within the
one-stage scenario as $\tau_{\rm dir}\sim R_2^3/(3D_b\rho)$ (in the
relevant case when $R_2\gg R_1\gg\rho$).  In fact, $\tau_{\rm dir}$
here is the mean diffusion time to a sphere of radius $\rho$ placed at
the geometric centre of the outer sphere and thus this form does not
take into account the fact that the target is situated on a reflecting
sphere of radius $R_1$, which effectively screens it
\cite{9}. Moreover, for a relevant geometrical setting in which the
ligand starts from the outer sphere, $\tau_{\rm dir}$ will be
evidently bigger. Not least, this reasoning clearly applies only to an
idealised situation when the confining domain $\Omega$ does not
contain "obstacles" and the motion of the ligand is not subject to any
kind of molecular crowding effects \cite{McGuffee10,Ghost16}.
However, if $\Omega$ is supposed to mimic the interior of a cell, it
should represent a complex spatial environment, filled with
impermeable organelles, filaments and proteins, which impose steric
constraints on the dynamics and may screen the target. As a
consequence, diffusion of a ligand often takes place effectively in a
tortuous labyrinthine spatial domain.  Recent analyses have provided
evidence that the FPT to the target within the direct search scenario
is essentially increased as compared to the one in which the cytosol
is treated as a homogeneous liquid-filled region \cite{ma}. Hence, the
above estimate can be rather inaccurate (in fact, representing a lower
bound on the actual diffusion time) but is still instructive for
understanding the time scales involved.

In contrast, within the ADS a ligand first finds any (random) point on the
inner sphere, which takes the typical time $\tau_b\sim R_2^3/(3D_bR_1)$.
Then, it nonspecifically binds to the surface and diffuses along with the
diffusion coefficient $D_s$ until it eventually locates the target. The
mean diffusion time $\tau_s$ for the latter process was calculated
\cite{adam,berg} and scales as $\tau_s\sim2R_1^2\ln(R_1/\rho)/D_s$. The
total mean diffusion time $\tau_{\rm AD}$ within the ADS is thus a sum of
two contributions, $\tau_{\rm AD}=\tau_b+\tau_s$, and the ratio 
\begin{align}
\label{ADS}
\delta=\frac{\tau_{\rm AD}}{\tau_{\rm dir}}=\frac{\rho}{R_1}\left(1+6\frac{
D_b}{D_s}\left(\frac{R_1}{R_2}\right)^3\ln\left(\frac{R_1}{\rho}\right)\right)
\end{align} 
shows whether the reduction-of-dimensionality scenario is advantageous (for
$\delta<1$) or not. One observes from equation \eqref{ADS} that the efficiency
of the ADS here is entirely controlled by the aspect ratios $\rho/R_1$ and $R
_1/R_2$ and the ratio $D_s/D_b$ of the diffusion coefficients. Clearly, $\tau
_b$ can also be larger, when obstacles are present in the bulk volume $\Omega$
but intuitively, its increase should not be as pronounced as the one for the
direct search of the target---within the ADS it suffices to find any point on
the inner sphere while for the direct search scenario a prescribed target
location is to be found.
  
In essence, the ADS takes advantage of the very slow logarithmic divergence
of $\tau_{\rm AD}$ in the limit $\rho\to0$, as opposed to a much faster $1/
\rho$-divergence of $\tau_{\rm dir}$ for the direct search scenario. At the
same time, there is a penalty to pay: the gain due to the reduction of the
singularity is counter-balanced by the reduction of the value of the
diffusion coefficient. For instance, the diffusion coefficient $D_b$ of a
ligand in cellular cytoplasm is typically of the order of a few tens of
$\mu\mathrm{m}^2/\mathrm{s}$, while once it gets associated to the inner
sphere, it experiences at best a $20$-fold reduction of the diffusivity
\cite{schmick}. Most often, however, such a reduction is more pronounced
and may amount to two or sometimes even three orders of magnitude
\cite{mac}. Such an extreme reduction is, however, rarely seen in
biophysical systems because most of compartment-separating surfaces are
soft and hence, the barriers against lateral diffusion are typically lower
than the ones specific to diffusion on hard solid surfaces \cite{ehr}. In
any case, a reduction of the diffusivity is detrimental for the ADS, in
virtue of equation \eqref{ADS}. As an example, consider a typical mammalian
cell in which a class I nuclear receptor\footnote{This example represents
an intermediate step within a complicated intracellular signal transduction
pathway in which a hormone penetrating from an extracellular medium into the
cells binds to a nuclear receptor at a random location within the cytoplasm
and causes it to undergo several chemical transformations. The reaction
product subsequently finds the entrance to the nuclear pore and penetrates
into the nucleus, where it binds to the hormone response element on nuclear
DNA \cite{100}.} searches for an entrance into a pore (of radius $\rho\sim3
{\rm nm}$) in the nuclear envelope. For eukaryotes, the karyoplasmic ratio
(KR) typically is of the order $(R_1/R_2)^3\sim0.08$ (see, e.g., \cite{pap}),
but the scatter around this value can be quite significant. In particular,
essential departures from the value $0.08$ are encountered in metastatic
tumours and are, in fact, used in both diagnosis and prognosis for several
tumour types \cite{rag}. Recent systematic analysis \cite{mag} of the
reported values of the KR, which compiled data for almost $900$
species---from yeast to mammals---provided evidence that the larger cells
almost invariably have relatively smaller nuclei, such that the nucleus
of a larger cell may occupy as little as few percent of the cell volume
across all scales of biological organisation, yielding lower values of KR.
For plant cells, which can often grow to larger sizes than animal cells,
it has been known for a long time that KR can be even smaller and amount to
just a fraction of $1$ percent (see, e.g., Table I in \cite{plant}).
Concurrently, $R_1$ may vary in size in different species but is usually
within the range $1\ldots5\mu\mathrm{m}$. Setting $R_1=3\mu\mathrm{m}$ and
$\rho=3{\rm nm}$ and assuming that KR is equal to its average value $0.08$,
i.e., $R_2=7\mu\mathrm{m}$, we thus find $\delta=3.4\times10^{-3}D_b/D_s$,
which signifies that for these particular values of the system parameters
the ADS is advantageous when $D_s/D_b\gtrsim 10^{-2}$. For larger cells or
some plant cells, for which KR is lower, and also in situations in which
the direct search for the entrance to the nuclear pore is obstructed by
the presence of other organelles, one may expect that the ADS is efficient
even for smaller values of the ratio $D_s/D_b$.

The above well-known arguments rely on estimates of the {\it mean\/}
FPT (MFPT), used as a proxy for the efficiency. In this work we
revisit the ADS from the broader perspective of the full statistics of
FPTs. The point is that, regardless of how a search process
proceeds---via a single or two stages---a ligand may follow a variety
of different paths from the starting point to the target, thus
resulting in a large variability of realisation-dependent values of
the FPT. The mean FPT, which is only the first moment of the
corresponding probability density function (PDF) averaged over the
initial position of the ligand, is instructive---yet it is evidently
insufficient to fully characterise neither the ADS nor the direct
search scenario. In fact, it is well-known that for bulk diffusion in
a bounded confining domain towards a perfect sink (see, e.g.,
\cite{carlos1,carlos2,alj1,alj2}), or a partially reactive target
(see, e.g., \cite{dist1,dist4,dist2}), or even a target with more
sophisticated surface reaction mechanism (see, e.g.,
\cite{dg1,dg2,dg3}), the PDF of the FPTs can be very broad. In other
words, large sample-to-sample fluctuations with disproportionally
different values of $\T$ are inherent and fluctuations around its mean
value $\langle\T\rangle$ can be comparable to it or even exceed the
mean value. It is known from standard statistical analysis that if the
PDF is centred around its mean value (e.g., in the case of a Gaussian
distribution), the mean value is representative of the actual
behaviour. In contradistinction, if the mean is well away from the
most probable value (as shown, e.g., in
\cite{dist1,dist4,dist2,alj1,alj2}) it is likely that it is associated
with the tail of the corresponding PDF and hence, is supported by some
rare realisations of the diffusive paths.  In this case,
$\langle\T\rangle$ may be orders of magnitude larger than the typical
time encountered in a considerable fraction of realisations of the
search process (and thus times shorter than the mean time are
typically observed in any given experiment).\footnote{In fact, the
most likely time is connected to "direct" trajectories moving
relatively straight to the target. The distance between the point of
release and the target defines the peak of the initial,
L{\'e}vy-Smirnov part, of the PDF, an effect that is also referred to
as "geometry-control" \cite{alj2}.  Longer times correspond to
"indirect" trajectories, in which the initial distance becomes
irrelevant \cite{alj2,dist1}.} This would also imply that one indeed
needs an extensive statistical sample in order to obtain a reliable
value for $\langle\T\rangle$ for comparison with theoretical
predictions. This is precisely the case for the direct search scenario
in such a geometrical setting, as evidenced recently in \cite{9}.
Concurrently, the full PDF of the FPT for the two-stage search process
is not known as yet and consequently, one does not know anything about
its broadness and other characteristic time scales (e.g., the typical
FPT), that is indispensable in order to obtain a fully comprehensive,
global picture of the search dynamics in the ADS. The knowledge of the
full PDF will also permit us to use as a robust characteristic the
survival probability, i.e., the probability that the target is not
found up to some prescribed time $t$.
  
In this paper, we first consider the case of a single ligand starting from
an arbitrary fixed position within the spherical-shell domain $\Omega$
and a single perfect target placed on the inner sphere. The term "perfect"
here means that we assume that there is no (energetic or entropic) barrier
against the reaction, such that the ligand binds to the target upon the first
arrival---the classical Smoluchowski setting. In the ADS, we calculate the
PDF of the FPTs to the target exactly, which permits us to analyse the spread
of the realisation-dependent FPTs, as well as to understand the contribution
of the typical first-passage time and of extreme events associated with the
tails of the distribution. Moreover, we compare our result against the PDF
for the direct search scenario, which was evaluated for the same geometrical
setting recently \cite{9}, in an exact spectral form as well as in approximate
but remarkably accurate form based on the self-consistent approximation (see
\cite{szabo,q2,q3,q4} for more details). Such a comparison between the ADS and
a one-stage search under identical conditions allows us to provide a complete
picture of the actual efficiency of the ADS which extends beyond the previous
analyses. Next, we consider the situation with an "amplified signal", in which
$N$ ligands are launched either from the same location on the outer sphere,
or from distinct locations on this surface. The former case is realised in
situations when some other ligand (e.g., a first messenger), which is moving
diffusively in the extracellular medium, arrives to a particular site on
the outer part of the plasma membrane and opens an ion channel spanning the
membrane. In this way, the first messenger effects $N$ ions to release inside
the cell from the same position. In turn, the latter case corresponds to a
situation when the plasma membrane hosts multiple receptors each interacting
with the first messengers moving in the extracellular medium and launching
the second messengers that diffuse now within the intracellular medium, all
of them seeking a single target on the inner sphere \cite{purves,alberts}. For
such amplified signals we also provide a comparative analysis of the PDFs for
the one-stage and two-stages scenarios. We finally remark that the results
presented here can be generalised rather straightforwardly for the analysis of
the PDFs of the terminal FPT in intracellular signal transduction processes
involving more than two nested domains (see \cite{10}), presenting in this
way a full stochastic description of the latter important processes within
the reduction-of-dimensionality scenario.

The paper is outlined as follows: In section \ref{model}, we describe the
geometrical and physical parameters of our model, introduce basic notations,
and derive the PDF of the first-passage times within the ADS. Section \ref{pdf}
is devoted to the analysis of our general result and of its asymptotic
behaviour, and also presents a comparison of the form of the PDF for the
ADS against the results for the direct, one-stage scenario derived in
\cite{9}. In section \ref{N} we consider the situation of an amplified signal
with $N$ independent ligands. Finally, in section \ref{conc} we
conclude with a brief recapitulation of our results and a
discussion. Details of intermediate derivations, as well as an
analysis of some limiting situations are relegated to Appendices. In
particular, {\clr \ref{sec:j_moments} considers the moments of the
conditioned first-passage times; \ref{sec:surface_diffusion} is
devoted to a discussion of different aspects of extremal values of
surface diffusion; in \ref{sec:AD} we discuss the form of the PDF for
the Adam-Delbr\"uck's scenario, while in \ref{direct} we present the
PDF for one-stage search process.}

\section{The ADS in a spherical shell domain}
\label{model}

\subsection{Model and basic notations}

We consider a spherical shell domain $\Omega$ (see figure \ref{fig1})
enclosed by impermeable boundaries of two concentric nested spheres with
radii $R_1$ (the inner sphere) and $R_2$ (the outer, perfectly reflecting
sphere). An immobile target with the shape of a spherical cap (dome) of
radius $\rho$ is located at the North pole of the inner sphere. In
spherical coordinates $(r,\theta,\phi)$, the target is defined by $r=R_1$
and $0\leq\theta\leq\ve$, where $\ve=\arcsin(\rho/R_1)$ is the angular
size of the target. For the above example of a nuclear receptor seeking
for a nuclear pore (as well as in many other realistic cases) $\ve\approx
10^{-3}$, but the value may be smaller or larger in other applications.
We note parenthetically that values $\ve\approx10^{-1}$ can be indicative
of the behaviour in situations when there are many small targets present on the
surface of the inner sphere (e.g., the nuclear membrane may host about
$10^2$ nuclear pores). As a consequence, we here consider $\ve$ as an
independent parameter and explore the dependence of the PDFs on its value.
Moreover, we also consider the ratio $D_b/D_s$ as an independent parameter
and analyse how the shape of the PDF itself along with the survival
probability vary with this ratio.

\subsection{General expressions}

Consider now a ligand that starts from a fixed position $\x$, performs a
diffusive motion with bulk diffusivity $D_b$ inside the spherical-shell domain,
until it hits the inner sphere for the first time at the random time $\T^{\rm
bulk}$. Then, it diffuses along the surface for the random time $\T^{\rm surf}$
after which the ligand hits the perimeter of the target for the first time.
The fact that the search process consists of two consecutive independent
stages of durations $\T^{\rm bulk}$ and $\T^{\rm surf}$ permits us to write
the PDF $H^{\rm AD}(t;\x)$ of the event that it took the ligand exactly time
$\T=\T^{\rm bulk}+\T^{\rm surf}$ to arrive to the target for the first time
as the convolution
\begin{align}
\label{int}
H^{\rm AD}(t;\x)=\int_{\partial\Omega}d\s\int^t_0dt_1\,H^{\rm surf}(t-t_1;
\s)\,j(t_1;\s|\x),
\end{align} 
where the first integral over $d\s$ is taken over the surface $\partial
\Omega$ of the inner sphere. Here $j(t_1;\s|\x)$ denotes the joint PDF of
the event that a ligand starting from $\x$ at time $t=0$ arrived for the
\textit{first time\/} at the inner sphere at time $t_1$ and that this
first arrival occurred at point $\s$. In turn, $H^{\rm surf}(t_2|\s)$
denotes the PDF that the duration of the surface diffusion, starting from
the point $\s$ and terminating when the ligand arrives at any point on
the perimeter of the target, is equal to $t_2$. An analogous expression
for more general $n$-stage processes in one-, two- and three-dimensional
unbounded domains has been recently studied in \cite{10}.

The form of expression \eqref{int} suggests that it can be conveniently
studied by resorting to the Laplace domain with respect to $t$. Let
\begin{align}
\tilde{H}^{\rm AD}(p;\x)=\int^{\infty}_0 dt\,e^{-pt}\,H^{\rm AD}(t;\x),\qquad 
\tilde{H}^{\rm surf}(p;\s)=\int^{\infty}_0dt\,e^{-pt}H^{\rm surf}(t;\s)
\end{align}
and
\begin{align}
\tilde{j}(p;\s|\x)=\int^{\infty}_0dt\,e^{-pt}\,j(t;\s|\x) 
\end{align}
denote the Laplace-transformed PDFs. Then, multiplying both sides of
equation \eqref{int} by $\exp(-pt)$ and taking the integral, we find 
\begin{align}
\label{z}
\tilde{H}^{\rm AD}(p;\x)&=\int_{\partial\Omega}d\s\,\tilde{H}^{\rm surf}(p;
\s)\,\tilde{j}(p;\s|\x) \nonumber\\
&=R_1^2\int\limits_0^\pi d\theta'\,\sin\theta'\int\limits_0^{2\pi}d\phi'\,
\tilde{j}(p,(R_1,\theta',\phi')|(r,\theta,\phi))\,\tilde{H}^{\rm surf}(p;
\theta'),
\end{align}
where the spherical coordinates $\theta'$ and $\phi'$ define the position of
the point $\s$, at which the ligand first lands on the inner sphere, while
the coordinates $(r,\theta,\phi)$ denote the position of the starting point
$\x$. Note that due to the axial symmetry, $\tilde{H}^{\rm surf}(p;\theta')$
is independent of the azimuthal angle $\phi'$.  
 
Equation \eqref{z} is the basis for our further analysis, but it requires
knowledge of the corresponding FPTs of the intermediate stages. The exact
form of $\tilde{j}(p;\x|\s)$ has been recently studied in \cite{Grebenkov20c}
and obeys
\begin{equation}
\label{eq:jtild}
\tilde{j}(p,\s|\x)=\frac{1}{4\pi R_1^2}\sum\limits_{n=0}^\infty(2n+1)P_n
\left(\frac{(\s\cdot\x)}{|\s||\x|}\right)g_n^{(p)}(r),
\end{equation}
where  
\begin{equation}
\label{eq:gnI}
g_n^{(p)}(r)=\frac{k'_n(\alpha R_2)i_n(\alpha r)-i'_n(\alpha R_2)k_n(\alpha r)}
{k'_n(\alpha R_2)i_n(\alpha R_1)-i'_n(\alpha R_2)k_n(\alpha R_1)}  
\end{equation}
are the radial functions, $\alpha=\sqrt{p/D_b}$,
$i_n(z)=\sqrt{\pi/(2z)}I_{ n+1/2}(z)$, and $k_n(z)=\sqrt{2/(\pi
z)}\,K_{n+1/2}(z)$ are the modified spherical Bessel functions of the
first and second kind, respectively. The prime denotes a derivative
with respect to the argument, and $P_n(z)$ are Legendre polynomials
whose argument is the cosine of the angle between the vectors $\x$ and
$\s$. In particular, equation \eqref{eq:jtild} defines the moments of
the FPT on the inner sphere, conditioned by the arrival point $\s$
(see \ref{sec:j_moments}).

The above expression for $\tilde{j}(p,\s|\x)$ is valid even in the limit
$R_2\to\infty$ (e.g., when $R_2$ becomes macroscopically large, as is often the
case in cell-to-cell communication processes) when the outer reflecting
boundary practically goes to infinity and one retrieves a common setting
of a spherical target, accessed by a particle diffusing in the unbounded
space $\Omega$. Even though the domain itself is unbounded, its boundary
$\pa$ is bounded, and the above solution is still applicable. From the
asymptotic behaviour of the modified spherical Bessel functions $i_n(z)$
and $k_n(z)$, one gets immediately that the radial functions $g_n^{(p)}(
r)$ converge to
\begin{equation}
g_n^{(p)}(r)\to \frac{k_n(\alpha r)}{k_n(\alpha R_1)}.
\end{equation}
In the following, we keep focusing on the setting with a finite $R_2$.

\subsection{Surface diffusion stage}

Despite the fact that the first-passage statistics for a particle diffusing
on the surface of a spherical domain toward a target of an arbitrary size
has been studied in the past, former works focused mostly on the MFPT
\cite{adam,berg} (see also \cite{Chao81,Sano81,Prustel13,Grebenkov19sphere}
and references therein). Notable exceptions are the Refs.
\cite{Chao81,Sano81}, in which the spectral expansion of the survival
probability in time domain was derived.  While its Laplace transform
allows one to access $ \tilde{H}^{\rm surf}(p;\s)$ as a spectral
expansion, as well, we get a more compact form for this function that
is suitable for further analysis of $\tilde{H}^{\rm AD}(p;\x)$ in
equation \eqref{z}. Relegating the details of calculations to
\ref{sec:surface_diffusion} below we merely display the final results
for $ \tilde{H}^{\rm surf}(p;\s)$ and $H^{\rm surf}(t;\s)$.  When
diffusion starts within the target, one has
\begin{align}
\tilde{H}^{\rm surf}(p;\theta')=1\quad(0\leq\theta'\leq\ve);
\end{align}
in turn, for $\ve\leq\theta'\leq\pi$, one gets the remarkably compact
form
\begin{align}
\label{q}
\tilde{H}^{\rm surf}(p;\theta')=\frac{P_{\mu}(-\cos(\theta'))}{P_{\mu}(-
\cos(\ve))},\qquad\mu=\frac{-1+\sqrt{1-4pR_1^2/D_s}}{2},
\end{align}
where $P_{\mu}(z)$ is the Legendre function of order $\mu$ which
depends on the Laplace parameter $p$. Note that for $p\ll
D_s/(4R_1^2)$ (corresponding to the long-$t$ tail of the PDF), the
parameter $\mu$ is a real number, while in the opposite limit $p\gg
D_s/(4 R^2_1)$ (corresponding to the short-$t$ tail of the PDF), $\mu$
is a complex number with a real part equal to $-1/2$, such that
$P_{\mu}(z)$ is related to the so-called conical (or Mehler)
function. Expression \eqref{q} can be inverted (see
\ref{sec:surface_diffusion} for more details) to produce the spectral
expansion (see also \cite{Chao81,Sano81})
\begin{align}
\label{qqz}
H^{\rm surf}(t;\theta')&=\frac{D_s\sin^2(\ve)}{R_1^2}\sum_{n=0}^{\infty}
\left(\int^1_{-\cos(\ve)}dx\left[P_{\nu_n}(x)\right]^2\right)^{-1/2}P_{
\nu_n}(-\cos(\theta'))P'_{\nu_n}(-\cos(\ve))\nonumber\\ 
&\qquad\times e^{-\nu_n (\nu_n+1) D_s t/R^2_1},
\end{align}
where the prime, as in equation \eqref{eq:gnI}, denotes the derivative with
respect to the argument, while $\nu_n$ are the solutions of
\begin{align}
\label{roots}
P_{\nu_n}(-\cos(\ve))=0
\end{align}
organised in an ascending order. Evidently, $\nu_n$ are functions of $\ve$.
Eventually, we note that the moments of $H^{\rm surf}(t;\theta')$ can be
found directly from expression \eqref{q} by differentiating it with
respect to the Laplace parameter $p$. In particular, differentiating
equation \eqref{q} once and twice with respect to $p$ and setting $p=0$,
one finds first two moments of the FPT $\T^{\rm surf}$ for $\ve\leq\theta'
\leq\pi$,
\begin{align}
\label{eq:Tsurf1}
\langle\T^{\rm surf}\rangle&=\frac{R_1^2}{D_s}\ln\left(\frac{1-\cos(\theta')
}{1-\cos(\ve)}\right),\\
\langle(\T^{\rm surf})^2\rangle&=\frac{2R_1^4}{D_s^2}\left[{\rm Li}_2\left(
\frac{1+\cos(\ve)}{2}\right)-{\rm Li}_2\left(\frac{1+\cos(\theta')}{2}
\right)+\ln\left(\frac{1-\cos(\ve)}{1- \cos(\theta')}\right)\right.
\nonumber\\
\label{eq:Tsurf2}
&\left.+\ln^2\left(\frac{1-\cos(\ve)}{2}\right)-\ln\left(\frac{1-\cos(\ve)}
{2}\right)\ln\left(\frac{1-\cos(\theta')}{2}\right)\right],
\end{align}
where ${\rm Li}_2(z)=-\int\nolimits_0^zdx\,\ln(1-x)/x$ is the dilogarithm.
Note that the MFPT was known (see, e.g., \cite{Sano81}), while the result
for the second moment has not been reported.

\begin{figure}
\begin{center}
\includegraphics[width=10cm]{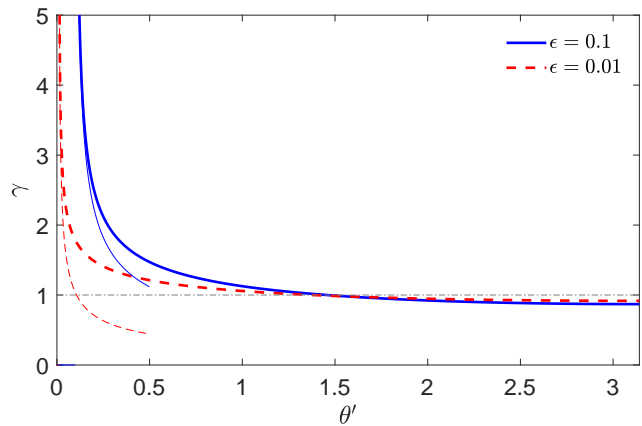}
\end{center}
\caption{Coefficient of variation $\gamma=\sqrt{{\rm var}(\T^{\rm surf})}
/\langle\T^{\rm surf}\rangle$ as function of $\theta'$ for the FPT to the
target of angular size $\ve$ by surface diffusion, for two values of $\ve$.
Thick lines represent the exact result from equations \eqref{eq:Tsurf1} and
\eqref{eq:Tsurf2}, while the thin lines show the asymptotic relation
\eqref{eq:gamma_asympt}.}
\label{fig:gamma}
\end{figure}

These expressions allow us to determine the variance ${\rm var}(\T^{\rm surf})
=\langle(\T^{\rm surf})^2\rangle-\langle\T^{\rm surf} \rangle^2 $ of the PDF,
and also to characterise its broadness (see \cite{carlos1,carlos2}) from the
corresponding coefficient of variation $\gamma=\sqrt{{\rm var}(\T^{\rm surf})
}/\langle\T^{\rm surf}\rangle$, which is defined as the ratio of the standard
deviation around the mean value and the mean value itself. Inspecting the
behaviour of $\langle\T^{\rm surf}\rangle$ and $\langle(\T^{\rm surf})^2
\rangle$, we realise that for fixed (sufficiently small) $\ve$ and $\theta'$
sufficiently close to $\ve$, $\langle \T^{\rm surf}\rangle\simeq2R_1^2(\theta'
-\ve)/(\ve D_s)$ and $\sqrt{{\rm var}(\T^{\rm surf})}\simeq2R_1^2\sqrt{(2\ln
2-1-2\ln\ve)(\theta'-\ve)}/(\sqrt{\ve}D_s)$. Combining these expressions, we
find that for $\theta'$ close to (small) $\ve$ the coefficient of variation
obeys
\begin{align}
\label{eq:gamma_asympt}
\gamma\sim\sqrt{\frac{(2\ln2-1-2\ln\ve)\,\ve}{\theta'-\ve}},
\end{align}  
and hence, it diverges when $\theta'\to\ve$. Overall, $\gamma$ is a
monotonously decreasing function of $\theta'$ that attains its minimal
value (for small enough $\ve$)
\begin{equation}
\gamma_{\rm min}\approx\frac{\sqrt{\pi^2/3-1+(1-2\ln(2/\ve))^2}}{2\ln(2/\ve)}
\end{equation}
for $\theta'=\pi$, i.e., when the starting point $\s$ of the particle is
located on the South pole (figure \ref{fig:gamma}). In the limit $\ve\to0$,
this minimal value approaches unity from below, i.e., $\gamma_{\rm min}\leq1$.
As a consequence, for any $\ve>0$, there exists a value $\theta^*_\ve$, at
which $\gamma=1$. For $\theta'<\theta^*_\ve$, the standard deviation of the
FPT is larger than the mean value $\langle\T^{\rm surf}\rangle$ such that
the latter cannot be used to characterise the search process exhaustively well.

\subsection{Asymptotic behaviour of equation \eqref{q}}
\label{sec:Hsurf_asympt}

Before we turn to the discussion of the ADS it is instructive to discuss
the asymptotic behaviour of its constituents: $\tilde{j}(p,\s|\x)$ in
equation \eqref{eq:jtild} and $\tilde{H}^{\rm surf}(p;\theta')$ in
equation \eqref{q}. We will show that both exhibit some non-trivial and
universal behaviour, which will permit us to reach several conclusive
statements about the asymptotic behaviour of the full PDF $H^{\rm AD}(t;
\x)$.

We start with $\tilde{H}^{\rm surf}(p;\theta')$ and consider first the
limit $p\to\infty$, which corresponds to the short-$t$ tail of the
associated PDF. In \ref{sec:surface_diffusion} we show that in this limit
$\tilde{H}^{\rm surf}(p;\theta')$ obeys
\begin{align}
\label{zq}
\tilde{H}^{\rm surf}(p;\theta')\simeq\sqrt{\frac{\sin(\ve)}{\sin(\theta')}}
\,e^{-R_1(\theta'-\ve)\sqrt{p/D_s}}\quad(p\to\infty)\,,
\end{align}
where the symbol $\simeq$ denotes that we consider the leading order of
this limiting behaviour. Inverting the Laplace transform in the latter
expression, we find that the short-$t$ asymptotic behaviour of the PDF
$H^{\rm surf}(t;\theta')$ follows
\begin{align}
\label{ls}
H^{\rm surf}(t;\theta')\simeq\sqrt{\frac{\sin(\ve)}{\sin(\theta')}}\,
\frac{(\theta'-\ve)R_1}{\sqrt{4\pi D_st^3}}\,e^{-(\theta'-\ve)^2R_1^2
/(4D_st)}.
\end{align}
Remarkably, this result coincides (up to the factor $\sqrt{\sin(\ve)/\sin
(\theta')}$) with the celebrated L\'evy-Smirnov distribution \cite{150,151},
defining the exact FPT PDF from a point $x_0=(\theta'-\ve)R_1>0$ to a perfect
target located at the origin of a \textit{one-dimensional\/} semi-infinite
line. Noticing that $R_1(\theta' -\ve)$ is, in fact, the geodesic distance
between the point $\s$ and the boundary of the target region, i.e., the
\textit{shortest\/} path along the surface of the inner sphere from the
starting point $\s$ to the target, its physical significance becomes
apparent: the short-$t$ tails of the PDF are dominated by such
trajectories which move diffusively along the shortest distance to the
target.  We note that such a L\'evy-Smirnov-like form, which is
specific to one-dimensional situations, has been previously evidenced
for rather different two- and three-dimensional geometrical settings
(see, e.g., \cite{dist1,dist4,dist2,alj1,alj2,baruch}) and therefore
seems to be quite generic. In particular, this result is analogous to
the "geometry control" unveiled previously in the case of pure bulk
diffusion in $\Omega$ \cite{alj1,dist1}.

In turn, the leading long-$t$ behaviour of the PDF $H^{\rm surf}(t;\theta')$
can be readily deduced from the series representation in equation \eqref{qqz}.
Recalling that the solutions $\nu_n$ form an increasing sequence, we realise
that at long times the dominant contribution corresponds to the smallest root
$\nu_0$ of equation \eqref{roots} and, therefore, the long-$t$ behaviour reads
\begin{align}
\label{q3}
H^{\rm surf}(t;\theta')&\simeq\frac{\sin^2(\ve)D_s}{R_1^2}\left(\int^1_{-\cos
(\ve)}dx\left[P_{\nu_0}(x)\right]^2\right)^{-1/2}P_{\nu_0}(-\cos(\theta'))
P'_{\nu_0}(-\cos(\ve))\nonumber\\
&\quad\times e^{-\nu_0(\nu_0+1)D_st/R^2_1}.
\end{align}
We note that $\nu_0$ depends on $\ve$ but is independent of $\theta'$; in
particular when $\ve$ is small, one has $\nu_0\approx1/(2\ln(2/\ve))$
\cite{Grebenkov19sphere}. As a consequence the longest relaxation time
\begin{equation}
\label{eq:tau_surf}
\tau^{\rm surf}=\frac{R^2_1}{\nu_0(\nu_0+1)D_s}\approx\frac{2R_1^2\ln(2
/\ve)}{D_s}\quad(\ve\ll1)
\end{equation}
is independent of the starting point, i.e., corresponds to such long times
at which a particle arrives to the target region after extensive
target-avoiding excursions and thus loses information about its initial
location. The latter is kept only through the amplitude in equation \eqref{q3}.

\subsection{Asymptotic behaviour of equation \eqref{eq:jtild}}

\begin{figure}
\begin{center}
\includegraphics[width=7.6cm]{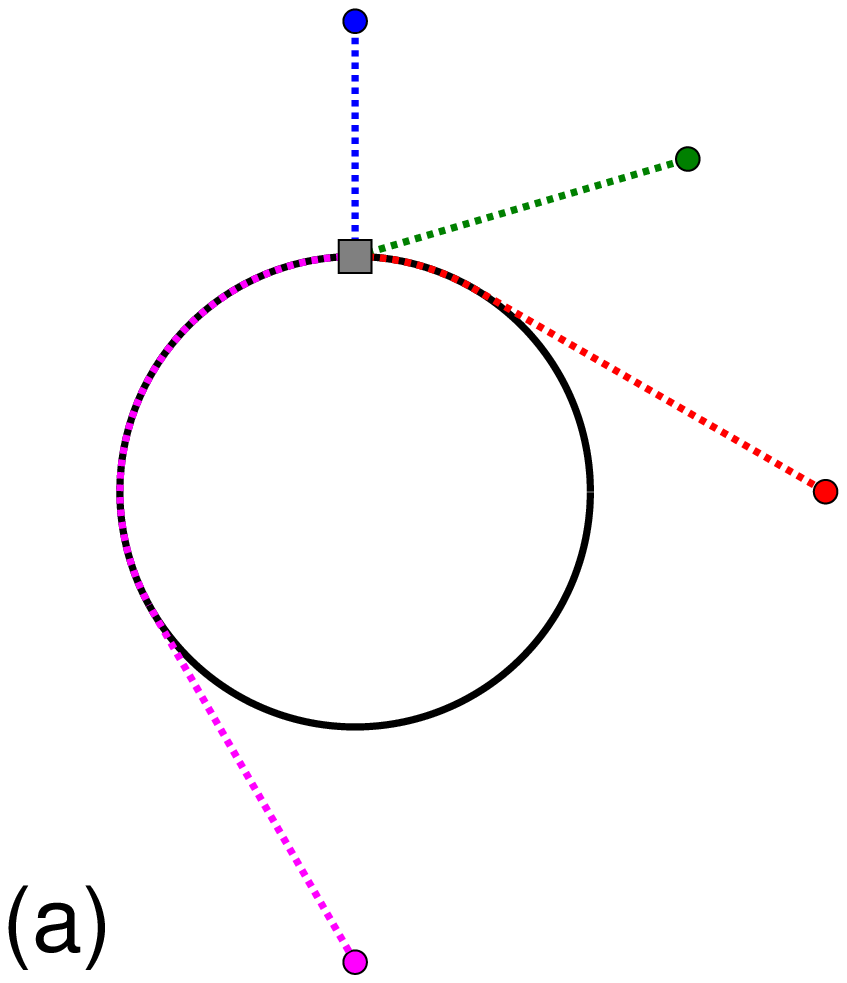}
\includegraphics[width=7.6cm]{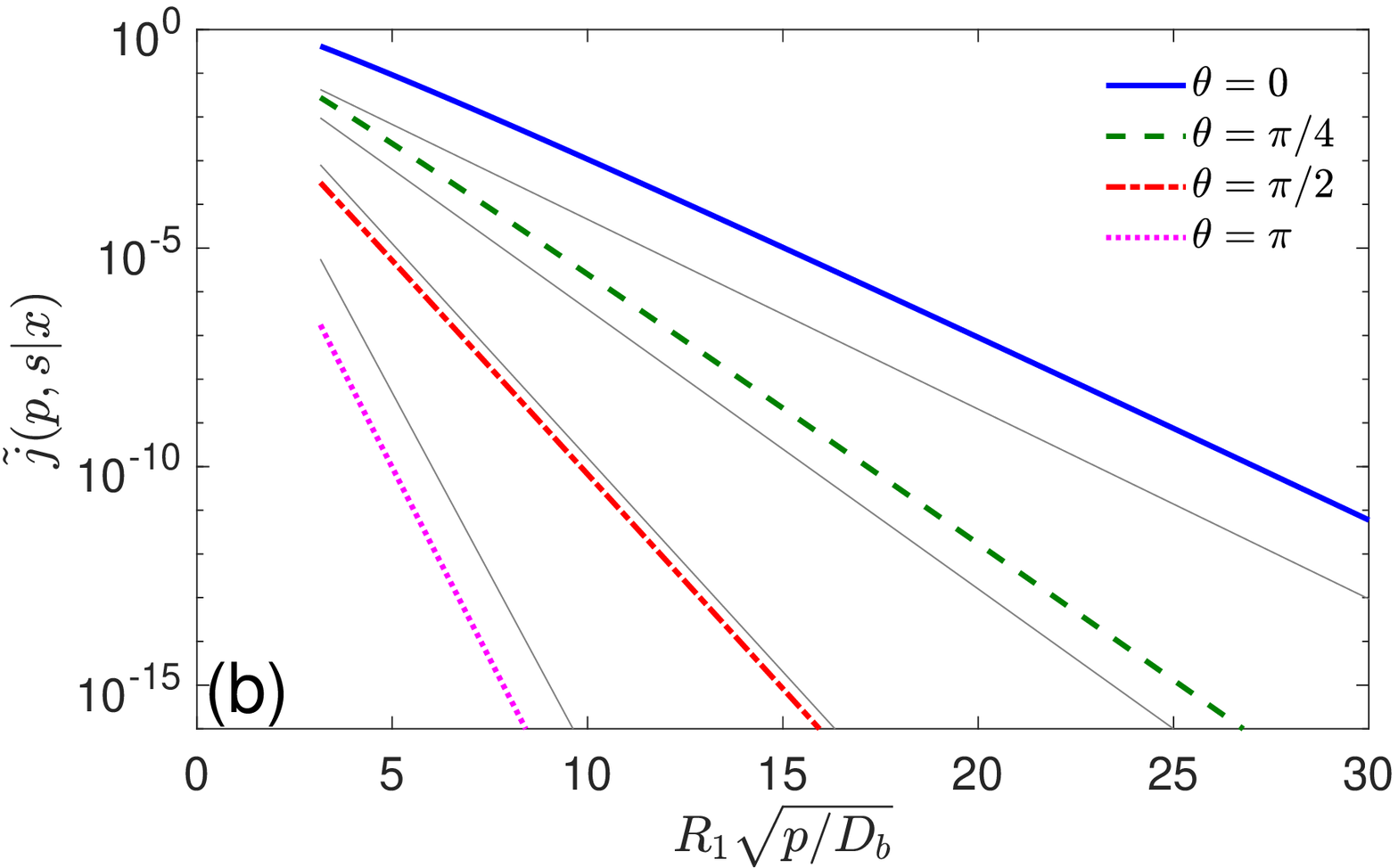}
\end{center}
\caption{Large-$p$ asymptotic behaviour of $\tilde{j}(p,\s|\x)$ in equation
\eqref{eq:jtild} for $\phi=\phi'$. Panel (a): Shortest distances to the arrival
point $\s$ located on the North pole (filled squares) from different starting
points $\x$ (filled circles). Blue ($\theta=0$) and green ($\theta = \pi/4$)
circles correspond to the case when $\s$ is directly visible from $\x$,
such that the shortest distance ${\ell}_{\rm vis}$ (see the corresponding
dotted lines) is given by the straight segment connecting the initial point
and the arrival point $\s$. Red ($\theta=\pi/2$) and magenta ($\theta=\pi$)
circles correspond to the situation in which the target appears on the
"dark" side of the inner sphere with respect to the initial point. In this
case the shortest distance ${\ell}_{\rm invis}$ consists of two parts:
A straight segment of length $h=\sqrt{r^2- R_1^2}$ from the initial point
to the point closest to the target on the horizon, and an arc from this
point to the target. Panel (b): Numerical analysis of $\tilde{j}(p,\s|\x)$
in equation \eqref{eq:jtild} as function of $R_1 \sqrt{p/D_b}$. Coloured
thick curves present $\tilde{j}(p,\s|\x)$ obtained by summing the series
in equation \eqref{eq:jtild} truncated at $n_{\rm max}=100$. As we are
interested in the large-$p$ behaviour, we set $R_2=\infty$ here to ease the
numerical computation of the radial functions $g_n^{(p)}(r)$.  Thin grey
lines show the asymptotic forms in equation \eqref{vis} and \eqref{invis}
with the shortest distances calculated for the corresponding geometrical setup
(note that the unknown functions $a(\theta,\theta')$ and $b(\theta,\theta')$,
controlling a vertical offset of these lines, were set to $1$).}
\label{fig2}
\end{figure}  

Let us now turn to the joint probability density $j(t,\s|\x)$. A
theoretical analysis of the large-$p$ asymptotic behaviour of expression
\eqref{eq:jtild} appears to be a difficult task; we realise that, in fact,
in order to determine the corresponding asymptotic form one has to perform
the sum in this equation exactly, which requires a cumbersome analysis. We
thus resort instead to a numerical analysis of expression \eqref{eq:jtild}
aiming to verify the following conjecture: we assume that, in line with the
"geometrical optics" arguments presented earlier in \cite{smit} (see section
II) and similarly to the above considered case of a search process on the
surface of a sphere, the short-$t$ tail of the associated PDF corresponds
to the diffusive motion along the shortest path (with length ${\ell}$) from
$\x=(r,\theta,\phi)$ to $\s=(R_1,\theta',\phi')$.  This path, however, has
a different form depending on whether the point $\s$ is directly "visible"
from $\x$, i.e., is located within a spherical cap region delimited by the
horizon, or whether it is located outside of this area being situated on the
"dark" side of the inner sphere and therefore invisible from 
$\x$. In case when $\s$ is visible from $\x$ we thus expect that
\begin{align}
\label{vis}
\tilde{j}(p,\s|\x)\simeq a(\theta,\theta')e^{-{\ell}_{\rm vis}\sqrt{p/D_b}}
\quad(p\to\infty),
\end{align}
where $a(\theta,\theta')$ is some unknown amplitude (recall that the result
in equation \eqref{ls} differs from the standard L\'evy-Smirnov density
by some function) and ${\ell}_{\rm vis}$ is the Euclidean distance from
$\x$ to $\s$. In turn, for the case when $\s$ is located on the "dark"
side, the shortest path ${\ell}_{\rm invis}$ consists of two parts (see
also \cite{smit}): the particle diffuses along a straight line of length
$h=\sqrt{r^2-R_1^2}$ connecting $\x$ and the point on the horizon, which
is closest to $\s$, and then travels diffusively in the immediate vicinity
of the inner sphere along the arc connecting this point on the horizon
and $\s$. Consequently, in this case we expect that $\tilde{j}(p,\s|\x)$
behaves in the limit $p\to\infty$ as
\begin{align}
\label{invis}
\tilde{j}(p,\s|\x)\simeq b(\theta,\theta')e^{-{\ell}_{\rm invis}\sqrt{p/D_b}}
\quad(p\to\infty),
\end{align}
where $b(\theta,\theta')$ is an unknown amplitude. Note that ${\ell}_{\rm
vis}$ and ${\ell}_{\rm invis}$ can be simply expressed through the Cartesian
coordinates of both points.

%{\clr [DG: Should we discuss whether or not the amplitudes $a$ and $b$
%depend on $p$?  Our notation $a(\theta,\theta')$ and
%$b(\theta,\theta')$ suggests that there is no dependence on $p$.
%However, the exponential form $e^{-\ell \sqrt{p/D_b)}}$ only captures
%the leading behavior, while power-law corrections are possible.  Our
%numerical results suggest that there might be no correction, but it is
%hard to state with certainty.  Or just leave it as it is?]}

In figure \ref{fig2} we compare our conjectured equations \eqref{vis} and
\eqref{invis} (thin grey lines) and the exact expression (thick coloured
curves) obtained from series truncation in equation \eqref{eq:jtild} for
two situations in which the target is visible from the starting point, and
two situations in which it is located on the dark side. We observe that
in all four cases the slopes of thin grey lines and of the thick coloured
curves are almost identical---which thus confirms our conjecture of the
large-$p$ behaviour of $\tilde{j}(p;\s|\x)$. Inverting the expressions in
equations \eqref{vis} and \eqref{invis}, we thus find that the short-$t$
behaviour of the corresponding PDF $j(t;\s|\x)$ has the L\'evy-Smirnov form
${\ell}\exp(-{\ell}^2/(4D_bt))/\sqrt{4\pi D_bt^3}$, up to a yet unknown
amplitude factor $a(\theta,\theta')$ or $b(\theta,\theta')$, respectively.

Lastly, we discuss the long-$t$ behaviour of the PDF $j(t,\s|\x)$.
As diffusion occurs in a bounded domain, $j(t,\s|\x)$ clearly exhibits
an exponential decay which is controlled by the pole $p_0^{\rm bulk}$ of
$\tilde{j}(p,\s|\x)$ with the smallest absolute value. This is actually the
pole of the radial function $g_0^{(p)}(r)$, which can be written as
\begin{equation}
\label{eq:p0_bulk}
p_0^{\rm bulk}=-\frac{1}{\tau^{\rm bulk}}<0,\mbox{ with }\tau^{\rm bulk}=
\frac{(R_2-R_1)^2}{D_b\alpha_0^2},
\end{equation}
where $\alpha_0$ is the smallest root of the transcendental equation (see,
e.g., \cite{dist1})
\begin{equation}
\frac{\tan(\alpha_0)}{\alpha_0}=\frac{R_2}{R_2-R_1}.
\end{equation}
When $R_1\ll R_2$, this solution behaves as $\alpha_0^2\approx R_1/(3(R_2-
R_1))$ and thus 
\begin{equation}
\tau^{\rm bulk}\approx\frac{R_2^3}{3D_bR_1}. 
\end{equation}
Therefore, in the long-$t$ limit the PDF $j(t,\s|\x)$ decays as
\begin{align}
\label{eq:jt_long}
j(t,\s|\x)\propto e^{-t/\tau^{\rm bulk}},
\end{align}
where $\tau^{\rm bulk}$ is the longest time characterising the FPT PDF
to the inner sphere conditioned by the constraint that this event took
place at point $\s$. Note that $\tau^{\rm bulk}$ is independent of
both the initial position $\x$ and the precise location of the arrival
point $\s$.  {\clr Some other properties of the probability flux
density $j(t,\s|\x)$ were discussed in a recent paper
\cite{Antoine22}.}

\section{Probability density function of the first-passage times
within the Adam-Delbr\"uck scenario}
\label{pdf}

In this section we focus on the statistics of the FPTs in the ADS. We
start from our main relation \eqref{z}, in which the integral over the
first arrival point $\s$ on the inner sphere can be further simplified
(see details in \ref{sec:AD}) to get the exact and fully explicit
solution {\clr in the Laplace domain}
\begin{align}
\nonumber
\tilde{H}^{\rm AD}(p;\x)&=\sum\limits_{n=0}^\infty g_n^{(p)}(r)P_n(\cos
\theta)\biggl\{\frac{P_{n-1}(\cos\ve)-P_{n+1}(\cos\ve)}{2}\\
\label{eq:Hp_final}
&-(1-\cos^2\ve)\frac{n+1/2}{\frac{pR^2}{D_s}+n(n+1)}\,\biggl(P'_n(\cos\ve)
+P_n(\cos\ve)\frac{P'_{\mu}(a)}{P_{\mu}(a)}\biggr)\biggr\}.
\end{align}
We first discuss the short-$t$ and long-$t$ asymptotic behaviour of the PDF
$H^{\rm AD}(t;\x)$. Next, numerically inverting the Laplace transform in
equation \eqref{eq:Hp_final} by help of the Talbot algorithm we discuss the
behaviour of the PDF in the time domain for different values of the system
parameters and compare it against the recently obtained PDF for the direct
search scenario in precisely the same geometrical settings \cite{9}. This
will give us a general idea of the shapes of two PDFs. Moreover, we turn
to the integrated characteristic---the survival probability
\begin{align}
\label{surv}
S^{\rm AD}(t;\x)=\int\limits_t^{\infty}dt'\,H^{\rm AD}(t';\x),
\end{align}
i.e., the probability that the target is not found up to time $t$. The
analysis of the survival probability for both search scenarios provides a
full understanding of the actual efficiency of each scenario and, hence, 
gives an idea which scenario is more successful. In the Laplace domain,
the survival probability in equation \eqref{surv} reads $\tilde{S}^{\rm
AD}(p;\x)=(1-\tilde{H}^{\rm AD}(p;\x))/p$, allowing us to determine its
asymptotic properties from those for the Laplace-transformed PDF. The
analysis of the survival probability will permit us to make several
conclusive statements. Lastly, we consider the particular, experimentally
relevant case when the starting point is uniformly distributed on a
spherical surface of radius $r$ such that $0<R_1\leq r\leq R_2$. Moreover,
in \ref{sec:AD} we present additional figures illustrating the behaviour
of $H^{\rm AD}(t';\x)$ for several \textit{fixed\/} starting points of the
ligand.

\subsection{Asymptotic behaviour}

To access the short-$t$ behaviour of $H^{\rm AD}(t;\x)$ we focus
on expression \eqref{z} together with equations \eqref{zq}, \eqref{vis},
and \eqref{invis}. Combining these expressions, we get the
Laplace-transformed PDF $ \tilde{H}^{\rm AD}(p;\x)$ in the limit $p
\to\infty$ in the form
\begin{align}
\label{mm}
\tilde{H}^{\rm AD}(p;\x)&\approx\sqrt{\sin(\ve)}\int_{\partial\Omega_{\rm
vis}}d\s\frac{a(\theta,\theta')}{\sqrt{\sin(\theta')}}\exp\left(-\left(
\frac{{\ell}_{\rm vis}(r,\theta,\theta')}{\sqrt{D_b}}+\frac{(\theta'-\ve)
R_1}{\sqrt{D_s}}\right)\sqrt{p}\right)+\nonumber\\
&\sqrt{\sin(\ve)}\int_{\partial\Omega_{\rm invis}}d\s\frac{b(\theta,
\theta')}{\sqrt{\sin(\theta')}}\exp\left(-\left(\frac{{\ell}_{\rm invis}
(r,\theta,\theta')}{\sqrt{D_b}}+\frac{(\theta'-\ve)R_1}{\sqrt{D_s}}\right)
\sqrt{p}\right),
\end{align}
where $\partial\Omega_{\rm vis}$ and $\partial\Omega_{\rm invis}$ denote
the two parts of the inner sphere, those that are "visible" and "invisible"
as seen from the starting point $\x$. In the limit $p\to\infty$ the
integrands in equation \eqref{mm} vanish exponentially fast with $p$. This
signifies that the dominant contribution to the integrals comes from such
values of the position $\s$ of the landing point onto the inner sphere for
which the coefficients in front of $\sqrt{p}$ are minimal, i.e.,
\begin{align}
\label{mm2}
\tilde{H}^{\rm AD}(p;\x)\sim\exp\left(-\ell_{\rm min}\sqrt{p/D_b}\right),
\end{align}
where
\begin{equation}
\ell_{\rm min}=\min\limits_{\theta'}\left\{\ell(r,\theta,\theta')+R_1
(\theta'-\ve)\sqrt{D_b/D_s}\right\},
\end{equation}
where $\ell(r,\theta,\theta')$ is either $\ell_{\rm vis}$ or $\ell_{\rm
invis}$, depending on the mutual arrangement of $\x$ and $\s$. As a
consequence, we expect that the short-$t$ tail of $H_{\rm AD}(t;\x)$ has
a universal L\'evy-Smirnov form, yet with a position-specific prefactor.

As diffusion occurs in a bounded domain, the PDF $H^{\rm AD}(t;\x)$
decays exponentially fast at long times, and the decay rate is determined
by the pole $p_0<0$ of $\tilde{H}^{\rm AD}(p;\x)$ with the smallest
absolute value. Since $\tilde{H}^{\rm AD}(p;\x)$ is obtained in equation
\eqref{z} by integrating the product of $\tilde{j}(p,\s|\x)$ and $\tilde{
H}^{\rm surf}(p;\s)$ over $\s$, $p_0$ is the pole of one of these two
functions. Skipping technical details presented in \ref{sec:long-time} we
conclude that
\begin{align}
\label{mm3}
H^{\rm AD}(t;\x)\simeq C_\ve(\x)\exp\left(-\frac{t}{{\rm max}\left\{\tau^{
\rm bulk},\tau^{\rm surf}\right\}}\right),
\end{align}
where the amplitude $C_\ve(\x)$ is explicitly computed in equations
\eqref{eq:Cve_surf} and \eqref{eq:Cve_bulk}. Here ${\rm max}\left\{\tau^{
\rm bulk},\tau^{\rm surf}\right\}$ signifies that the onset and the decay
of the long-$t$ asymptotic form is entirely controlled by the longest of
the two characteristic times $\tau^{\rm bulk}$ and $\tau^{\rm surf}$, but
\textit{not by their sum}, which, in contrast, is assumed in the conventional
criterion of the applicability of the ADS. We finally note that the
decay rate in equation \eqref{mm3} is independent of the starting point
$\x$ for any relation between $\tau^{\rm bulk}$ and $\tau^{\rm surf}$.

\subsection{Random starting point}
\label{sec:random}

In many situations of practical interest the launch of the ligand does not
take place from a fixed prescribed position. Instead there are many points
on the outer sphere from which the ligand can start its search for the target.
In this regard it is instructive to consider the case when $\x$ can be any
(uniformly-distributed) point on the spherical surface of radius $r$ (not
necessarily the outer boundary, $r=R_2$). The behaviour of the FPT PDF
corresponding to the case when the ligand starts from a prescribed position
turns out to be quite similar, as discussed in \ref{starting}. 

The Laplace-transformed PDF for a random starting point, which we denote as
$\overline{\tilde{H}_{\rm AD}(p;r)}$, is obtained by integration of equation
\eqref{eq:Hp_final} over the angular coordinates of $\x$, yielding
\begin{align}
\overline{\tilde{H}^{\rm AD}(p;r)}&=\frac{1}{4\pi r^2}\int_{|\x|=r}d\x\,
\tilde{H}^{\rm AD}(p;\x)\nonumber\\
\label{eq:Hsurfav}
&=\frac{1-\cos(\ve)}{2}g_0^{(p)}(r)\left(1-\frac{D_s (1+\cos(\ve))}{pR_1^2}
\frac{P'_{\mu}(-\cos(\ve))}{P_{\mu}(-\cos(\ve))}\right),
\end{align} 
where $g_0^{(p)}(r)$ and $\mu$ were defined in equations \eqref{eq:gnI} and
\eqref{q}, respectively. We stress that this surface-averaged quantity
differs from the volume average over a uniformly distributed starting point
inside the confining domain $\Omega$.

The asymptotic behaviour of $\overline{\tilde{H}^{\rm AD}(p;r)}$ is derived
in \ref{sec:AD} and reads
\begin{align}
\nonumber
\overline{H^{\rm AD}(t;r)}&\simeq\frac{R_1}{r}\,\frac{(r-R_1)}{\sqrt{4\pi
D_bt^3}}\exp\left(-\frac{(r-R_1)^2}{4D_bt}\right)\\
\label{lm00}
&\times\biggl(\frac{1-\cos(\ve)}{2}+\frac{\sin(\ve)\sqrt{D_sD_bt^2}}{R_1
(r-R_1)}+\frac{\cos(\ve)D_bD_st^2}{R_1^2(r-R_1)^2}\biggr).
\end{align}
The prefactor in the first line is the PDF of the FPT to a perfectly
reactive sphere of radius $R_1$ by bulk diffusion in a
three-dimensional unbounded domain.  At very short times, the first
term is dominant, and one gets, up to a geometric prefactor $(1-\cos(
\ve))R_1/(2r)$, the L\'evy-Smirnov function of $t$,
\begin{align}
\label{lm}
\overline{H^{\rm AD}(t;r)}\simeq\frac{(1-\cos(\ve))R_1}{2r}\,\frac{(r-
R_1)}{\sqrt{4\pi D_bt^3}}\exp\left(-\frac{(r-R_1)^2}{4D_bt}\right).
\end{align}
Interestingly, this short-$t$ tail of the PDF is independent of the
surface diffusion coefficient $D_s$. Apparently, this is associated
with the fact that when we average over the starting point, the major
contribution to the PDF, which entirely defines its asymptotic
behaviour, comes from those initial locations that are placed directly
over the target site. The contributions from other starting points,
for which the search process will include a tour of surface diffusion,
is only sub-dominant in this limit. However, this leading-order
contribution is attenuated by the prefactor $1-\cos(\ve)\approx
\ve^2/2$, which can be very small for small targets. In turn, the second and
third terms in equation \eqref{lm00}, although being subleading in powers of
time, are weighted by $\sin(\ve)\approx\ve$ and by $\cos(\ve)\approx1$,
respectively. This means that when $\ve$ is sufficiently small, there exists
an intermediate range of times, for which the third term is actually the
dominant one,
\begin{equation}
\overline{H^{\rm AD}(t;r)}\simeq\frac{D_s\sqrt{D_bt}}{\sqrt{4\pi}\,rR_1(r-
R_1)}\exp\left(-\frac{(r-R_1)^2}{4D_bt}\right).
\label{lm000}
\end{equation}
Curiously, this expression is independent of the target size (as soon
as $\ve$ is small).  This is due to the fact that a decrease of $\ve$
reduces the relative contribution of trajectories that arrive to the
target directly from the bulk and thus avoid surface diffusion.
Hence, for small $\ve$ the search process will necessitate surface
diffusion.

In the opposite long-time limit, there is no difference between the
fixed and random starting point so that our former asymptotic
relation \eqref{mm3} is valid for $\overline{H^{\rm AD}(t;r)}$.

\begin{figure}
\begin{center}
\includegraphics[width=100mm]{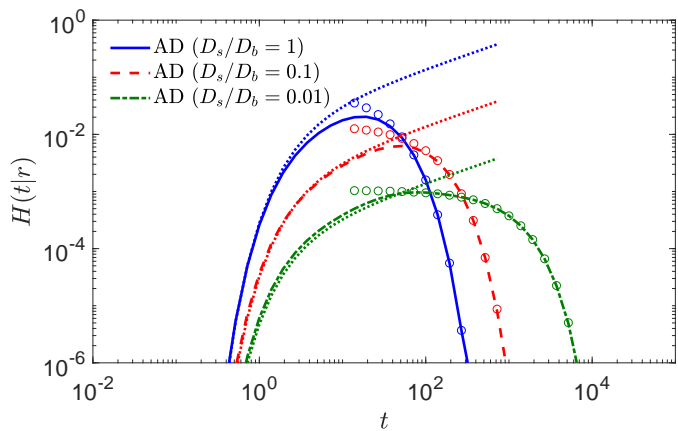}
\end{center}
\caption{
FPT PDF $\overline{H^{\rm AD}(t;r)}$ to a target of the small angular
size $\ve=0.01$ within the ADS, with $R_2/R_1=5$, for which the
starting point is uniformly distributed over the outer sphere
($r/R_1=5$). The three thick coloured curves show the PDF obtained
from the exact solution \eqref{eq:Hsurfav} via numerical inversion of
the Laplace transform.  Dotted lines represent the short-time
asymptotic relation \eqref{lm00}, whereas circles show the long-time
asymptotic relation \eqref{mm3}.  {\clr The leading-order
L\'evy-Smirnov-type relation \eqref{lm} is not shown because it lies
below the bottom limit $10^{-6}$ of the plot; in fact, its maximal
amplitude is $2.89\times10^{-7}$, i.e., this approximation fails by
4 to 5 orders of magnitude.}  The length and time units are fixed by
setting $R_1=1$ and $D_b=1$.}
\label{fig:Ht_asympt}
\end{figure}

Figure \ref{fig:Ht_asympt} illustrates the asymptotic behaviour of the
surface-averaged PDF $\overline{H^{\rm AD}(t;r)}$ within the ADS.  One
sees that the asymptotic relation \eqref{lm00} accurately describes
the short-time behaviour. As the target is rather small, the dominant
contribution comes from the third term in equation
\eqref{lm00}; in particular, the three asymptotic curves only differ
by the multiplicative factor $D_s$. We stress that the first term in
equation \eqref{lm00} in this setting is totally irrelevant here (if
one only kept the first term, the asymptotic curves would be below
$10^{-6}$ and thus invisible in this figure; for this reason, they are
not shown). {\clr In other words, the leading-order
L\'evy-Smirnov-type relation \eqref{lm} fails, and one has to rely on
our asymptotic relation \eqref{lm00}, in which the third, sub-leading
term is dominant.}  The long-time asymptotic relation \eqref{mm3} is
also very accurate. Note that the amplitude $C_\ve(r)$ is given by
equation \eqref{eq:Cve_bulk} for the curve with $D_s/ D_b=1$, for
which $\tau^{\rm bulk}>\tau^{\rm surf}$; in turn, it is given by
equation \eqref{eq:Cve_surf} for the two other curves with $D_s/
D_b=0.1$ and $D_s/D_b=0.01$. We conclude that both short- and
long-time asymptotic relations are accurate.

The derivative of $\overline{\tilde{H}^{\rm AD}(p;r)}$ in equation
\eqref{eq:Hsurfav} with respect to $p$ determines the surface-averaged
MFPT. Skipping the details of this computation (see \ref{sec:AD}), the
final result reads
\begin{equation}
\label{eq:Tsurfav}
\overline{T^{\rm AD}}=\frac{(r-R_1)(2R_2^3-rR_1(r+R_1))}{6rR_1D_b}
+\frac{R_1^2}{2D_s}\biggl(2\ln\biggl(\frac{2}{1-\cos\ve}\biggr)-(1+
\cos\ve)\biggr).
\end{equation}
Here, the first term is the MFPT to the inner sphere, which is
evidently independent of the target size. In turn, the second term is
the contribution from the surface diffusion towards the target,
averaged over the distribution of the first arrival point onto the
surface (the so-called harmonic measure density). This contribution is
independent of the radius $R_2$ of the outer sphere, as well as of the
radial coordinate $r$ of the starting point.  Expectedly, the second
term vanishes as $\ve\to\pi$ (the target covers the whole inner
sphere) and diverges logarithmically as the target shrinks,
$\ve\to0$. Note that if $r\gg R_1$, the diffusing ligand has
sufficient time before hitting the inner sphere to loose the memory on
its starting point, so that $T^{\rm AD}(\x)\approx
\overline{T^{\rm AD}}$. In other words, the explicit relation
\eqref{eq:Tsurfav} can be used to estimate the MFPT from a fixed
starting point $\x$ when it is located far from the inner sphere.
Setting $r=R_2$, $R_1\ll R_2$ and $\ve\ll1$, one can easily check
that $\overline{T^{\rm AD}}$ coincides, to the leading order, with
$\tau_{\rm AD}$, which was discussed in section \ref{sec:intro}.

\subsection{Comparison of the two search scenarios}
\label{comp}

We now compare the efficiency of the two search processes: ADS versus
the direct search scenario. We focus on the case when the starting point
is uniformly distributed over the outer sphere, for three reasons. (i)
As discussed earlier, we keep in mind applications to microbiology, in
which a particle enters the cytosol from the plasma membrane and searches
for a nuclear pore; here, the spatial (angular) locations of the entrance
channel and the nuclear pore are generally not known and can thus be modelled
as random. (ii) When the inner sphere is small in comparison to the outer
sphere (i.e, $R_1\ll R_2=r$), the particle has enough time to diffuse before
hitting the inner sphere, and the information on the precise location of a
fixed starting point is generally lost (except for very short trajectories
determining the left, short-$t$ tail of the PDF)---in other words, in this
setting, there is no notable difference between fixed and random starting
points. (iii) The average over angular coordinates of the starting point
eliminates all terms in the series representations \eqref{eq:Hp_final} of
the PDF, except for $n=0$, that facilitates its numerical computation and
reduces eventual truncation errors (when an infinite series is replaced by a
finite sum)---this is particularly relevant in case of a small target when
one would have to keep a large number of terms to get accurate results,
whereas the computation of the radial functions $g_n^{(p)}(r)$ may be
problematic for large $n$. In \ref{sec:AD}, we briefly discuss the case of
a fixed starting point---which is conceptually important---and compare it
to our main conclusions here for a random starting point.

The precise form of the Laplace transform $\overline{\tilde{H}^{\rm
AD}(p; r)}$ is given in equation \eqref{eq:Hsurfav}. While the exact
solution $\overline{\tilde{H}^{\rm dir}(p;r)}$ for the direct search
scenario was derived in \cite{9}, we use the approximate relations
\eqref{eq:J_inner} and \eqref{eq:Happ_inner_surfav} for our
discussion {\clr (see details in \ref{direct}).}  Note that the latter
have a much simpler and explicit form, as compared to the exact
solution (which requires a numerical inversion of matrices), and are
remarkably accurate, as shown in
\cite{9}. In both cases the PDFs in the time domain are obtained via
numerical inversion of the corresponding Laplace transforms using the
Talbot algorithm. We set the angular size of the target equal to
$\ve=0.01$, which is approximately ten times bigger than the angular
size of a typical nuclear pore. This choice of a larger value of $\ve$
is due to some numerical limitations.  In fact, going to very small
values of $\ve$ necessitates taking into account too many terms in
equation \eqref{eq:J_inner} for the direct search, {\clr whose
numerical accuracy cannot be properly controlled by our algorithm.}
Nevertheless, the considered value $\ve=0.01$ allows us to illustrate
the main features of the FPT PDF and compare the two search
scenarios. We fix length and time units by setting $R_1=1$ and $D_b=1$
and investigate the effect of other parameters onto the PDFs and the
survival probabilities.

\begin{figure}
\begin{center}
\includegraphics[width=7.6cm]{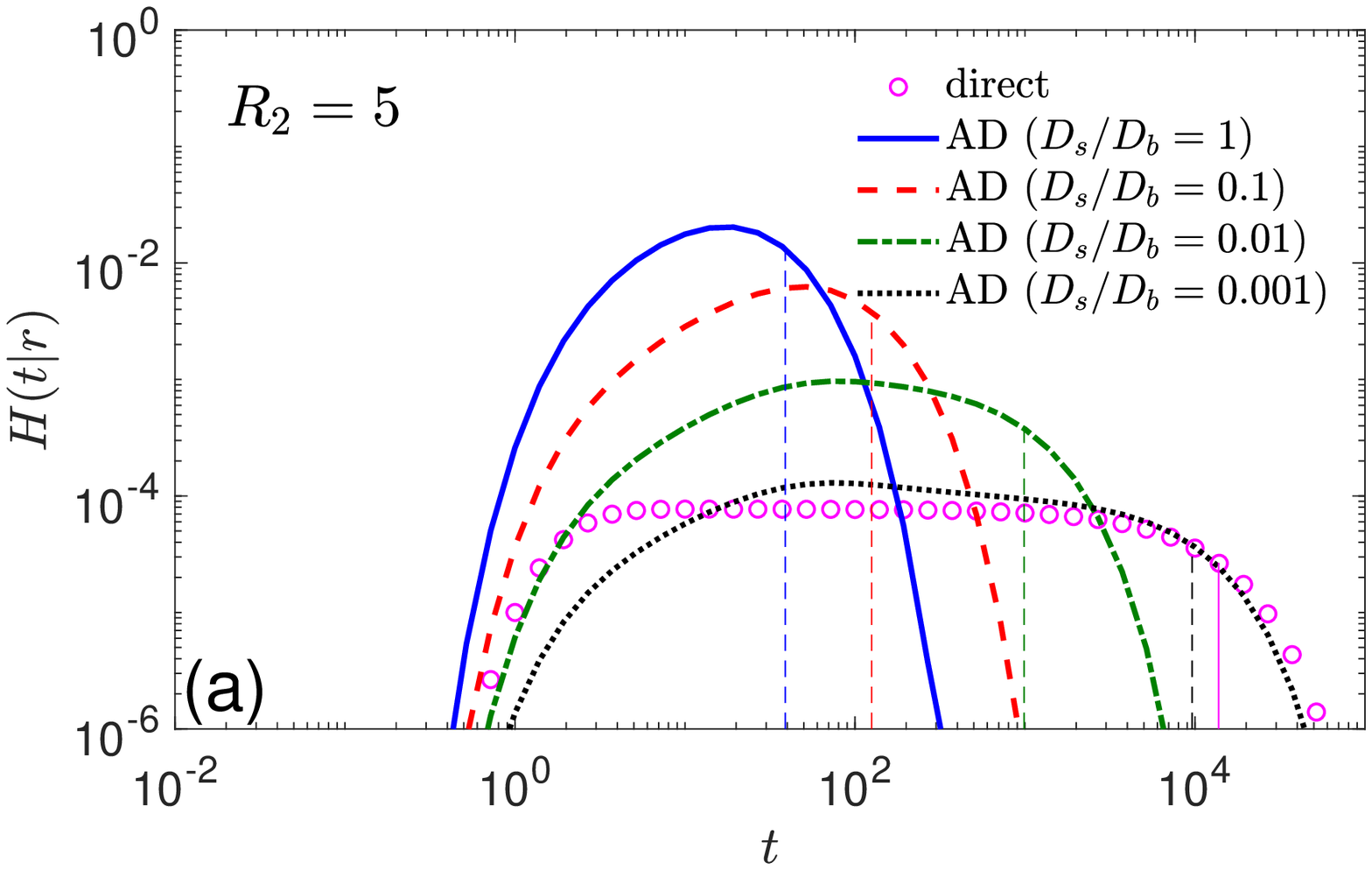}
\includegraphics[width=7.6cm]{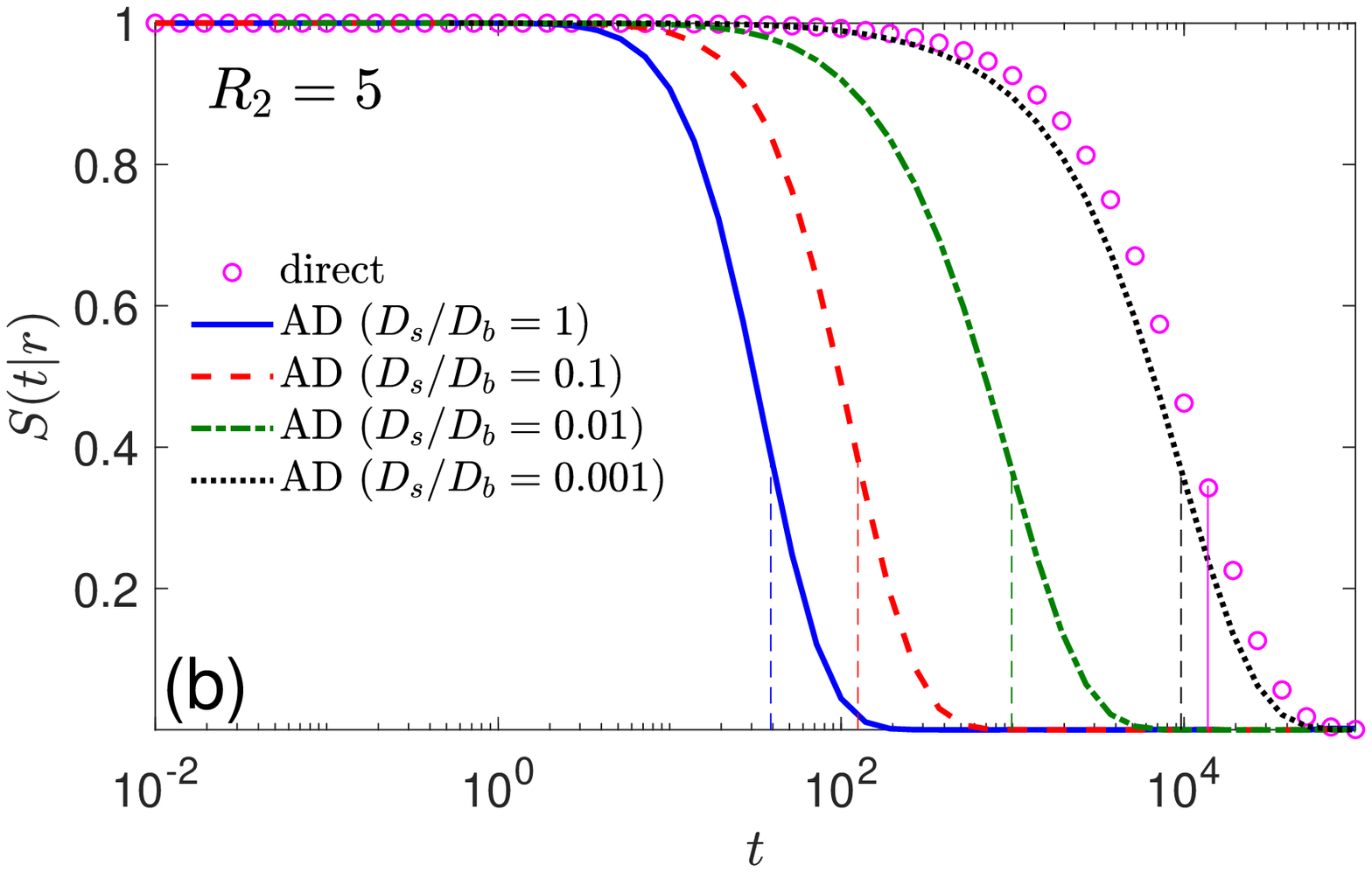}
\includegraphics[width=7.6cm]{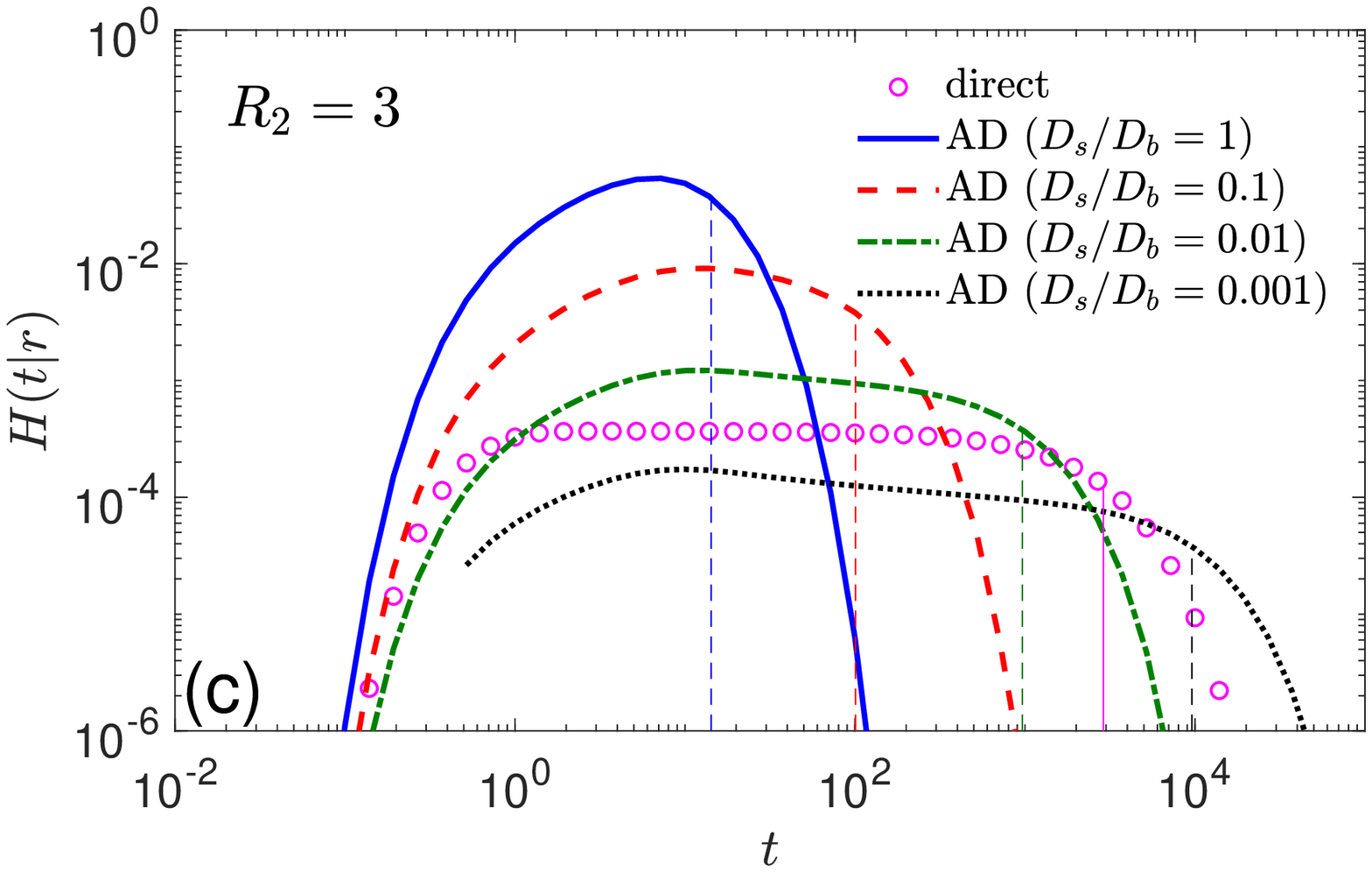}
\includegraphics[width=7.6cm]{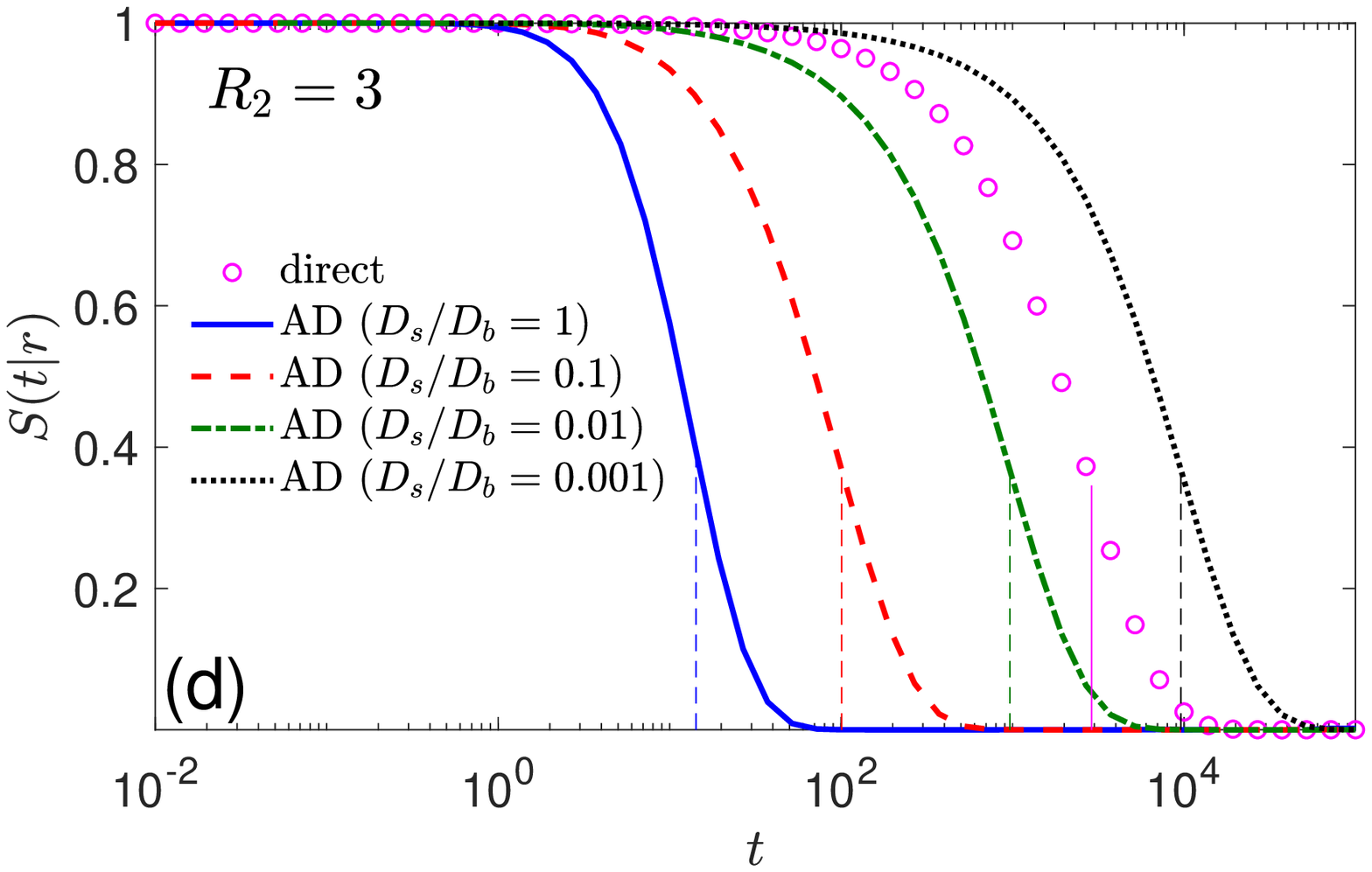}
\end{center}
\caption{PDF {\bf (a,c)} and associated survival probability {\bf (b,d)}
determining the FPT to a target of small angular size $\ve=0.01$, with
$R_2/R_1=5$ {\bf (a,b)} (KR=0.008) or $R_2/R_1=3$ {\bf (c,d)} (the KR=
0.037), and with a starting point uniformly distributed over the outer
sphere ($r/R_1=5$). Comparison between the direct search (open circles,
\ref{direct}) and the ADS (coloured curves corresponding to different
values of the ratio $D_s/D_b$, see the legend). Vertical lines indicate
the MFPTs given by equation \eqref{eq:Tsurfav} for the ADS and by equation
\eqref{eq:Tdir_surfav} for the direct search. The numerical results for
the direct search were obtained by {\clr using the approximate relation
\eqref{eq:J_inner_app} with $n_{\rm inter} = 50$ and $n_{\rm
max}=250$ and by truncating} the series
\eqref{eq:c_series} at $n_{\rm max}=2000$.  Note that $\tau^{\rm
surf}\approx5.2/D_s$, whereas $\tau^{\rm bulk}\approx 29.3$ for
$R_2/R_1=5$ and $\tau^{\rm bulk}\approx4.7$ for $R_2/R_1=3$, i.e.,
$\tau_{\rm surf}$ is dominant for all cases except for
$D_s/D_b=1$. Note that length and time units are fixed by setting
$R_1=1$ and $D_b=1$.}
\label{fig:Ht}
\end{figure}

Figure \ref{fig:Ht} illustrates our main results. Panels {\bf (a,b)}
correspond to the geometric setting with a small karyoplasmic ratio
$\mathrm{KR}=0.008$ (with $R_2/R_1=5$), while panels {\bf (c,d)} refer
to the larger value $\mathrm{KR}=0.037$ (with $R_2/R_1=3$). Panels
{\bf (a)} and {\bf (c)} present the FPT PDFs for both scenarios, with
different values of the ratio $D_s/D_b$ (see the legends) and fixed
angular size $\ve=0.01$ of the target, {\clr offering insight into}
the functional form of the full PDFs, as well as of the locations of
the most probable and mean FPTs (see the vertical dashed lines). For
the ADS, the PDF is broadening with decreasing surface diffusion
coefficient $D_s$. In fact, according to equation \eqref{lm}, the left
short-time tail of the PDF is mainly controlled by bulk diffusion and
thus does not change much with $D_s$. In turn, the right long-time
tail is directly affected by $D_s$ via the time $\tau^{\rm surf}$
given by equation \eqref{eq:tau_surf} which becomes dominant in
equation \eqref{mm3} when $D_s$ is small enough. For instance, when
$D_s/D_b=10^{-3}$ (which is appropriate for diffusion on hard solid
surfaces), the PDF shown in panel {\bf (a)} spans over five orders of
magnitude in time. In turn, for larger $D_s$, the PDF of the ADS is
more compact and attains substantially larger values in the vicinity
of the most probable FPT than its counterpart for the direct search
scenario.  Decreasing $R_2/R_1$ from $5$ to $3$ (panels {\bf (c,d)})
which results in the increase of the KR from $\mathrm{KR}\approx0.008$
to $\mathrm{KR} \approx0.037$, shifts the left tail of the PDF to the
left (towards shorter times), because the target is now closer to the
starting point. In turn, if $D_s$ is sufficiently small, the right
tail of the distribution is still controlled by $\tau^{\rm surf}$,
which is independent of $R_2$. As a consequence, the PDF is getting
even broader.

While panels {\bf (a)} and {\bf (c)} present a basic conceptual
understanding of the structure of the PDFs, panels {\bf (b)} and {\bf
(d)} display the survival probability, which indeed proves to be a
very robust measure of the relative efficiency of both search
scenarios. We observe that for a small ratio $\mathrm{KR}=0.008$ and
{\clr all considered values of $D_s/D_b$,} the survival probabilities
for the ADS, at any fixed $t$, are {\clr smaller} than the survival
probability of the target within the direct search scenario. In this
situation, the ADS can be qualified as a more efficient search
strategy as compared to the direct search. {\clr This conclusion also
agrees with the fact that the MFPT for the direct search is longer
than the MFPT within the ADS.}  Upon increase of KR (see the panel
{\bf (d)}), we notice that the ADS still outperforms the direct search
scenario for {\clr $D_s/D_b \gtrsim 0.01$,} and it becomes less
efficient at {\clr $D_s/D_b = 0.001$}.  Recall, however, that these
curves are calculated for $\ve=0.01$, which is somewhat higher than
the typically encountered values of $\ve$.  On intuitive grounds, we
may thus expect that for a smaller target with $\ve=0.001$, as it is
realised in the case of a nuclear pore, the ADS will perform better
even in this case.  Therefore, we demonstrate that the ADS can indeed
be more efficient search scenario for quite reasonable values of the
system parameters. We finally note that the survival probability at
the MFPT appears to be quite universally equal to $0.35$, which
signifies that {\clr two-thirds} of searching trajectories find the
target before this time, and only the remaining {\clr one-third}
arrive to the target location at longer times.

\begin{figure}
\begin{center}
\includegraphics[width=7.6cm]{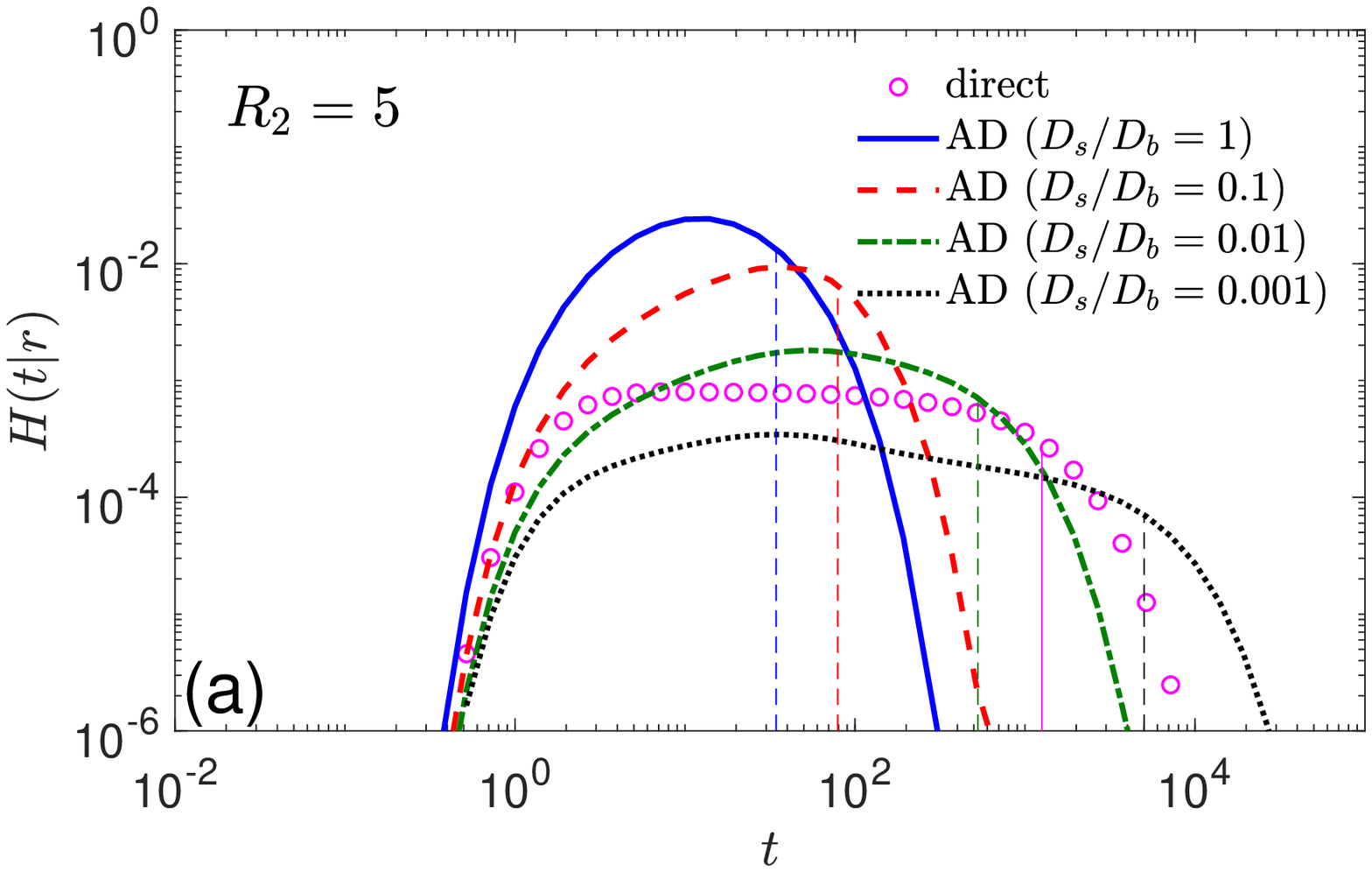}
\includegraphics[width=7.6cm]{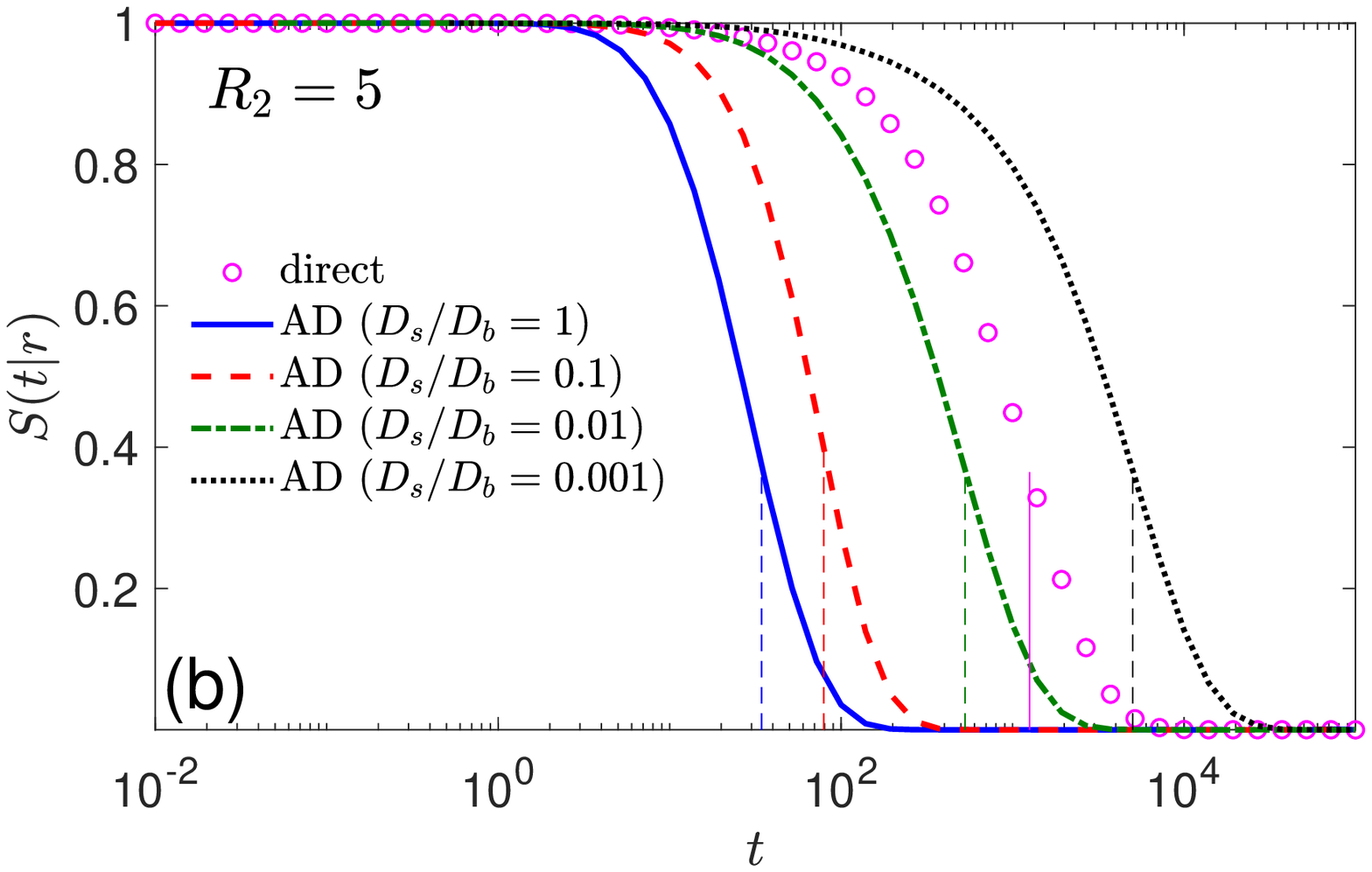}
\includegraphics[width=7.6cm]{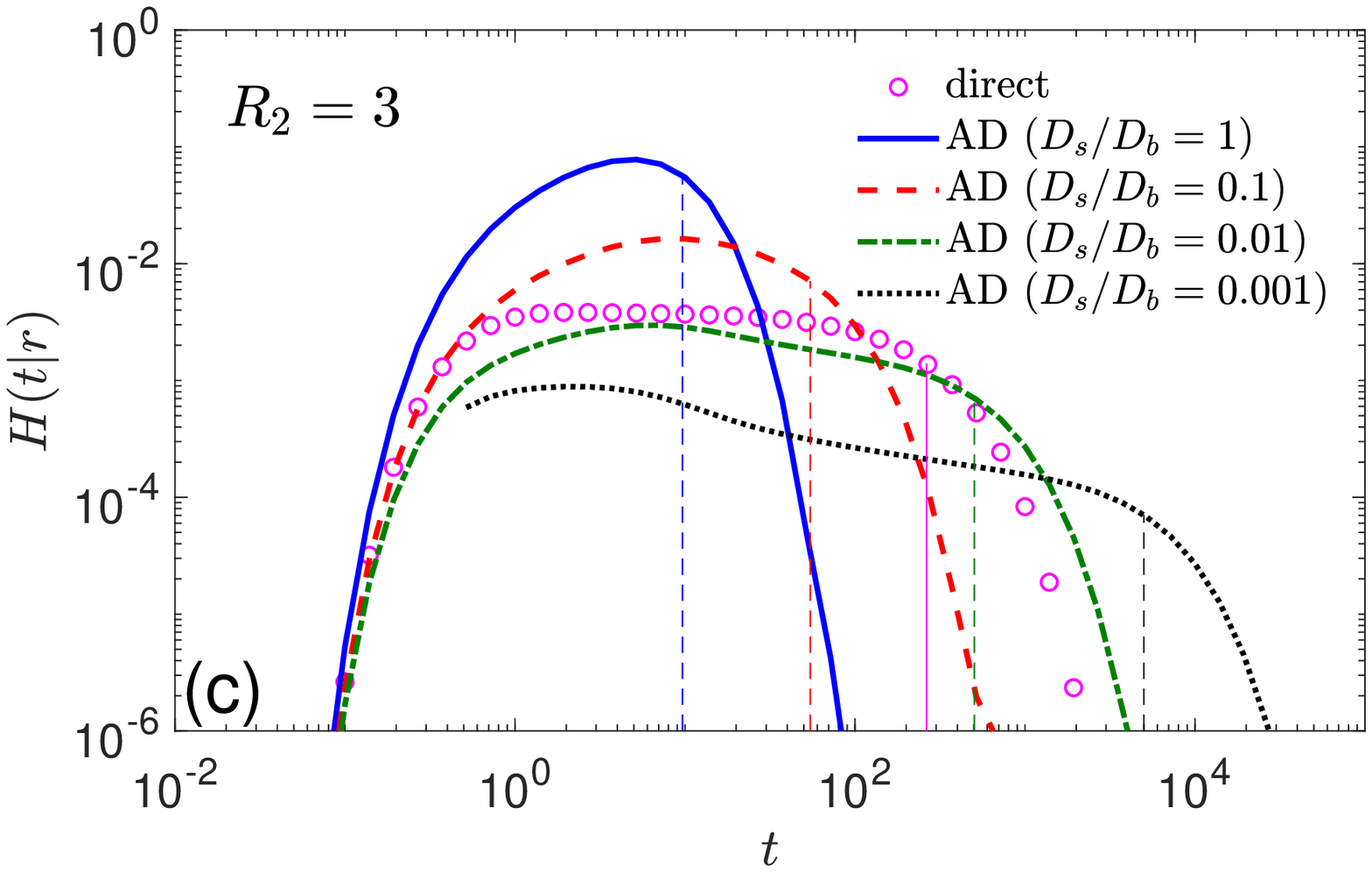}
\includegraphics[width=7.6cm]{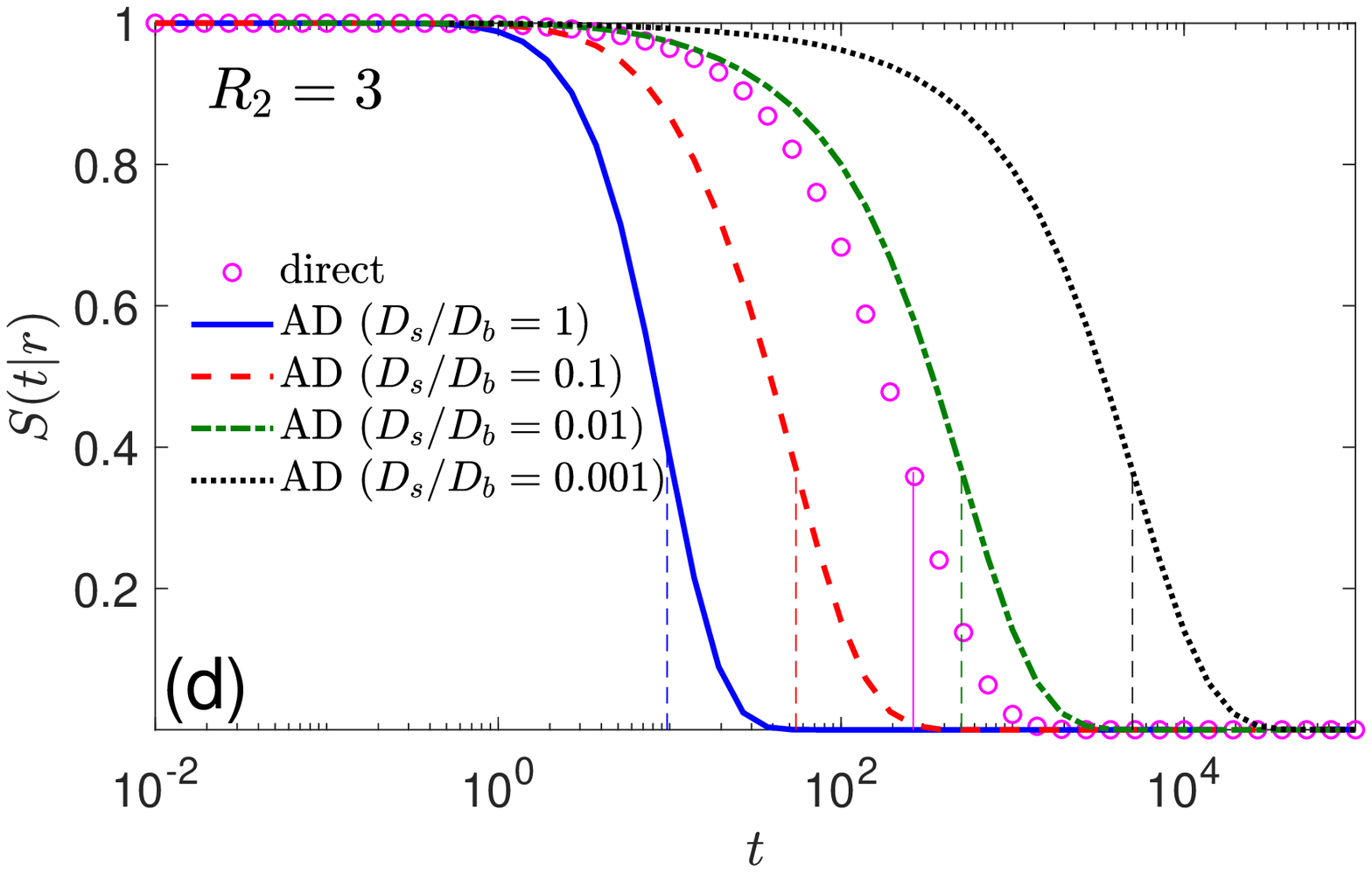}
\end{center}
\caption{Same as in figure \ref{fig:Ht} but for a larger target with $\ve=0.1$.}
\label{fig:Ht2}
\end{figure}

We finish this section by the analysis of the effect of the target
size on the shapes of the PDFs and the survival probabilities. Figure
\ref{fig:Ht2} presents these functions in the same setting, except
that now the target is tenfold larger, $\ve=0.1$. As we already
remarked, such a large value of $\ve$ can be indicative of the
behaviour in situations in which approximately $100$ nuclear pores are
present on the surface of the nucleus. Remarkably, the PDFs within the
ADS remain nearly the same as for a smaller target without any visible
change, as compared to figure \ref{fig:Ht}. At first thought, this is
a counter-intuitive behaviour. However, we recall that the right tail
of the PDF is mainly determined by $\tau^{\rm surf}$, which changes
logarithmically slowly at small $\ve$. In fact, as discussed in
\cite{Grebenkov19sphere}, $\nu_0\approx1/(2\ln(2/\ve))$ as $\ve\to0$
so that $\tau^{\rm surf}\approx 2R_1^2\ln(2/\ve)/D_s$. Moreover, as we
saw in section \ref{sec:random}, the left tail of the distribution is
also weakly dependent on the target size.  For instance, the PDFs for
even smaller target size $\ve=0.001$, which is representative of the
nuclear pore, are expected to be similar to those shown in figure
\ref{fig:Ht}. In contrast, the FPT PDF for the direct search scenario
depends much more strongly on the target size. In fact, its right tail
is characterised by the decay time, which is of the order of the MFPT
and thus scales as $1/\ve$. The amplitude of the left tail is also
affected by $\ve$. We thus observe that the efficiency of the ADS is
weakly dependent on the target size and mainly controlled by two
ratios $R_1/R_2$ and $D_s/D_b$.  Overall, we conclude that in the case
of larger targets {\clr the ADS outperforms the direct search scenario
for $D_s/D_b \gtrsim 0.01$ (when $R_2/R_1 = 5$) and $D_s/D_b\gtrsim
0.1$ (when $R_2/R_1 = 3$),} and is less efficient otherwise.

\section{Amplified signals}
\label{N}

We now extend our analysis to the important case of so-called
"amplified" signals, when $N>1$ independently diffusing ligands search
for a single target. We note that the ensuing speed-up of the search
process has been intensively studied in rather diverse geometrical
settings (different from our setting here) for the direct search
scenario
\cite{Weiss83,Basnayake18,Basnayake19,Schuss19,Lawley20a,Lawley20b,Grebenkov20,Majumdar20,Grebenkov22}. We
start by introducing auxiliary notations and formulating some basic
general results.

\begin{figure}
\begin{center}
\includegraphics[width=7.6cm]{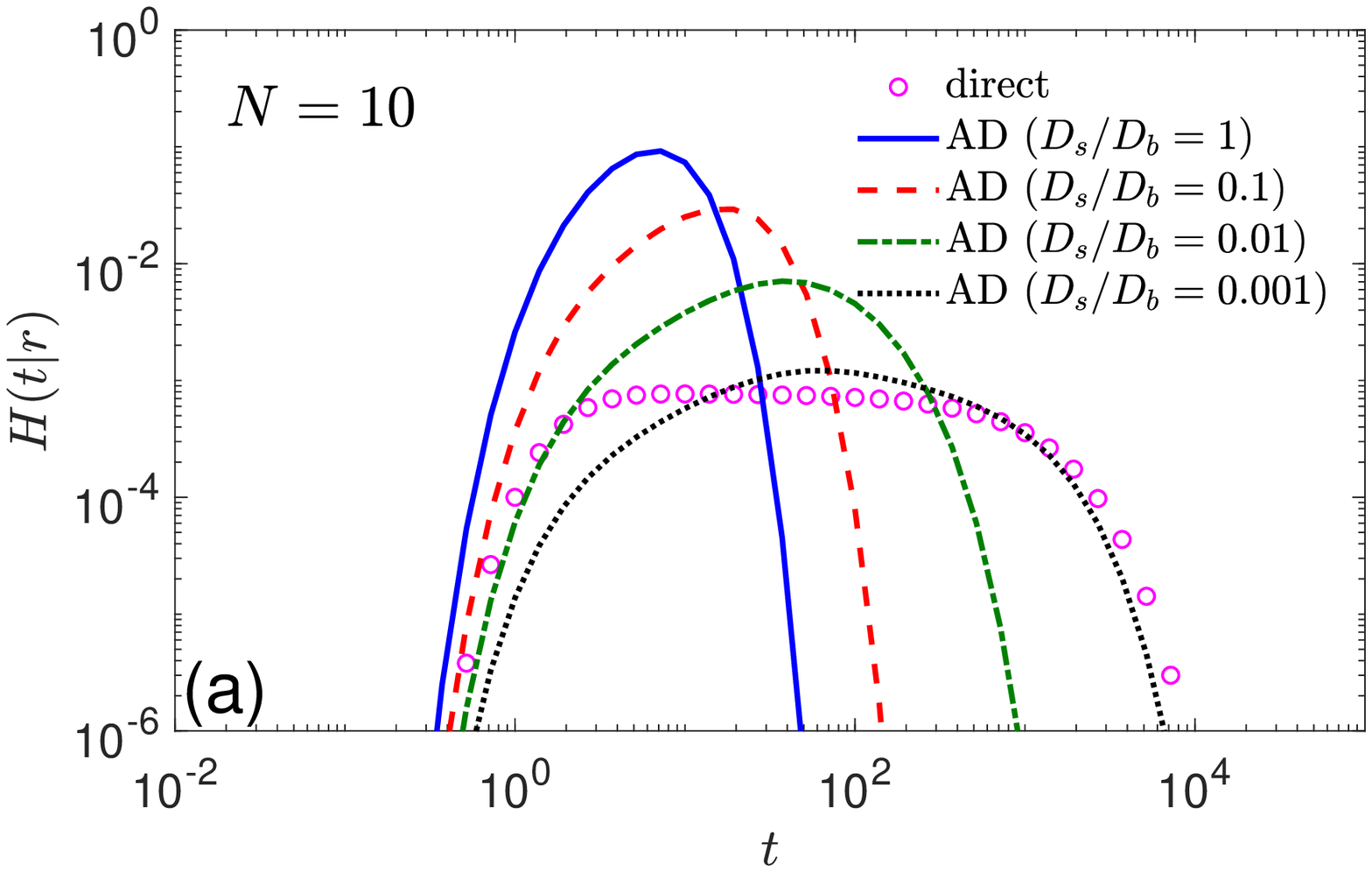}
\includegraphics[width=7.6cm]{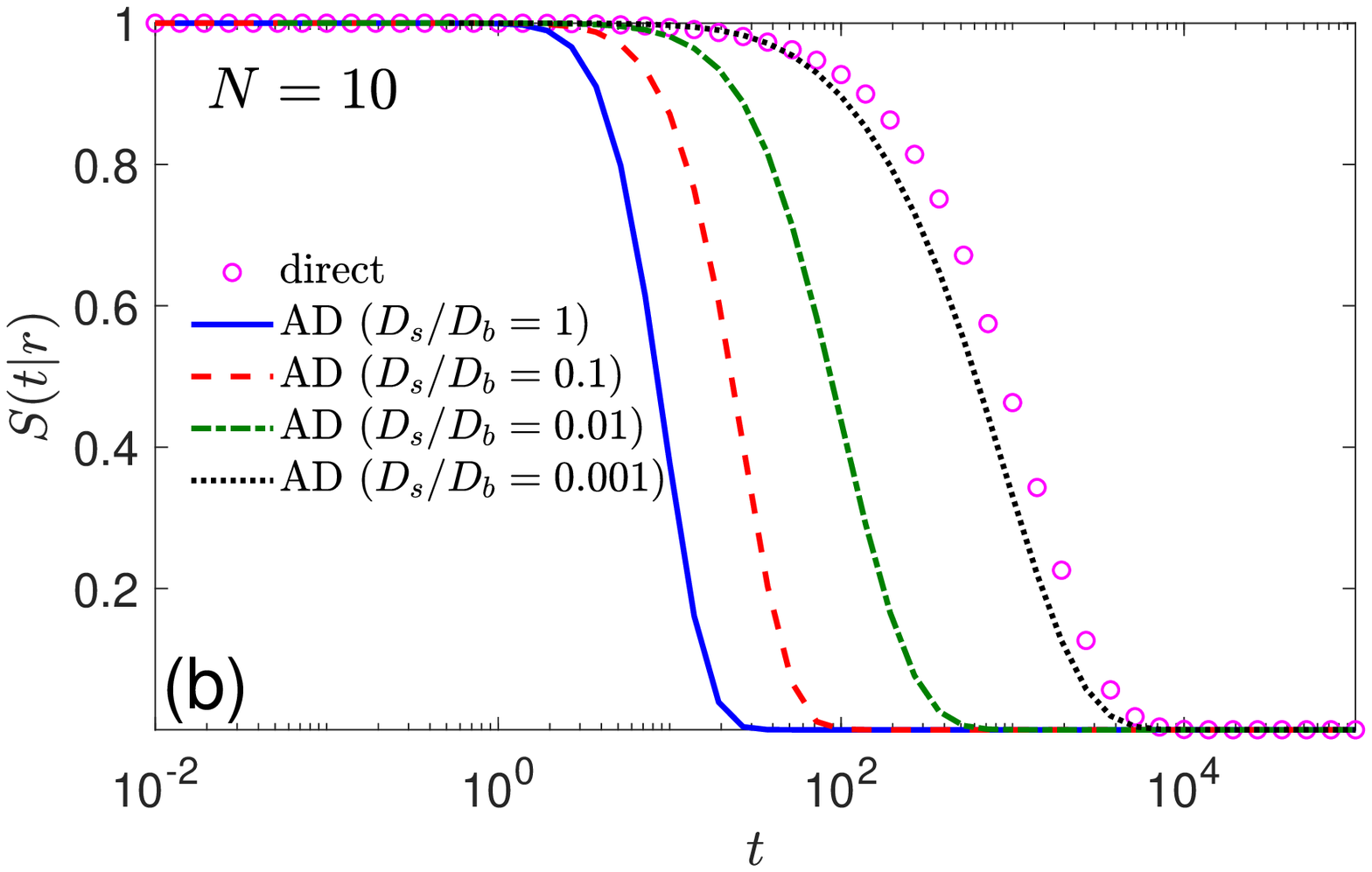}
\includegraphics[width=7.6cm]{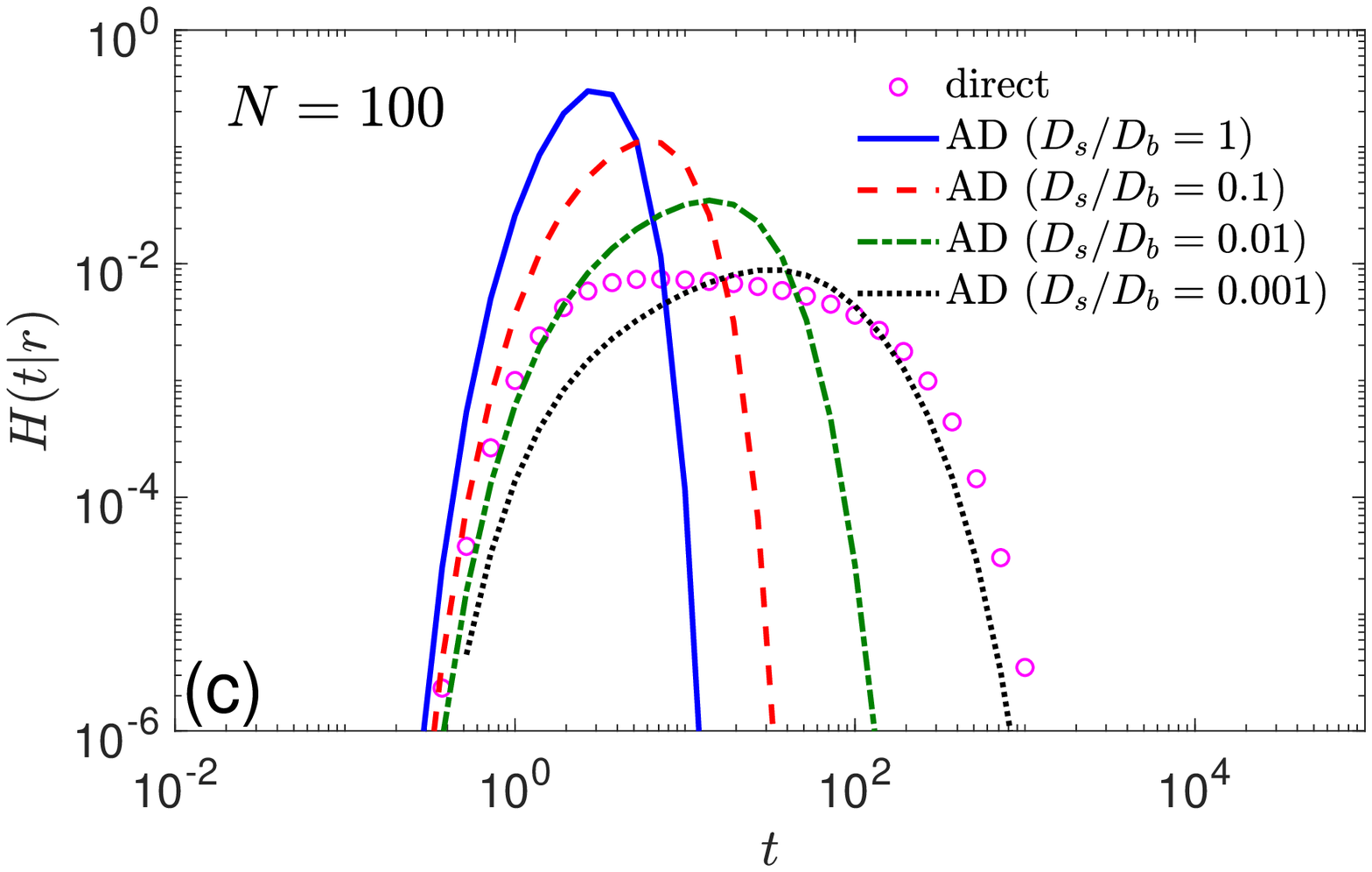}
\includegraphics[width=7.6cm]{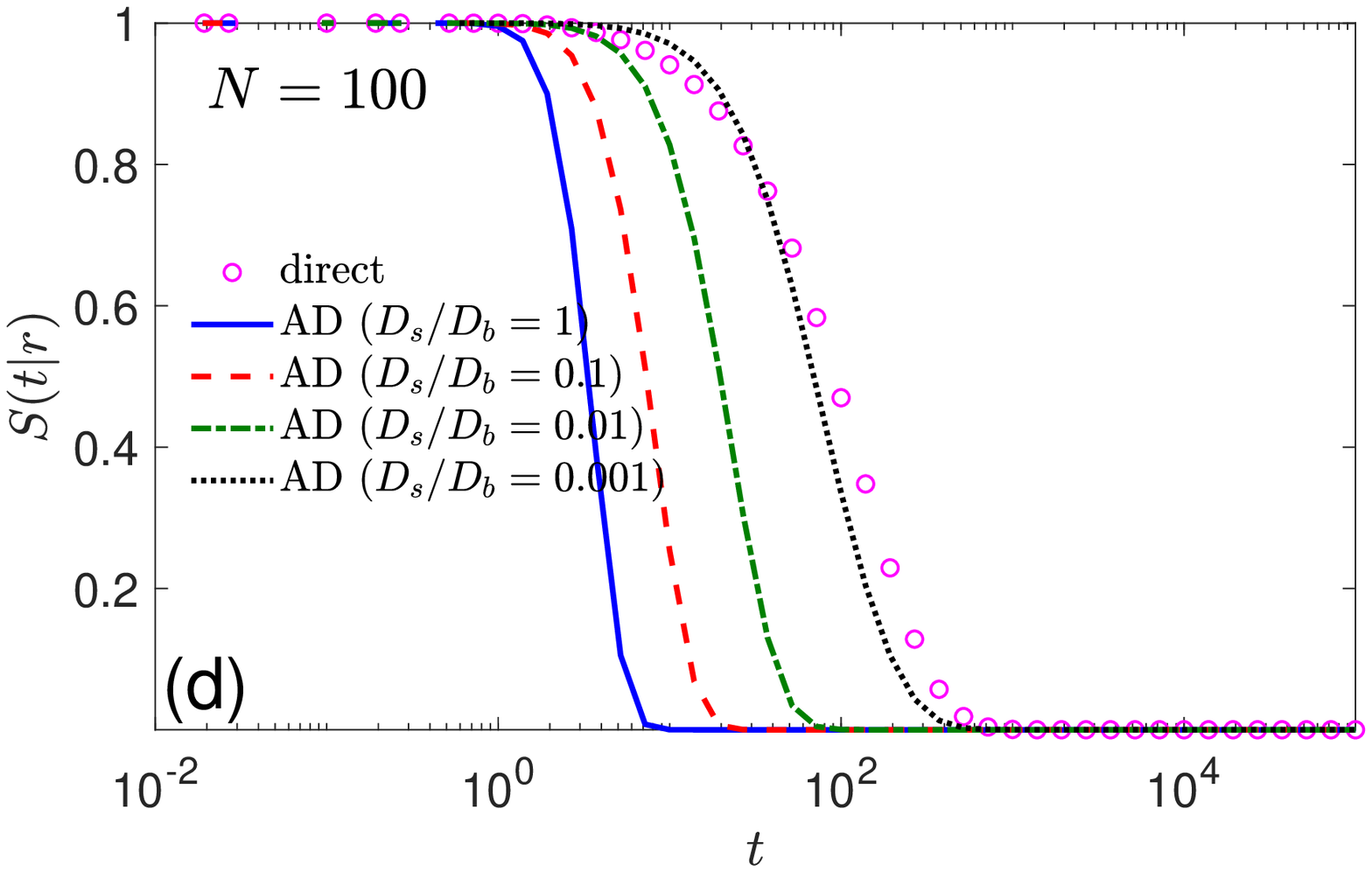}
\end{center}
\caption{
PDF {\bf (a,c)} and associated survival probability {\bf (b,d)}
determining the fastest FPT to a target of small angular size
$\ve=0.01$, with $N = 10$ {\bf (a,b)} and $N = 100$ {\bf (c,d)}
particles, $R_2/R_1=5$ (KR=0.008), and starting positions
independently and uniformly distributed over the outer sphere
($r/R_1=5$). Comparison between the direct search (open circles,
\ref{direct}) and the ADS (coloured curves corresponding to different
values of the ratio $D_s/D_b$, see the legend).  The numerical results
for the direct search were obtained by {\clr using the approximate relation
\eqref{eq:J_inner_app} with $n_{\rm inter} = 50$ and $n_{\rm
max}=250$.}  }
\label{fig:Ht_N}
\end{figure}

Let $\T_i$ (with $i=1,2,\ldots,N$) denote the time instant when
the $i$th ligand arrives to the target for the first time. Consequently,
the target is considered to be "found" when the {\it fastest\/} of the $N$
particles arrives to the target location, i.e., $\T=\min\{\T_1,\T_2,\ldots,
\T_N\}$. Moreover, if all ligands start from the same fixed point $\x$,
the {\clr survival probability that determines the probability law of}
the fastest FPT (fFPT) is given by
\begin{equation}
S^{\rm AD}_N(t;\x)=\P\{\T>t\}=\P\{\T_1>t,\T_2>t,\ldots,\T_N>t\}=[S^{\rm
AD}(t;\x)]^N,
\end{equation}
where we took advantage of a physically plausible assumption that all
ligands move independently of each other. If, in contrast, each ligand
starts from a random position, which is uniformly distributed on a
sphere of radius $r$, independently from the positions of the other
ligands, then we have
\begin{equation}
\overline{S^{\rm AD}_N(t;r)}=\P\{\T>t\}=\P\{\T_1>t,\T_2>t,\ldots,\T_N>t\}
=\left[\overline{S^{\rm AD}(t;r)}\right]^N .
\end{equation}
In both cases, the associated PDFs are given by the time derivative
\begin{equation}
\begin{split}
H^{\rm AD}_N(t;\x)=N[S^{\rm AD}(t;\x)]^{N-1}\,H^{\rm AD}(t;\x),\\ 
\overline{H^{\rm AD}_N(t;r)}=N\left[\overline{S^{\rm AD}(t;r)}\right]^{
N-1}\,\overline{H^{\rm AD}(t;r)}.
\end{split}
\label{amppdf}
\end{equation}
These two expressions allow for a detailed comparison of both
scenarios in the amplified signal case, as presented in figure
\ref{fig:Ht_N}. Expressions
\eqref{amppdf} signify that, once the survival probability for a
single ligand is found in the time domain---as evaluated in the
previous sections---one has ready access to the statistics of the fFPT
$\T$. Note, however, that the large-$N$ asymptotic analysis of the
moments of $\T$ is more cumbersome (see, e.g.,
\cite{Weiss83,Lawley20b} in the direct search case).

{\clr Panels {\bf (a)} and {\bf (c)} of figure \ref{fig:Ht_N} depict}
the PDF $\overline{H^{\rm AD}_N(t;r)}$ for the case when the starting
points of the ligands are uniformly distributed over the outer
sphere. The geometric setting is the same as for figure \ref{fig:Ht}
{\clr with $R_2/R_1 = 5$.}  We see that the qualitative comparison
between the ADS and the direct search does not change much for $N=10$
and $N=100$; in particular, the PDF for the direct search is
relatively close to that for the ADS with $D_s/D_b=10^{-3}$. As
expected, all PDFs are narrowing with increasing $N$. This is expected
from the asymptotic behaviour for a single ligand. In fact, the
long-time relation
\eqref{mm3} implies that
\begin{equation}
\overline{H^{\rm AD}_N(t;r)}\simeq N[C_\ve(r)]^N\exp\bigl(-N t/\max\{\tau^{
\rm bulk},\tau^{\rm surf}\}\bigr)\qquad(t\to\infty),
\end{equation}
i.e., the decay time is decreased by the factor $N$, and thus the right tail
is shifted towards the left (to shorter times). In contrast, as the survival
probability $\overline{S^{\rm AD}_N(t;r)}$ is close to unity at short times,
the left tail is just multiplied by $N$ and does not shift, i.e.,
\begin{equation}
\overline{H^{\rm AD}_N(t;r)}\simeq N\,\overline{H^{\rm AD}(t;r)}\qquad(t\to0).
\end{equation}
This relation is valid for moderate values of $N$. In turn, when $N$ is very
large, we expect the emergence of an intermediate range of times for which
the factor $\left[\overline{S^{\rm AD}(t;r)}\right]^{N-1}$ cannot be replaced
by unity and starts to affect the distribution of the fFPT.

In panels {\bf (b)} and {\bf (d)} we depict the corresponding survival
probabilities. We observe that they exhibit essentially the same
behaviour as those in the case $N=1$, except for the fact that an
abrupt decay to zero starts at progressively earlier times. We also
infer from figure \ref{fig:Ht_N} that, maybe somewhat
counter-intuitively, the relative efficiency of both search scenarios
is the same as in the case of a single particle. Thus, the ADS appears
to be more efficient for {\clr all considered values of $D_s/D_b$.  We
recall, however, that the relative efficiency also depends on the
ratio $R_2/R_1$, see section \ref{comp}.} Similar conclusions hold for
$N$ particles started from a fixed point located far from the target
(not shown).

\section{Conclusions}
\label{conc}

For the simple geometrical setting of two nested, concentric and
impenetrable spheres with an immobile target site being placed at the
North pole of the inner sphere, we presented a detailed comparison of
two search scenarios: (i) the direct search mode, when the particle
needs to locate the target solely by bulk diffusion in the volume
between the concentric spheres, with the surface of the inner sphere
being perfectly reflecting, except for the target (see \cite{9} for
more details); and (ii) the Adam-Delbr\"uck reduction-of-dimensionality
scenario \cite{adam}, for which the particle first attaches
non-specifically to the reactive surface of the inner sphere and
locates the target by diffusive search on the 2D surface. For both
scenarios we calculate the FPT PDF for the searching ligand, that is
initially released from a fixed point, or from a random position on the
surface of the outer sphere. We also considered the case of "amplified" signals, when
initially $N$ ligands are launched from the same fixed or random
positions, and the search is completed when the fastest out of $N$
particles arrives to the target. Such settings indeed correspond to
various situations and problems encountered in cellular biology,
biophysics, and biochemistry. In particular, they appear as a crucial
intermediate step in many signal transduction pathways in cellular
environments.

To show that the reduction-of-dimensionality scenario may be
beneficial in certain situations, Adam and Delbr\"uck originally
presented a comparison of the efficiency of both direct and ADS
diffusive searches for a single ligand, judging solely from the
\textit{mean\/} times of diffusion towards the target site. In the
present work confronting both scenarios, we went beyond the analysis
of the MFPT---which by now is known to be non-representative in even
simple geometries and sometimes even plainly misleading---and
therefore focused our analysis on the PDF of the random individual
FPTs from a fixed or a random starting position to the target. We
compared the behaviour of the full FPT PDF as well as of their
integrated characteristic, the survival probability, which appeared to
be a robust measure of the actual efficiency of both search scenarios.
{\clr In fact, if 
\begin{equation} \label{eq:criterionS}
S_1(t) \leq S_2(t)  \qquad \forall~t\geq 0
\end{equation}
for all times, the first search process (described by $S_1(t)$) is
more efficient than the second one.  As the MFPT is the integral of
the survival probability, this inequality implies the conventional
efficiency criterion,
\begin{equation} \label{eq:criterionT}
T_1 \leq T_2,
\end{equation}
in terms of the MFPTs $T_1$ and $T_2$ of two search processes.  In
this situation, both criteria are equivalent.  However, the opposite
claim is not true, namely, \eqref{eq:criterionT} does not imply
\eqref{eq:criterionS}.  For instance, one may have $S_1(t) \leq
S_2(t)$ at short times and $S_1(t) \geq S_2(t)$ at long times, even
though $T_1 = T_2$.  In other words, one search process can be more
efficient at short times and less efficient at long times. We
conclude that the comparison of survival probabilities provides a more
systematic and comprehensive insight into the search efficiency.} On this
basis, we specified realistic physical conditions in which the
reduction-of-dimensionality scenario outperforms the direct search for
both cases when a single ligand is present and when the signal is
amplified.

Our analysis can be generalised in several directions. First, we
considered exclusively the case of standard Brownian motion. In
reality, both diffusion in the bulk and the surface diffusion may be
(transiently) anomalous.  {\clr These anomalous features can be
incorporated in a standard way, e.g., via a subordination technique.}
Second, we supposed that the spherical-shell domain is a homogeneous
liquid-filled region and does not contain any obstacles or
"crowders". In turn, especially in cellular environments, this is not
the case due to the presence of organelles, proteins and etc, which
will certainly affect the dynamics \cite{ma}. Moreover, in some cases
the cellular cytoplasm is actively moved by non-equilibrium action of
molecular motors
\cite{christine,christine1,Witzel19}, thus changing the mixing dynamics inside the
cell and creating dynamics heterogeneities \cite{Chechkin17,Lanoiselee18}.
Consequently, these effects may alter the relative efficiency of
both search scenarios in the realistic cellular setting. Third, both
scenarios considered here pertain to two somewhat extreme situations:
the particle is either perfectly reflected by the surface of the inner
sphere or, in the ADS, non-specifically adsorbs to it upon first
encounter. In reality, the situation can be more complex. As discussed
recently in \cite{q2} within the context of a narrow-escape problem,
there are distance-dependent potential interactions with the surface
on which the target is located and the presence of such interactions
modifies the search processes. As realised in \cite{q2}, the most
efficient interactions, i.e., resulting in the shortest FPT to the
target, are neither too long-ranged not too short-ranged (as
corresponds to the Adam-Delbr\"uck picture) such that the optimal
trajectories are intermittent---upon arrival to the inner sphere a
ligand performs a finite surface diffusion tour and then desorbs to
the bulk, arrives to the surface again, performing a new tour of
surface diffusion, and so on, until the target is finally found.  Even
though such intermittent search strategies have been studied in the past
(see, e.g.,
\cite{Benichou11a,Benichou10,Benichou11,Rupprecht12a,Rupprecht12b,Rojo11,Rojo13}
and references therein), former works were almost exclusively focused
on the mean FPT and the search optimality was qualified by its
minimisation.  In our future work, we will explore these possible
generalisations of our present analysis.

\section*{Acknowledgments}

The authors acknowledge helpful discussions with Baruch Meerson, Stas
Shvartzman and Leonid Mirny. DG acknowledges the Alexander von
Humboldt Foundation for support within a Bessel Prize award. RM
acknowledges support from the German Science Foundation (DFG Grant ME
1535/12-1) as well as from the Foundation for Polish Science Fundacja
na rzecz Nauki Polskiej, FNP) through an Alexander von Humboldt Polish
Honorary Research Scholarship.

\appendix

\section{Moments of the conditioned first-passage time}
\label{sec:j_moments}

The Laplace transform of the joint probability density $j(t,\s|\x)$ in
\eqref{eq:jtild} determines all positive integer moments of the FPT
$\T_{\s}$ to the inner sphere, conditioned by the arrival point $\s$,
\begin{equation}
\langle\T_{\s}^k\rangle=\int\limits_0^\infty dt\,t^k\,j(t,\s|\x)=(-1)^k
\lim\limits_{p\to0}\frac{\partial^k}{\partial p^k}\tilde{j}(p,\s|\x).
\end{equation}
One sees that it is sufficient to compute the Taylor expansion in powers
of $p$ of the radial functions $g_n^{(p)}$ from \eqref{eq:gnI}. For instance,
the first moment is given by
\begin{equation}
\label{eq:Ts}
\langle\T_{\s}\rangle=-\frac{1}{4\pi R_1^2}\sum\limits_{n=0}^\infty(2n+1)
P_n\left(\frac{(\s\cdot\x)}{|\s|\,|\x|}\right)\lim\limits_{p\to0}\frac{
\partial g_n^{(p)}(r)}{\partial p}.
\end{equation}
A lengthy but straightforward computation produces
\begin{align}
\nonumber 
&\lim\limits_{p\to0}\frac{\partial g_n^{(p)}(r)}{\partial p}=\frac{R_2^2}{
4D_b}\,\frac{(n+1)(r/R_2)^n+n(r/R_2)^{-n-1}}{(n+1)(R_1/R_2)^n+n(R_1/R_2)^{
-n-1}}\biggl\{\frac{1}{(n+1)(r/R_2)^n+n(r/R_2)^{-n-1}}\\
\nonumber
&\times\biggl[\frac{n+1}{n+3/2}(r/R_2)^{n+2}-\frac{n-1}{n-1/2}(r/R_2)^n
+\frac{n+2}{n+3/2}(r/R_2)^{-n-1}-\frac{n}{n-1/2}(r/R_2)^{-n+1}\biggr]\\
\nonumber
&-\frac{1}{(n+1)(R_1/R_2)^n+n(R_1/R_2)^{-n-1}}\biggl[\frac{n+1}{n+3/2}(R_1/
R_2)^{n+2}\\
\label{eq:dgn}
&-\frac{n-1}{n-1/2}(R_1/R_2)^n+\frac{n+2}{n+3/2}(R_1/R_2)^{-n-1}-\frac{n}{
n-1/2}(R_1/R_2)^{-n+1}\biggr]\biggr\}.
\end{align}
The average of \eqref{eq:Ts} over $\s$ results in the well-known expression
for the MFPT to the inner sphere,
\begin{equation}
\label{eq:Ts_averaged}
\int\limits_{\pa}d\s\,\langle\T_{\s}\rangle=-\lim\limits_{p\to0}\frac{
\partial g_0^{(p)}(r)}{\partial p}=\frac{(r-R_1)(2R_2^3-rR_1(R_1+r))}{6D_b
rR_1}.
\end{equation}
Similarly, if the starting point $\x$ is uniformly distributed on a sphere of
radius $r$, the integral over the angular coordinates in \eqref{eq:Ts} cancels
all terms with $n > 0$, yielding
\begin{align}
\nonumber
\frac{1}{4\pi r^2}\int\limits_{|\x| = r}d\x\,\langle\T_{\s}\rangle&=-\frac{1}{
4\pi R_1^2}\lim\limits_{p\to0}\frac{\partial g_0^{(p)}(r)}{\partial p}\\
\label{eq:Ts_surfav}
&=\frac{1}{4\pi R_1^2}\,\frac{(r-R_1)(2R_2^3-rR_1(R_1+r))}{6D_brR_1},
\end{align}
which does not depend on the arrival point $\s$.

Note that when the starting point is located on the outer sphere, $r=R_2$, the
first term in \eqref{eq:dgn} is cancelled, and one gets the simpler expression
\begin{align*}
&\lim\limits_{p\to0}\frac{\partial g_n(R_2)}{\partial p}=-\frac{R_2^2}{4D_b}\,
\frac{(2n+1)(R_1/R_2)^{n+1}}{\bigl[n+(n+1)(R_1/R_2)^{2n+1}\bigr]^2}\\
&\times\biggl[\frac{n+2}{n+3/2}-\frac{n}{n-1/2}(R_1/R_2)^{2}-\frac{n-1}{n-1/2}
(R_1/R_2)^{2n+1}+\frac{n+1}{n+3/2}(R_1/R_2)^{2n+3}\biggr].
\end{align*}

\section{Surface diffusion on the sphere}
\label{sec:surface_diffusion}

The problem of surface diffusion on a sphere has been intensively studied in
the past (see \cite{Chao81,Sano81,Prustel13,Grebenkov19sphere} and references
therein). In particular, the exact solution for the concentration profile in
the presence of a perfect target was given in \cite{Chao81,Sano81}, whereas
its extension to a partially reactive target with reversible binding was
provided in \cite{Grebenkov19sphere}. Here we extend the derivation from
\cite{Grebenkov19sphere} to get two equivalent representations of the FPT
PDF for diffusion on the sphere.

Following \cite{Grebenkov19sphere}, we consider diffusion on a sphere of
radius $R$ (which is set to $R_1$ in the main text) towards a circular
target of angular size $\ve$ located on the {\it South} pole. This location
is specific to this Appendix and differs from our consideration of the
target on the North pole throughout the remaining text. This choice is
taken for the fact that the solution will be given in terms of Legendre
functions $P_\nu(x)$ with non-integer index $\nu$, which are regular at
$x=1$ (the North pole) and singular at $x=-1$ (the South pole). As a
consequence, it is natural to locate the target on the South pole to
eliminate such singularities. Alternatively, one could search for the
solution in terms of $P_\nu(1-x)$, which is equivalent to exchanging
the North and South poles (i.e., by replacing the angle $\theta$ by
$\pi-\theta$). Either way, once the derivation is completed, we will
use such a swap to reformulate the final results for the target located on
the North pole, to be consistent with the remainder of the paper.

We search for the Laplace-transformed survival probability $\tilde{S}^{\rm
surf}(p;\s)$ satisfying the boundary value problem
\begin{align}
(p-D_s\Delta_{\s})\tilde{S}^{\rm surf}(p;\s)&=1\quad(\theta>\pi-\ve),\\
D_s\partial_n\tilde{S}^{\rm surf}(p;\s)+\kappa_s\tilde{S}^{\rm surf}(p;\s)&=0
\quad(\theta=\pi-\ve),
\end{align}
where $\s=(\theta,\phi)$ is a point on the sphere, $\pi-\ve$ is the angular
coordinate of the target region (here, it is a circle around the South pole),
$\partial_n$ is the normal derivative oriented towards the South pole, $\Delta
_{\s}$ is the Laplace-Beltrami operator, and $\kappa_s$ is the reactivity of
the target. Even though the main text deals with a perfect target ($\kappa_s=
\infty$), we here consider the more general case of a partially reactive
target with a finite reactivity $\kappa_s$, from which the perfect target
limit will follow. The axial symmetry of the problem implies that the
survival probability depends only on the angle $\theta$ so that $\Delta_{\s}
=\tfrac{1}{R^2}\L$, where
\begin{equation}
\L=\partial_x (1-x^2)\partial_x\quad\mbox{ with }x=\cos\theta.  
\end{equation}
In the following, we will therefore write functions in terms of $x$, e.g.,
$\tilde{S}^{\rm surf}(p;x)$.

\subsection{Two representations for the Green's function}

We start by considering the Green's function for diffusion on the sphere,
satisfying
\begin{equation}
\label{eq:surface_G_def}
(s-\L)\tilde{G}(x,y|s)=\delta(x-y),\quad\bigl(\partial_x\tilde{G}(x,y|s)
\bigr)|_{x=a}=\frac{qR}{\sqrt{1-a^2}}\tilde{G}(a,y|s),
\end{equation}
where $s=pR^2/D_s$, $q=\kappa_s/D_s$, and $a=\cos(\pi-\ve)=-\cos(\ve)$ is
the location of the target. The second relation is the Robin boundary
condition on the partially reactive target. While $\tilde{G}(x,y|s)$ should
be well known, we provide here the main steps of its derivation for the
reader's convenience.

On one hand, the Green's function admits the spectral expansion over the
Legendre functions $P_\nu(x)$ of the first kind which are the eigenfunctions
of the operator $\L$, $\L P_{\nu}(x)=-\nu(\nu+1)P_\nu(x)$. One thus gets
(see \cite{Grebenkov19sphere} for details)
\begin{equation}
\label{eq:Green_sphere1}
\tilde{G}(x,y|s)=\sum\limits_{n=0}^\infty\frac{b_n^2\,P_{\nu_n}(x)\,P_{\nu_n}
(y)}{s+\nu_n(\nu_n+1)}\quad(a\leq x\leq1,a\leq y\leq 1),
\end{equation}
where
\begin{equation}
\label{eq:bn}
b_n=\left(\int\limits_a^1 dx\,[P_{\nu_n}(x)]^2\right)^{-1/2} 
\end{equation}
are the normalisation constants, and $\nu_n$ are the solutions of the equation
\begin{equation}
\label{eq:sphere_BC}
P'_\nu(a)=\frac{qR}{\sqrt{1-a^2}}\,P_\nu(a)
\end{equation}
that ensures the Robin boundary condition, and $P'_\nu(a)=\bigl(\tfrac{
\partial P_\nu(x)}{\partial x}\bigr)_{|x = a}$. We emphasise that the index
$\nu$ of the Legendre function $P_\nu(x)$ is the unknown to be determined
here. In the case of a perfect target ($q=\infty$), this relation becomes
\begin{equation}
\label{eq:sphere_BCD}
P_\nu(a)=0.  
\end{equation}
One can easily check that the spectral expansion in \eqref{eq:Green_sphere1}
is the solution of \eqref{eq:surface_G_def}. Indeed, the application of the
operator $(s-\L)$ to $\tilde{G}(x,y|s)$ yields 
\begin{equation}
\label{eq:sphere_completeness}
\sum\nolimits_{n=0}^\infty b_n^2\,P_{\nu_n}(x)\,P_{\nu_n}(y)=\delta(x-y)
\end{equation}
due to the completeness relation for the eigenfunctions $P_{\nu_n}(x)$.

On the other hand, the Green's function can be found as a combination of two
linearly independent solutions of the homogeneous equation $(s-\L)u=0$ which
are given as $P_{\mu}(x)$ and $Q_{\mu}(x)$, where $Q_\nu(x)$ is the Legendre
function of the second kind, and
\begin{equation}
\mu=\frac{-1+\sqrt{1-4s}}{2}
\end{equation}
is the solution of the equation $s=-\nu(\nu+1)$ which satisfies $\mu\simeq-s$
in the limit $s\to0$. One searches a solution of the form
\begin{equation}
\tilde{G}(x,y|s)=\left\{\begin{array}{ll}AP_{\mu}(x)&(y<x<1),\\ 
B\bigl(\hat{Q}P_{\mu}(x)-\hat{P}Q_{\mu}(x)\bigr)&(a<x<y),\end{array}\right. 
\end{equation}
with the shortcut notations $\hat{P}=P_{\mu}(a)-hP'_{\mu}(a)$ and $\hat{Q}=
Q_{\mu}(a)-hQ'_{\mu}(a)$, where $h=\sqrt{1-a^2}/(qR)$. The first relation
ensures the regularity of the solution at $x=1$, while the second relation
takes care of the boundary condition \eqref{eq:sphere_BC} at $x=a$. The
coefficients $A$ and $B$ are determined by requiring the continuity of
$\tilde{G}(x,y|s)$ at $x=y$ and the drop by $1/(1-y^2)$ of its derivative
at $x=y$. One then finds
\begin{align}
A&=-\frac{\hat{Q}P_{\mu}(y)-\hat{P}Q_{\mu}(y)}{\hat{P}},\\
B&=-\frac{P_{\mu}(y)}{\hat{P}},
\end{align}
where the Wronskian
\begin{equation}
P_\nu(y)Q'_\nu(y)-P'_\nu(y)Q_\nu(y)=\frac{1}{1-y^2}
\end{equation}
was used. One finally finds
\begin{equation}
\label{eq:Green_sphere2}
\tilde{G}(x,y|s) = 
\left\{\begin{array}{ll}P_{\mu}(x)\bigl(Q_{\mu}(y)-\frac{\hat{Q}}{\hat{P}}
P_{\mu}(y)\bigr)&(y<x<1),\\ 
P_{\mu}(y)\bigl(Q_{\mu}(x)-\frac{\hat{Q}}{\hat{P}}P_{\mu}(x)\bigr)&(a<x<y). \end{array} \right. 
\end{equation}
Equating the right-hand sides of equations \eqref{eq:Green_sphere1} and
\eqref{eq:Green_sphere2}, one gets an identity which allows one to
evaluate various series involving the zeros $\nu_n$ (see the review
\cite{Grebenkov21_review} for other examples and applications).

\subsection{FPT distribution}

In order to get the Laplace-transformed survival probability, it is sufficient
to integrate the Green's function $\tilde{G}(x,y|s)$ over $y$ from $a$ to $1$.
For instance, the integral of \eqref{eq:Green_sphere1} yields 
\begin{equation}
\label{eq:tildeS_spectral}
\tilde{S}^{\rm surf}(p;x)=\frac{R^2}{D_s}\int\limits_{a}^1dy\,\tilde{G}(x,y|s)
=\frac{R^2}{D_s}\sum\limits_{n=0}^\infty\frac{b_n^2\,P_{\nu_n}(x)(1-a^2)P'_{
\nu_n}(a)}{\nu_n(\nu_n+1)(\frac{pR^2}{D_s}+\nu_n(\nu_n+1))},
\end{equation}
where we used the identity
\begin{align} \nonumber
\int\limits_a^bdx\,P_\nu(x) &=\frac{P_{\nu-1}(a)-P_{\nu+1}(a)-P_{\nu-1}(b)+ P_{\nu+1}(b)}{2\nu+1} \\  \label{eq:Pnu_int}
& =\frac{1-a^2}{\nu(\nu+1)}P'_\nu(a)-\frac{1-b^2}{\nu(\nu+1)}P'_\nu(b),
\end{align}
and a similar relation holds for $Q_\nu(x)$. Expression
\eqref{eq:tildeS_spectral} was considered, e.g., in \cite{Grebenkov19sphere}.
It can also be written as
\begin{align}
\nonumber
\tilde{S}^{\rm surf}(p;x)&=\frac{1}{p}\sum\limits_{n=0}^\infty b_n^2\,P_{\nu
_n}(x)(1-a^2) P'_{\nu_n}(a)\biggl(\frac{1}{\nu_n(\nu_n+1)}-\frac{1}{\frac{
pR^2}{D}+\nu_n(\nu_n+1)}\biggr)\\
&=\frac{1}{p}\biggl(1-(1-a^2)\sum\limits_{n=0}^\infty\frac{b_n^2\,P_{\nu_n}(x)
\,P'_{\nu_n}(a)}{\frac{pR^2}{D}+\nu_n(\nu_n+1)}\biggr),
\end{align}
where we evaluated explicitly the first sum by integrating
\eqref{eq:sphere_completeness} and using again \eqref{eq:Pnu_int}.
As a consequence, the Laplace-transformed PDF then becomes
\begin{equation}
\label{eq:Hp_sphere}
\tilde{H}^{\rm surf}(p;x)=1-p\tilde{S}^{\rm surf}(p;x)=(1-a^2)\sum\limits_{n=0}
^\infty\frac{b_n^2P_{\nu_n}(x)P'_{\nu_n}(a)}{\frac{R^2p}{D_s}+\nu_n(\nu_n+1)}.
\end{equation}
The inverse Laplace transform can be easily found to be
\begin{equation}  
H^{\rm surf}(t;x)=\frac{(1-a^2)D_s}{R^2}\sum\limits_{n=0}^\infty b_n^2P_{\nu_n}
(x)P'_{\nu_n}(a)\,e^{-\nu_n(\nu_n+1)D_st/R^2}.
\end{equation}
This is a rare example of an exact explicit representation of the FPT PDF,
except for a numerical step required to find the roots $\nu_n$. The MFPT can
be formally deduced from the above equations, but it admits the simple
closed-formed expression \cite{Sano81}
\begin{equation}
T^{\rm surf}(x)=\frac{R^2}{D}\ln\left(\frac{1+x}{1+a}\right)+\frac{R}{\kappa
_s}\sqrt{\frac{1-a}{1+a}}.
\end{equation}

While the series representation \eqref{eq:Hp_sphere} of $\tilde{H}^{\rm surf}
(p;x)$ was convenient for getting $H^{\rm surf}(t;x)$, another representation
is needed for the analysis of the asymptotic behaviour. For this purpose, one
can calculate $\tilde{S}^{\rm surf}(p;x)$ by integrating $\tilde{G}(x,y|s)$
from \eqref{eq:Green_sphere2}. After simplifications, we get
\begin{equation}
\tilde{S}^{\rm surf}(p;x)=\frac{1}{p}\biggl(1-\frac{P_{\mu}(x)}{P_{\mu}(a)-
\frac{\sqrt{1-a^2}}{qR}P'_{\mu}(a)}\biggr) 
\end{equation}
and thus
\begin{equation}
\label{eq:Hpsurf_general}
\tilde{H}^{\rm surf}(p;x)=\frac{P_{\mu}(x)}{P_{\mu}(a)-\frac{\sqrt{1-a^2}}{qR}
P'_{\mu}(a)}.
\end{equation}
For a perfect target ($q=\infty$), one simply gets
\begin{equation}
\label{eq:Hp_sphere_Dir}
\tilde{H}^{\rm surf}(p;x)=\frac{P_{\mu}(x)}{P_{\mu}(a)}.
\end{equation}
Note that these solutions could be obtained directly by solving the related
boundary value problems with the operator $\L$.

\subsection{Short-time asymptotic behaviour}

The above representation for $\tilde{H}^{\rm surf}(p;x)$ is suitable for
getting the short-time asymptotic behaviour of the PDF. This corresponds to
the large-$p$ limit, for which one can apply the asymptotic behaviour of the
Legendre functions. Indeed, as $p\to\infty$, $\mu\simeq-\tfrac12+i\sqrt{s}$
and we use
\begin{equation*}
P_\mu(\cos\theta)\simeq\sqrt{\frac{\theta}{\sin\theta}}J_0((\,u+1/2)\theta)
\approx\sqrt{\frac{\theta}{\sin\theta}} J_0(i\sqrt{s}\theta)=\sqrt{\frac{
\theta}{\sin\theta}}I_0(\sqrt{s}\theta)\simeq\frac{e^{\sqrt{s}\,\theta}\,
s^{-1/4}}{\sqrt{2\pi\sin\theta}},
\end{equation*}
from which equation \eqref{eq:Hp_sphere_Dir} yields for a perfect target,
\begin{equation}
\tilde{H}^{\rm surf}(p;\theta)\simeq\sqrt{\frac{\sin\ve}{\sin\theta}}\,
e^{-(\theta-\ve)R\sqrt{p/D_s}}\quad(p\to\infty).
\end{equation}
The leading exponential term here is expected, given that $(\theta-\ve)R$
is the geodesic distance between the starting point and the target. As a
consequence, we get
\begin{equation}
H^{\rm surf}(t;\theta) \simeq \sqrt{\frac{\sin\ve}{\sin\theta}}\,\frac{R(
\theta-\ve)e^{-R^2(\theta-\ve)^2/(4D_st)}}{\sqrt{4D_st^3}}\quad(t\to0).
\end{equation}

\subsection{Green's function in absence of the target}

As mentioned in the main text, the self-consistent approximation
\eqref{eq:self_surface} can also be expressed in terms of Legendre
functions. For this purpose, we need some identities that can be
derived from the Green's function without the target. In the limit
$\ve\to0$ (or $a\to-1$), the Green's function approaches
\begin{equation}
\tilde{G}(x,y|s)=\left\{\begin{array}{ll}P_{\mu}(x)\bigl(Q_{\mu}(y)-
P_{\mu}(y)C_{\mu}\bigr)&(y<x<1),\\P_{\mu}(y)\bigl(Q_{\mu}(x)-P_{\mu}(x)
C_{\mu}\bigr)&(-1<x<y),\end{array}\right. 
\end{equation}
with
\begin{equation}
C_{\mu}=\lim\limits_{a\to-1}\frac{Q_{\mu}(a)}{P_{\mu}(a)}=\frac{\pi}{2\,
\tan(\pi\mu)},
\end{equation}
where we used the asymptotic behaviour of Legendre functions (see
\cite{Erdelyi}, p. 164). At the same time, this Green's function
admits the spectral expansion over Legendre polynomials $P_n(x)$,
\begin{equation}
\tilde{G}(x,y|s)=\sum\limits_{n=0}^\infty\frac{(n+1/2)\,P_{n}(x)\,P_{n}(y)}{s
+n(n+1)}.
\end{equation}

The integral of $\tilde{G}(x,y|s)$ over $y$ from $a$ to $1$, for $a<x<1$
yields the identity
\begin{equation}
\label{eq:identity2a}
\frac12\sum\limits_{n=0}^\infty\frac{P_{n}(x)\,(P_{n-1}(a)-P_{n+1}(a))}{s+n(
n+1)}=\frac{1}{s}\biggl(1-(1-a^2)P_{\mu}(x)\bigl(Q'_{\mu}(a)-C_{\mu}P'_{\mu}(a)
\bigr)\biggr),
\end{equation}
where we used the identity \eqref{eq:Pnu_int}. We emphasise that the condition
$a<x<1$ should be satisfied, otherwise the identity does not hold. In the
opposite case, one needs to integrate $\tilde{G}(x,y|s)$ over $y$ from $-1$ to
$a$, to get for $-1<x<a$ that
\begin{equation}
\label{eq:identity2b}
\frac12\sum\limits_{n=0}^\infty\frac{P_{n}(x)\,(P_{n-1}(a)-P_{n+1}(a))}{s+n(
n+1)}=-\frac{(1-a^2)P'_{\mu}(a)\bigl(Q_{\mu}(x)-C_{\mu}P_{\mu}(x)\bigr)}{s}.
\end{equation}

The integral of \eqref{eq:identity2a} over $x$ from $a$ to $1$ yields
another identity,
\begin{equation}
\label{eq:identity3}
\frac12\sum\limits_{n=0}^\infty\frac{(P_{n-1}(a)-P_{n+1}(a))^2}{(2n+1)(s+n
(n+1))}=\frac{(1-a^2)^2P'_{\mu}(a)\bigl(Q'_{\mu}(a)-C_{\mu}P'_{\mu}(a)
\bigr)}{s^2}+\frac{1-a}{s}. 
\end{equation}
Note that the derivatives $P'_\nu(x)$ and $Q'_\nu(x)$ can be rewritten using
the recurrence relations,
\begin{align}
(2\nu+1)(1-x^2)P'_{\nu}(x)&=\nu(\nu+1)(P_{\nu-1}(x)-P_{\nu+1}(x)),\\
(2\nu+1)(1-x^2)Q'_{\nu}(x)&=\nu(\nu+1)(Q_{\nu-1}(x)-Q_{\nu+1}(x)).
\end{align}

Finally, integrating the representation of the Dirac distribution,
\begin{equation}
\frac12\sum\limits_{n=0}^\infty(2n+1)P_n(x)P_n(y)=\delta(x-y)
\end{equation}
over $y$ from $a$ to $1$, yields
\begin{equation}
\frac12\sum\limits_{n=0}^\infty P_n(x)(P_{n-1}(a)-P_{n+1}(a))=\Theta(x-a),
\end{equation}
where $\Theta(x-a)$ is the Heaviside step function.

\section{Solution for the Adam-Delbr\"uck's scenario}
\label{sec:AD}

We here start from the convolution-like relation \eqref{z} and show how
the integral over the arrival point on the inner sphere can be evaluated.
We recall that the Laplace transforms $\tilde{j}(p,\s|\x)$ and $\tilde{H}
^{\rm surf}(p;\s)$ were given by the explicit relations \eqref{eq:jtild}
and \eqref{q}. To proceed, we use the addition theorem for (normalised)
spherical harmonics $Y_{mn}(\theta,\phi)$,
\begin{equation}
\label{eq:addition}
\frac{2n+1}{4\pi}P_n\left(\frac{(\s\cdot\x)}{|\s|\,|\x|}\right)=\sum
\limits_{m=-n}^nY_{mn}^*(\theta,\phi)Y_{mn}(\theta',\phi'),
\end{equation}
where $\x=(r,\theta,\phi)$ and $\s=(R_1,\theta',\phi')$. We substitute
this expression into equation \eqref{eq:jtild} and then into equation
\eqref{z}. As $\tilde{H}^{\rm surf}(p;\theta')$ does not depend on
$\phi'$ due to the axial symmetry, the integral over $\phi'$ cancels all
contributions in the sum over $m$, except for $m=0$:
\begin{align*}
\tilde{H}^{\rm AD}(p;\x)&=\int\limits_0^\pi d\theta'\,\sin\theta'\int
\limits_0^{2\pi} d\phi'\,\sum\limits_{n=0}^\infty g_n^{(p)}(r)\sum\limits
_{m=-n}^nY_{mn}^*(\theta,\phi)Y_{mn}(\theta',\phi')\,\tilde{H}^{\rm surf}
(p;\theta')\\
&=\sum\limits_{n=0}^\infty g_n^{(p)}(r)(n+1/2)P_n(\cos\theta)\,c_n^{(p)}(\ve),
\end{align*}
where
\begin{align*}
c_n^{(p)}(\ve)&=\int\limits_0^{\pi}d\theta'\,\sin\theta'\,P_n(\cos\theta')\,
\tilde{H}^{\rm surf}(p;\theta')\\
&=\int\limits_0^{\ve}d\theta'\,\sin\theta'P_n(\cos\theta')+\frac{1}{P_{\mu}
(a)}\int\limits_{\ve}^{\pi}d\theta'\,\sin\theta'\,P_n(\cos \theta')P_{\mu}
(\cos(\pi-\theta'))\\
&=\frac{P_{n-1}(\cos\ve)-P_{n+1}(\cos\ve)}{2n+1}+\frac{(-1)^n}{P_{\mu}(a)}
I_n^{(p)}(-\cos\ve),
\end{align*}
with 
\begin{equation}
I_n^{(p)}(a)=\int\limits_a^1 dx\,P_n(x)P_{\mu}(x),
\end{equation}
where we used $P_n(-x)=(-1)^nP_n(x)$. Multiplying the Legendre equation
for $P_\mu(x)$ by $P_n(x)$, subtracting its symmetrised version, and
integrating over $x$ from $a$ to $1$, one immediately finds
\begin{equation}
I_n^{(p)}(a)=(1-a^2)\frac{P'_n(a)P_{\mu}(a)-P'_{\mu}(a)P_n(a)}{n(n+1)-\mu
(\mu+1)}.
\end{equation}
Substituting this expression, we get
\begin{align}
c_n^{(p)}(\ve)&=\frac{P_{n-1}(\cos\ve)-P_{n+1}(\cos\ve)}{2n+1}-\frac{1-
\cos^2\ve}{\frac{pR^2}{D_s}+n(n+1)}\,\frac{P'_n(\cos\ve)P_{\mu}(a)+P_n
(\cos\ve)P'_{\mu}(a)}{P_{\mu}(a)},
\end{align}
where we used $P'_n(-x)=(-1)^{n+1}P'_n(x)$, and $-\mu(\mu+1)=pR^2/D_s$.  

The additive structure of the coefficients $c_n^{(p)}(\ve)$ suggests to
represent $\tilde{H}^{\rm AD}(p;\x)$ as a linear combination of the two
contributions
\begin{equation}
\label{eq:Hp_final0}
\tilde{H}^{\rm AD}(p;\x)=\tilde{H}_1(p;\x)-\tilde{H}_2(p;\x),
\end{equation}
where
\begin{align*}
\tilde{H}_1(p;\x)&=\frac12\sum\limits_{n=0}^\infty g_n^{(p)}(r)P_n(\cos
\theta)\bigl[P_{n-1}(\cos\ve)-P_{n+1}(\cos\ve)\bigr],\\
\tilde{H}_2(p;\x)&=(1-\cos^2\ve)\sum\limits_{n=0}^\infty g_n^{(p)}(r)
P_n(\cos\theta)\frac{n+1/2}{\frac{pR^2}{D_s}+n(n+1)}\\
&\times\frac{P'_n(\cos\ve)P_{\mu}(a)+P_n(\cos\ve)P'_{\mu}(a)}{P_{\mu}(a)}.
\end{align*}
If the particle starts on the inner sphere, $r=R_1$, one has $g_n^{(p)}(
R_1)=1$, and the first contribution is simply $\tilde{H}_1(p;\x)=\Theta(
\ve-\theta)$. Indeed, it is equal to unity when $\theta<\ve$ (i.e., the
start is on the target), and to $0$ otherwise. In general, for $r>R_1$,
this contribution corresponds to the direct arrival onto the target. This
can be seen by integrating the probability flux density $\tilde{j}(p,\s|
\x)$ from \eqref{eq:jtild} over the target, yielding precisely $\tilde{H}
_1(p;\x)$. Note that equation \eqref{eq:Hp_final0} is reproduced in the
main text as equation \eqref{eq:Hp_final}. When the starting point $\x$
is uniformly distributed over a sphere of radius $r$, all terms with $n>
0$ are cancelled, and one gets the much simpler expression \eqref{eq:Hsurfav}.

\subsection{Short-time behaviour}

It is instructive to analyse the large-$p$ asymptotic behaviour of
equation \eqref{eq:Hsurfav} that corresponds to the short-time
behaviour of $\overline{H^{\rm AD}(t;r)}$.  As $p\to\infty$
one has $\mu\approx-\tfrac{1}{2}+i\sqrt{pR_1 ^2/D_s}$, i.e.,
$P_{\mu}(z)$ is close to a conical function, for which
\begin{equation}
P_\mu(z)\approx P_{-\frac12+i\eta}(z)\approx\biggl(\frac{\theta}{\sin
(\theta)}\biggr)^{1/2}\,I_0(\eta\theta)\quad(\eta\to\infty),
\end{equation}
where $\eta=\sqrt{pR_1^2/D_s}$, $\theta=\acos(z)$, and $I_0(z)$ is the
modified Bessel function of the first kind. Using the large-$z$ behaviour
of $I_0(z)$, one gets
\begin{equation}
P_\mu(z)\approx\frac{\exp(\eta\,\acos(z))}{\sqrt{2\pi\eta}\,(1-z^2)^{1/4}}
\quad(\eta\to\infty).
\end{equation}
Evaluating the derivative with respect to $z$, we find
\begin{equation}
\frac{P'_\mu(z)}{P_\mu(z)}\approx-\frac{\eta}{\sqrt{1-z^2}}+\frac{z}{2
(1-z^2)}\quad(\eta\to\infty),
\end{equation}
where we kept the second (sub-leading) term. As a consequence,
\begin{equation}
\frac{P'_\mu(a)}{P_\mu(a)}\approx-\frac{\sqrt{pR_1^2/D_s}}{\sin(\ve)}-
\frac{\cos(\ve)}{2\sin^2(\ve)}\quad(p\to\infty).
\end{equation}
Substituting this expression into equation \eqref{eq:Hsurfav} and using
$g_0^{(p)}(r)\approx\frac{R_1}{r}e^{-(r-R_1)\sqrt{p/D_b}}$ as $p\to\infty$,
we get
\begin{align}
\nonumber
\overline{\tilde{H}^{\rm AD}(p;r)}&\simeq\frac{(1-\cos(\ve))R_1}{2 r}  
\exp\left(-(r-R_1)\sqrt{\frac{p}{D_b}}\right)\\
\label{lm0}
&\times\left(1+\frac{1+\cos(\ve)}{\sin(\ve)R_1}\sqrt{\frac{D_s}{p}}+\frac{
D_s\cos(\ve)}{2(1-\cos(\ve))pR_1^2}\right).
\end{align}
One can thus evaluate the inverse Laplace transform term by term, yielding
equation \eqref{lm00}.

\subsection{Long-time behaviour}
\label{sec:long-time}

The long-time behaviour of the PDF $H^{\rm AD}(t;\x)$ is determined by the
largest pole of the Laplace transform $\tilde{H}^{\rm AD}(p;\x)$ (we recall
that all poles are negative and thus the largest pole has the smallest
absolute value). In the special case of the starting point on the inner
sphere, $|\x|=R_1$, there is no bulk diffusion, and the ADS is reduced to
the surface tour alone, i.e., $H^{\rm AD}(t;\s)=H^{\rm surf}(t;\s)$, which
was studied earlier (see, e.g., \cite{Grebenkov19sphere}).  We exclude
therefore this special case and assume that $|\x|>R_1$.

According to equation \eqref{z}, the set of poles of this function is the
union of the poles $\{p_k^{\rm bulk}\}$ of $\tilde{j}(p,\s|\x)$ and of the
poles $\{p_k^{\rm surf}\}$ of $\tilde{H}^{\rm surf}(p;\s)$. The former are
determined as the poles of the functions $g_n^{(p)}(r)$; in particular, the
largest pole $p_0^{\rm bulk}$ is the pole of $g_0^{(p)}(r)$, which is given
by equation \eqref{eq:p0_bulk}. In turn, the largest pole $p_0^{\rm surf}$
is determined by the smallest zero $\nu_0$ of equation \eqref{roots} as
\begin{equation}
p_0^{\rm surf}=-\frac{1}{\tau^{\rm surf}}<0, 
\end{equation}
with $\tau^{\rm surf}$ given by equation \eqref{eq:tau_surf}. The
long-time behaviour of $H^{\rm AD}(t;\x)$ is determined by $p_0=\max\{p_0^{
\rm bulk},p_0^{\rm surf}\}$. As $p_0^{\rm surf}$ depends on the target size
$\ve$, whereas $p_0^{\rm bulk}$ does not, one has $p_0^{\rm bulk}\ne p_0^{\rm
surf}$ in the general case. While the analysis is similar for $p_0^{\rm bulk}
=p_0^{\rm surf}$, we do not discuss this special case.

The long-time behaviour of $H^{\rm AD}(t;\x)$ can be determined by evaluating
the inverse Laplace transform by the residue theorem and keeping only the
contribution from the largest pole. We separately consider two situations:

(i) If $p_0=p_0^{\rm surf}>p_0^{\rm bulk}$, the function $\tilde{j}(p,\s|\x)$
is not singular at $p=p_0$ and is thus kept in equation \eqref{z}; in turn,
we only keep the singular term with $n=0$ from expansion \eqref{eq:Hp_sphere}
of $\tilde{H}^{\rm surf}(p;\s)$. The residue theorem then yields
\begin{equation}
\label{eq:HAD_p0surf}
H^{\rm AD}(t;\x)\simeq e^{p_0t}\biggl(2\pi R_1^2\int\limits_a^1 dx'\,\tilde{
j}(p_0,(\acos(x'),0)|\x)\frac{D_s\sin^2(\ve)}{R_1^2}\,b_0^2P_{\nu_0}(x')
P'_{\nu_0}(a)\biggr),
\end{equation}
where $b_0^2$ is given by equation \eqref{eq:bn}, $a=-\cos(\ve)$, and the
integral over $\partial\Omega$ was reduced to the integral over $\theta'$,
which is written in terms of $x'=\cos(\theta')$. This integral can be
evaluated explicitly by use of equation \eqref{eq:jtild} and the addition
theorem \eqref{eq:addition}. However, it is more convenient to focus on the
average when the starting point is uniformly distributed on a sphere of
radius $r$, for which
\begin{equation}
\overline{\tilde{j}(p,\s|\x)}\equiv\frac{1}{4\pi r^2}\int\limits_{|\x|=r}d\x
\,\tilde{j}(p,\s|\x)=\frac{g_0^{(p)}(r)}{4\pi R_1^2},
\end{equation}
and thus equation \eqref{eq:HAD_p0surf} becomes
\begin{align}
\label{eq:HAD_p0surf_av}
\overline{H^{\rm AD}(t;r)}\simeq e^{p_0t}\biggl(\frac{g_0^{(p_0)}(r)}{2}
\int\limits_{a}^1dx'\frac{D_s\sin^2(\ve)}{R_1^2}\,b_0^2 P_{\nu_0}(x')P'_
{\nu_0}(a)\biggr)\simeq e^{p_0t}C_\ve(r),
\end{align}
with
\begin{equation}
\label{eq:Cve_surf}
C_\ve(r)=g_0^{(p_0)}(r)\frac{D_s\sin^4(\ve)}{2R_1^2}\,\frac{b_0^2[P'_{\nu_0}
(-\cos(\ve))]^2}{\nu_0(\nu_0+1)},
\end{equation}
where we used the identity
\begin{equation}
\label{eq:Pnu_identity}
\int\limits_a^1dx\,P_\nu(x)=\frac{1-a^2}{\nu(\nu+1)}P'_\nu(a).
\end{equation}
Note that equation \eqref{eq:gnI} implies for $p_0 < 0$ that
\begin{equation}
g_0^{(p_0)}(r)=\frac{R_1}{r}\,\frac{R_2\sqrt{|p_0|/D_b}\cos(\sqrt{|p_0|/
D_b}(R_2-r))-\sin(\sqrt{|p_0|/D_b}(R_2-r))}{R_2\sqrt{|p_0|/D_b}\cos(\sqrt{
|p_0|/D_b}(R_2-R_1))-\sin(\sqrt{|p_0|/D_b}(R_2-R_1))}.
\end{equation}

(ii) If $p_0=p_0^{\rm bulk}>p_0^{\rm surf}$, the function $\tilde{H}^{\rm
surf}(p;\s)$ is not singular at $p=p_0$ and is thus kept in equation
\eqref{z}; in turn, we only keep the singular term with $n=0$ from
expansion \eqref{eq:jtild} of $\tilde{j}(p,\s|\x)$. The residue theorem
then yields
\begin{equation}
\label{eq:HAD_p0bulk}
H^{\rm AD}(t;\x)\simeq e^{p_0 t}C_{\ve}(r),
\end{equation}
with
\begin{equation}
\label{eq:Cve_bulk}
C_\ve(r)=\frac12\res_{p_0}\{g_0^{(p)}(r)\}\int\limits_0^\pi d\theta'\,
\sin(\theta')\tilde{H}^{\rm surf}(p_0;\theta').
\end{equation}
As $p_0$ is a pole of $g_0^{(p)}(r)$, other radial functions disappeared,
and the result does not depend on the starting point $\x$.

Note that the integral over $\theta'$ can be evaluated explicitly by use of
equation \eqref{q},
\begin{align*}
&\int\limits_0^\pi d\theta'\,\sin(\theta')\tilde{H}^{\rm surf}(p_0;\theta')
=\int\limits_0^{\ve}d\theta'\,\sin(\theta')\,1+\int\limits_{\ve}^\pi d\theta'
\,\sin(\theta')\frac{P_{\mu_0}(-\cos(\theta'))}{P_{\mu_0}(a)}\\
&\quad=(1-\cos(\ve))+\int\limits_a^1 dx'\,\frac{P_{\mu_0}(x')}{P_{\mu_0}(a)}
=(1-\cos(\ve))-\frac{P'_{\mu_0}(a)(1-a^2)D_s}{P_{\mu_0}(a)\,p_0 R_1^2},
\end{align*}
where $\mu_0=\tfrac{1}{2}(-1+\sqrt{1-4p_0R_1/D_s})$, and we used the identity
\eqref{eq:Pnu_identity}, the relation $\mu_0(\mu_0+1)=-p_0R_1^2/D_s$ and
recalled that $\tilde{H}^{\rm surf}(p;\theta')=1$ for any $0\leq\theta'\leq
\ve$ (on the target). The residue of $g_0^{(p)}(r)$ can also be found by
direct computation,
\begin{align}
\nonumber
\res_{p_0}\{g_0^{(p)}(r)\}&=\frac{2D_b\alpha_0 R_1\sqrt{1+\beta^2\alpha_0^2}}{
r(R_2-R_1)^2(1+\beta(\beta\alpha_0^2-1))}\biggl[\beta\alpha_0\cos\biggl(
\alpha_0\frac{R_2-r}{R_2-R_1}\biggr)-\sin\biggl(\alpha_0\frac{R_2-r}{R_2
-R_1}\biggr)\biggr],
\end{align}
where $\beta=R_2/(R_2-R_1)$.

Combining equations \eqref{eq:HAD_p0surf_av} and \eqref{eq:HAD_p0bulk}, we
obtain relation \eqref{mm3}, in which the amplitude $C_\ve(r)$ is given
either by equation \eqref{eq:Cve_surf} for $p_0^{\rm surf} > p_0^{\rm
bulk}$, or by equation \eqref{eq:Cve_bulk} for $p_0^{\rm surf} < p_0^{\rm
bulk}$.

\subsection{Mean first-passage time}

Evaluating the derivative of $\tilde{H}^{\rm AD}(p;\x)$ at $p=0$ we obtain
the MFPT within the ADS. To get some qualitative picture we start from the
general representation \eqref{z},
\begin{align*}
T^{\rm AD}(\x)&=-\lim\limits_{p\to0}\frac{\partial\tilde{H}^{\rm AD}(p;\x)}{
\partial p}\\ 
&=\int\limits_{\pa}d\s\,\biggl[\biggl(-\partial_p\tilde{j}(p,\s|\x)\biggr)_{
p=0}\,\underbrace{\tilde{H}^{\rm surf}(0;\theta_{\s})}_{=1}+\tilde{j}(0,\s|\x)
\underbrace{\biggl(-\partial_p\tilde{H}^{\rm surf}(p;\s)\biggr)_{p=0}}_{=T^{
\rm surf}(\s)}\biggr].
\end{align*}
The first term can be interpreted as the MFPT to the inner sphere from a fixed
point $\x$, conditioned to arrive at point $\s$, and then averaged over all
locations $\s$. In turn, the second term is the average of the MFPT $T^{\rm
surf}(\s)$ on the surface, averaged over the arrival point $\s$ with the
harmonic measure density (factor $\tilde{j}(0,\s|\x)$). In practice, we can use the exact solution
\eqref{eq:Hp_final} to compute the MFPT. Even though the exact computation
of this limit is feasible, we consider the simpler case of the
surface-averaged MFPT
\begin{align*} 
\overline{T^{\rm AD}}&\equiv\frac{1}{4\pi r^2}\int\limits_{|\x|=r}d\x\,T^{
\rm AD}(\x)=-\lim\limits_{p\to0}\partial_p\overline{\tilde{H}^{\rm AD}(p;r)}\\
&=-\frac{1-\cos\ve}{2}\biggl\{\biggl(\partial_p g_0^{(p)}(r)\biggr)_{p=0}
\biggl(1-\frac{D_s(1+\cos\ve)}{pR_1^2}\,\frac{P'_{\mu}(a)}{P_{\mu}(a)}\biggr)
_{p=0}\\
&+\underbrace{\bigl(g_0^{(p)}(r)\bigr)_{p=0}}_{=1}\biggl(\partial_p\biggl(
\frac{D_s(1+\cos\ve)}{pR_1^2}\,\frac{P'_{\mu}(a)}{P_{\mu}(a)}\biggr)\biggr)
_{p=0}\biggr\}.
\end{align*}
In the first term, we have
\begin{equation}
\lim\limits_{p\to0}\biggl(\partial_pg_0^{(p)}(r)\biggr)=-\frac{(r-R_1)(2R_2^3
-rR_1(r+R_1))}{6rR_1D_b}.
\end{equation}
In the limit $p\to0$ we approximate $\mu\simeq-s-s^2+O(s^3)$ (with $s=pR_1^2/
D_s$), so that we can use the expansion (see \cite{Szmytkowski,Laurenzi})
\begin{equation}
P_\nu(x)=1+\nu\ln\biggl(\frac{1+x}{2}\biggr)-\nu^2\Li_2\biggl(\frac{1-x}{2}
\biggr)+O(\nu^3)\quad(\nu\to0),
\end{equation}
where $\Li_2(z)$ is the dilogarithm function of the second order, $L_2(z)=-
\int\nolimits_0^zdx\ln(1-x)/x$. We thus find
\begin{align}
P_{\mu}(a)&=1-s\ln\biggl(\frac{1+a}{2}\biggr)-s^2\biggl(\ln\biggl(\frac{1+
a}{2}\biggr)+\Li_2\biggl(\frac{1-a}{2}\biggr)\biggr)+O(s^3),\\
P'_{\mu}(a)&=-\frac{s}{1+a}-s^2\biggl(\frac{1}{1+a}+\frac{\ln((1+a)/2)}{1-a}
\biggr)+O(s^3),
\end{align}
so that
\begin{equation}
\frac{P_{\mu}(a)}{P'_{\mu}(a)}=-\frac{1+a}{s}\biggl(1-s\biggl(1+\frac{2}{1-a}
\ln\frac{1+a}{2}\biggr)+O(s^2)\biggr),
\end{equation}
and thus
\begin{equation}
\frac{D_s(1+\cos\ve)}{pR_1^2}\,\frac{P'_{\mu}(a)}{P_{\mu}(a)}=-\frac{1-a}{1+a}
\biggl(1+s\biggl(1+\frac{2}{1-a}\ln\frac{1+a}{2}\biggr)\biggr)+O(s^2).
\end{equation}
From this we conclude that
\begin{equation}
\overline{T^{\rm AD}}=(-\partial_p g_0(r))_{p=0}-\frac{(1-a)R_1^2}{2D_s}
\biggl(1+\frac{2}{1-a}\ln\frac{1+a}{2}\biggr),
\end{equation}
which can also be written more explicitly as equation \eqref{eq:Tsurfav}.

\subsection{Fixed vs random starting point}
\label{starting}

\begin{figure}
\includegraphics[width=8cm]{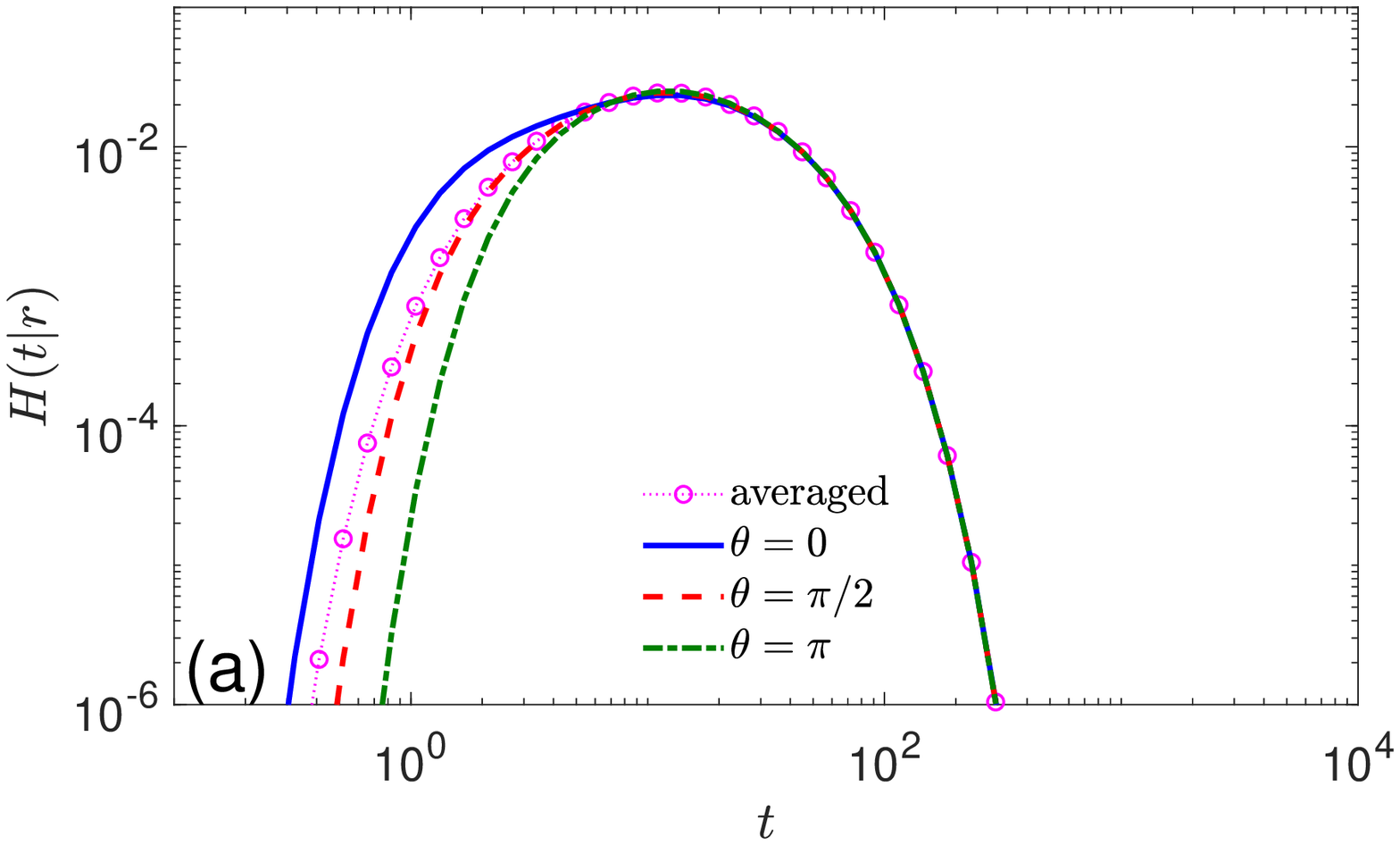}
\includegraphics[width=8cm]{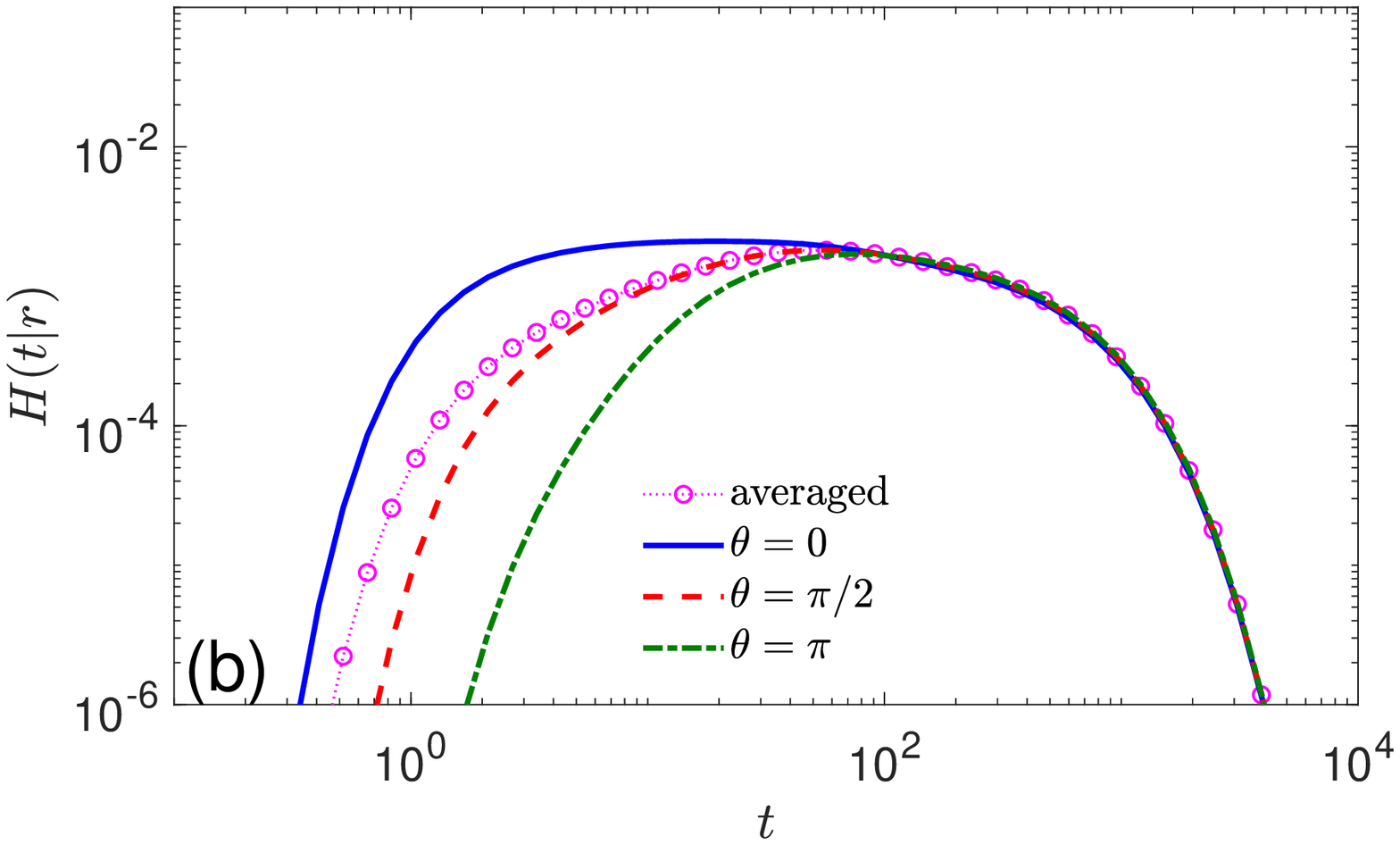}
\caption{FPT PDF to a circular target located on the North pole of the inner
sphere within the ADS, for $R_1=1$, $R_2=5$, $r=5$, $\ve=0.1$, $D_s/D_b=1$ {\bf
(a)}, and $D_s/D_b=0.01$ {\bf (b)}. The surface-averaged PDF $\overline{H^{\rm
AD}(t;r)}$ (circles) is compared to the PDF $H^{\rm AD}(t;\x)$ with a fixed
starting point $\x=(r,\theta,0)$, for three choices of $\theta$, as indicated
in the legend. All PDFs were obtained via numerical inversion of the Laplace
transform by the Talbot algorithm.}
\label{fig:Ht_starting}
\end{figure}

As argued in section \ref{comp}, the average of the PDF when the starting
point is uniformly distributed on a sphere of given radius $r$ is often more
representative than the case of a fixed starting point. Here we briefly
discuss the differences between these two situations. As expected, the
long-time behaviour of the PDF does not depend on the starting point, whereas
the short-time behaviour is generally much more sensitive to the distance
between the starting point and the target. Figure \ref{fig:Ht_starting}
illustrates this point for the geometric setting considered in the main
text (with $R_1=1$ and $r=R_2=5$), except that we take a larger target size,
$\ve=0.1$. Note that this larger value actually enhances possible differences
due to the starting point since a larger target is found faster and thus the
diffusing particle has less time to loose the information on the starting
point. When $D_s=D_b$ (panel (a)), the left tail of the PDF is most shifted
to the left (to shorter times) when the fixed starting point is located on
the North pole of the outer sphere, i.e., right above the target (located on
the North pole of the inner sphere). In other words, this setting favours the
fastest arrival onto the target, as expected. In particular, it is faster than
in the case of random starting point, which is a sort of weighted average
with different distances to the target. In turn, when the fixed target is
located on the equator ($\theta=\pi/2$) or on the South pole ($\theta=\pi$)
of the outer sphere, the surface averaged PDF provides the faster arrival at
short times. Overall, the difference between the four cases is moderate. This
difference is, however, increased when $D_s/D_b=0.01$ (panel (b)).

\section{PDF for one-stage search process}
\label{direct}

The exact spectral solution for the Laplace-transformed FPT PDF in the
conventional one-stage scenario was presented in \cite{9}.  This
solution involves an infinite-dimensional matrix with explicitly known
elements. In practice, one has to truncate this matrix and then
numerically invert it.

In order to avoid these numerical steps, we use the self-consistent
approximation developed and validated in \cite{9}. The approximate solution
reads
\begin{equation}
\label{eq:Happ_inner}
\tilde{H}_{\rm dir}^{\rm app}(p;\x)=J_p\sum\limits_{n=0}^\infty g_n^{(p)}(r)
\,P_n(\cos\theta)\,\frac{1}{\mu_n^{(p)}}\,\frac{P_{n-1}(\cos\ve)-P_{n+1}
(\cos\ve)}{2},
\end{equation}
where
\begin{equation}
\label{eq:J_inner}
J_p=\left(\frac{1}{q}+\frac{1}{2(1-\cos\ve)}\sum\limits_{n=0}^\infty\frac{
\bigl(P_{n-1}(\cos\ve)-P_{n+1}(\cos\ve)\bigr)^2}{(2n+1)\mu_n^{(p)}}\right)
^{-1},
\end{equation}
and
\begin{equation}
\label{eq:mupI}
\mu_n^{(p)}=-\biggl(\frac{\partial g_n^{(p)}(r)}{\partial r}\biggr)_{r = R_1}.
\end{equation}
Here, there is no need for matrix inversion as the expression is fully
explicit. Moreover, the limit $q\to\infty$ can be easily obtained by
simply setting $q=\infty$ in equation \eqref{eq:J_inner}. While the
inverse Laplace transform of equation \eqref{eq:Happ_inner} can be found
exactly via the residue theorem (see \cite{9} for details), we resort
to a numerical inversion by the Talbot algorithm. When the starting
point $\x$ is uniformly distributed over a sphere of radius $r$,
equation \eqref{eq:Happ_inner} is reduced to
\begin{equation}
\label{eq:Happ_inner_surfav}
\overline{\tilde{H}_{\rm dir}^{\rm app}(p;r)}=\frac{J_pg_0^{(p)}(r)}{2
\mu_0^{(p)}}(1-\cos\ve).
\end{equation}

The approximate solution \eqref{eq:Happ_inner} can also be used to compute
the moments of the FPT, in particular, the MFPT was derived in Appendix B
of \cite{9}. The surface average of equation (B.9) from \cite{9} reads
\begin{equation}
\label{eq:Tdir_surfav}
\overline{T_{\rm dir}^{\rm app}}=\frac{R_2^3-R_1^3}{3D_bR_1}\bigg[C_\ve+
\frac{5R_2^3-9R_2^2R_1+1}{5(R_2^3-R_1^3)}-R_1\frac{5r^3+10R_2^3-3r(R_1^2
+6R_2^2)}{10r(R_2^3-R_1^3)}\biggr],
\end{equation}
where
\begin{equation}
\label{eq:c_series}
C_\ve=\frac{2}{(1-\cos\ve)R_1q}+\frac{1}{(1-\cos\ve)^2}\sum\limits_{n=1}
^\infty\frac{(P_{n-1}(\cos\ve)-P_{n+1}(\cos\ve))^2}{(2n+1)R_1\mu_n^{(0)}},
\end{equation}
and
\begin{equation}
\mu_n^{(0)}=\frac{n(n+1)}{R_1}\,\frac{1-(R_1/R_2)^{2n+1}}{n+(n+1)(R_1/R_2)
^{2n+1}}.
\end{equation}
When $R_1\ll R_2$ the factor $(R_1/R_2)^{2n+1}$ can be neglected, so that
$\mu_n^{(0)}\approx(n+1)/R_1$. The asymptotic behaviour of series like in
equation \eqref{eq:c_series} was analysed in \cite{q2}. Since $R_1 \mu_n^
{(0)}$ can be interpreted as $Rg'_n(R)/g_n(R)$ from the Supplementary
Material of \cite{q2}, one has $1/(R_1 \mu_n^{(0)})\approx1/(n+1)=1/n-1/
n^2+O(n^{-3})$ such that one can apply equation (S76) with $\omega=3$
from equation (S56). This yields the small-$\ve$ behaviour
\begin{equation}
\label{eq:c_series_eps}
C_\ve\approx\frac{4}{R_1q}\ve^{-2}+\frac{32}{3\pi}\ve^{-1}-\ln(1/\ve)+O(1)
\quad(\ve\to0).
\end{equation}
When the target is partially reactive ($q<\infty$) the dominant term scales
as $\ve^{-2}$, showing that the direct search is reaction-limited. In turn,
for a perfect target, on which we focus in this paper, the first term
disappears, and one gets the $\ve^{-1}$-behaviour, with a logarithmic
correction. As discussed in \cite{q2} for the problem of the narrow escape
from a sphere, the numerical factor $32/(3\pi)\approx3.40$ differs by only
$8\%$ from the exact value $\pi$, which was known for that problem. This
minor difference comes from the self-consistent approximation. We expect
therefore that substitution of $C_\ve\approx\pi/\ve$ in equation
\eqref{eq:Tdir_surfav} would result in the correct leading-order
behaviour of $\overline{T_{\rm dir}}$ in our setting. In particular,
when $\rho\ll R_1\ll R_2$ (and thus $\ve\ll1$), we get $\overline{T_{\rm
dir}^{\rm app}} \approx\tau_{\rm dir}=R_2^3/(3D_b\rho)$ that was introduced
in section \ref{sec:intro}.

\subsection*{Limit $R_1\to R_2$}

This is an interesting limit in which the thickness of the shell region
between the outer and inner spheres shrinks, such that a particle is
deemed to perform a diffusive motion almost on the surface of a sphere
and seeking a circular target on that surface.

Setting $R_1=R-\delta$, $R_2=R$ and $r=R$, we get from \eqref{eq:mupI}, to
the leading order,
\begin{equation}
\mu_n^{(p)}\simeq\delta\,\biggl(\frac{p}{D}+\frac{n(n+1)}{R^2}\biggr)+O(
\delta^2).
\end{equation}
In this limit, the self-consistent approximation \eqref{eq:Happ_inner}
yields
\begin{align}
\nonumber
\tilde{H}^{\rm app}_{\rm dir}(p;\x)&=(1-\cos\ve)\biggl(\sum\limits_{n=0}^
\infty\frac{(P_{n-1}(\cos\ve)-P_{n+1}(\cos\ve))^2}{(2n+1)\bigl(\frac{R^2p}{
D}+n(n+1)\bigr)}\biggr)^{-1}\\
\label{eq:self_surface}
&\times\sum\limits_{n=0}^\infty\frac{P_n(\cos\theta)(P_{n-1}(\cos\ve)-P_{n
+1}(\cos\ve))}{\bigl(\frac{R^2p}{D}+n(n+1)\bigr)}.
\end{align}
One can easily check that $\tilde{H}^{\rm app}_{\rm dir}(p;\x)$
converges to unity as $p\to0$, as it should to fulfill the
normalisation condition of the PDF. Note that both series can be
evaluated explicitly by using the identities \eqref{eq:identity2b} and
\eqref{eq:identity3}.

\begin{figure}
\begin{center}
\includegraphics[width=10cm]{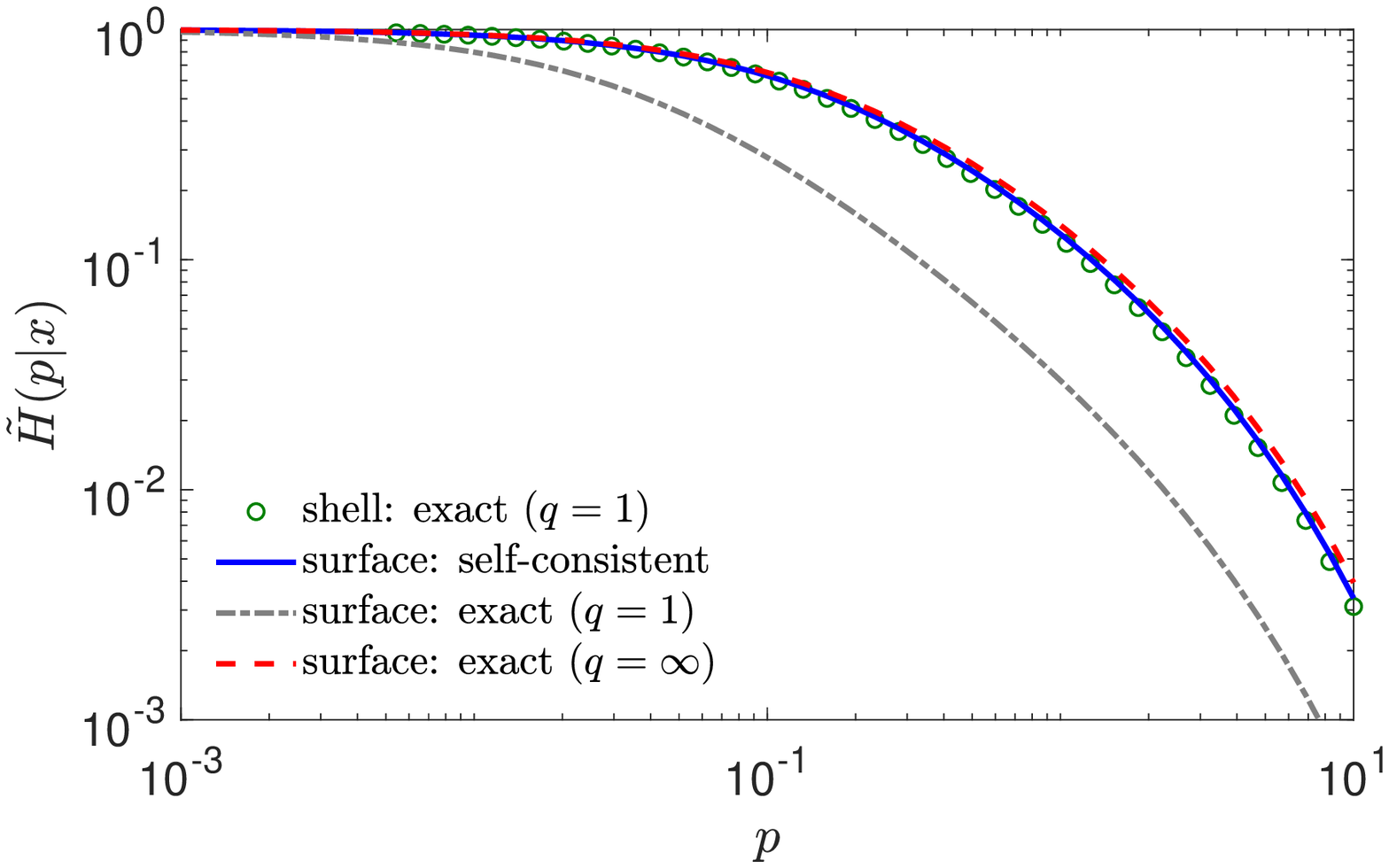}
\end{center}
\caption{Laplace-transformed PDF $\tilde{H}_{\rm dir}(p;\x)$ for diffusion
in a thin spherical shell between spheres of radii $R_1=0.999$ and
$R_2=1$ towards a circular target of angular size $\ve=0.1$ located at
the North pole, with $r=1$ and $\theta=\pi/2$. Green circles show the
exact solution from \cite{9} with $q=1$ (truncated at $n_{\rm
max}=100$) and the solid blue line represents equation
\eqref{eq:self_surface} (the self-consistent approximation in the
limit $\delta\to0$). For comparison, the Laplace-transformed PDF
$\tilde{H}^{\rm surf}(p;\theta)$ from equation
\eqref{eq:Hpsurf_general} for surface diffusion with $D_s=D_b$ towards the
same target is shown by the dash-dotted grey line for $q=1$ and by the
dashed red line for $q=\infty$.}
\label{fig:thin}
\end{figure}

Curiously, the approximation \eqref{eq:self_surface} does not depend on the
reactivity parameter $q=\kappa/D_b$. This behaviour is illustrated in figure
\ref{fig:thin}, in which we compare the Laplace-transformed PDF for
diffusion inside a thin shell (of width $0.001$) to that for surface
diffusion on the sphere. There is good agreement between the three curves:
the exact solution for a thin shell with $q=1$, the self-consistent
approximation \eqref{eq:self_surface} (independent of $q$), and the exact
solution $\tilde{H}^{\rm surf}(p;\theta)$ from \eqref{eq:Hpsurf_general} for
surface diffusion on the sphere towards a perfectly reactive target ($q=
\infty$). In contrast, the exact solution for surface diffusion on the
sphere towards a partially reactive target ($q=1$) stands out. We conclude
that the Laplace-transformed PDF in a very thin shell is close to that for
surface diffusion towards a perfect target.

To rationalise this behaviour, let us consider diffusion in a very thin
stripe, $\R\times(0,\delta)$, on which an interval $(-R,R)\times\{0\}$
represents the target. Once a particle enters the proximity of the target
(i.e., the rectangle $(-R,R)\times(0,\delta)$), it hits the target a very
large number of times, so that the reaction occurs very rapidly after
entering this zone. In the limit $\delta\to0$, the number of encounters
grows, such that the effective target on the line is perfectly reactive.
The same argument holds for three-dimensional diffusion between parallel
planes, $\R^2\times(0,\delta)$, when the target is a disk $\{(x,y,0)\in\R^3:
x^2+y^2<R^2\}$. This is equivalent to our setting because the curvature
of the spherical shell does not matter in the limit $\delta\to0$.

\subsection*{Numerical implementation}

{\clr
Even though the approximative solution \eqref{eq:Happ_inner} is fully
explicit, its numerical computation may be challenging, especially for
small $\ve$.  In fact, one needs to truncate the infinite series in
\eqref{eq:J_inner}, and the truncation order $\nmax$ has to be large
when $\ve$ is small.  At the same time, the numerical evaluation of
the radial function $g_n^{(p)}$ that involve $i_n(z)$ and $k_n(z)$ and
determine $\mu_n^{(p)}$, becomes unstable for large $n$.  Moreover, as
one needs to perform an inverse Laplace transform of $\tilde{H}_{\rm
dir}^{\rm app}(p;\x)$ to get back to the time domain, this computation
has to be realised for any $p\in\C$. In order to overcome these
numerical difficulties, we adopt the following two-step scheme.

In the first step, we approximate the radial functions with large $n$ as
\begin{equation}
g_n^{(p)}(r)\approx g_{n,\infty}^{(p)}(r)=\frac{k_n(\alpha r)}{k_n(\alpha
R_1)},
\end{equation}
where $\alpha=\sqrt{p/D}$ and $g_{n,\infty}^{(p)}(r)$ are the radial
functions for the problem without outer sphere (i.e., for $R_2 =
\infty$).  This approximation can be easily justified in the limit
$p\to \infty$ for any $r < R_2$ and any $n$ just by looking at the
asymptotic behaviour of the modified spherical Bessel functions
$i_n(z)$ and $k_n(z)$ in \eqref{eq:gnI}.  We checked numerically that
this approximation is also applicable for smaller $p$ if $r \ll R_2$
and $n \gg 1$.

In the second step, we employ the recurrence relations for $k_n(z)$ to
enable an iterative computation of the radial function.  In fact, one
has
\begin{equation}
k_{n+1}(z)=k_{n-1}(z)+\frac{2n+1}{z}k_n(z),\qquad k'_n(z)=-k_{n-1}(z)-
\frac{n+1}{z}k_n(z).
\end{equation}
Using the first relation, one gets
\begin{equation}
g_{n+1,\infty}^{(p)}(r)=\frac{k_{n-1}(\alpha r)+\frac{2n+1}{\alpha r}
k_n(\alpha r)}{k_{n-1}(\alpha R_1)+\frac{2n+1}{\alpha R_1}k_n(\alpha R_1)}
=g_{n,\infty}^{(p)}(r)\frac{R_1}{r}\frac{1+\frac{f_n(\alpha r)}{2n+1}}{1+
\frac{f_n(\alpha R_1)}{2n+1}},
\end{equation}
where
\begin{equation}
f_n(z)=z\frac{k_{n-1}(z)}{k_n(z)}.
\end{equation}
Using again the first recurrence relation and the explicit form $k_0(z)=
e^{-z}/z$ one finds
\begin{equation}
f_1(z)=\frac{z^2}{z+1},\qquad f_n(z)=\frac{z^2}{f_{n-1}(z)+2n-1}\quad
(n=2,3,\ldots).
\end{equation}
One sees that the $f_n(z)$ are rational functions, which can be computed
iteratively. These functions have the following asymptotic behaviours
\begin{equation}
f_n(z)\simeq\frac{z^2}{2n-1}-\frac{z^4}{(2n-1)^2(2n-3)}+O(z^6)\quad(z\to0)
\end{equation}
for $n>1$ (note that $f_1(z)=z^2-z^3+\ldots$ contains the $z^3$ term), and
\begin{equation}
f_n(z)\simeq z-n+\frac{n(n+1)}{2z}+O(z^{-2})\quad(z\to\infty).
\end{equation}

Finally, we can approximate
\begin{equation}
\mu_n^{(p)}\approx\mu_{n,\infty}^{(p)}=-\biggl(\partial_r g_{n,\infty}^{(p)}(r)
\biggr)_{r=R_1}=-\alpha\frac{k'_n(\alpha R_1)}{k_n(\alpha R_1)}.
\end{equation}
Using the second recurrence relation, one then gets
\begin{equation}
\label{eq:mun_approx}
R_1\mu_{n,\infty}^{(p)}=n+1+f_n(\alpha R_1)\qquad(n=1,2,\ldots)
\end{equation}
(note that $R_1\mu_{0,\infty}^{(p)}=1+\alpha R_1$). We then use the following
numerical approximation
\begin{align}
\nonumber
J_p&\approx\left(\frac{1}{q}+\frac{1}{2(1-\cos\ve)}\left\{\sum\limits_{n=0}^{
n_{\rm app}}\frac{\bigl(P_{n-1}(\cos\ve)-P_{n+1}(\cos\ve)\bigr)^2}{(2n+1)
\mu_n^{(p)}}\right.\right.\\
\label{eq:J_inner_app}
&\left.\left.+\sum\limits_{n=n_{\rm app}+1}^{\nmax}\frac{\bigl(P_{n-1}(\cos\ve)
-P_{n+1}(\cos\ve)\bigr)^2}{(2n+1)\mu_{n,\infty}^{(p)}}\right\}\right)^{-1},
\end{align}
in which we keep the first $n_{\rm app}+1$ terms with the exact form of $\mu_
n^{(p)}$ and replace $\mu_n^{(p)}$ by its approximation $\mu_{n,\infty}^{(p)}$
from \eqref{eq:mun_approx} for the terms with $n_{\rm app}<n\leq\nmax$. In this
way, the truncation order $\nmax$ can be made sufficiently large to ensure
accurate computations. We used $n_{\rm app}=50$ and $\nmax=250$.}


\section*{References}

\begin{thebibliography}{999}

\bibitem{adam} G. Adam and M. Delbr\"uck, 
{\it Reduction of Dimensionality in Biological Diffusion Processes},
In Structural Chemistry and Molecular Biology, A. Rich and
N. Davidson, Eds.  (W H Freeman and Co., San Francisco, 1968).

\bibitem{trur} H. J. Trurnit, {\it \"Uber monomolekulare Filme an
Wassergrenzfl\"achen und \"uber Schichtfilme}, 
In: Fortschritte der Chemie Organischer Naturstoffe, vol 4., A. Butenandt
et al. (eds) (Springer, Vienna, 1945).

\bibitem{buch} T. B\"ucher, {\it Probleme des Energietransports innerhalb
lebender Zellen}, Adv. Enzymol. {\bf 14}, 1 (1953).

\bibitem{eigen} M. Eigen, {\it Diffusion control in biochemical reactions}, 
in Quantum Statistical Mechanics in the Natural Sciences, S. L. Mintz and
S. N. Widmayer, eds.  (Plenum Press, New York, 1974), p. 37

\bibitem{richter} P. M. Richter and M. Eigen, {\it Diffusion controlled
reaction rates in spheroidal geometry. Application to repressor-operator 
association and membrane bound enzymes}, Biophys. Chem. {\bf 2}, 255 (1974).

\bibitem{kozak1} A. J. Frank, M. Gr\"atzel and J. J. Kozak, {\it On the
reduction of dimensionality in radical decay kinetics induced by micellar
systems}, J. Am. Chem. Soc. {\bf 98}, 3317 (1976).

\bibitem{berg} H. C. Berg and E. M. Purcell, {\it Physics of Chemoreception},
Biophys. J. {\bf 20}, 193 (1977).

\bibitem{ast} R. D. Astumian and Z. A. Schelly, {\it Geometric effects of
reduction of dimensionality in interfacial reactions}, J. Am. Chem. Soc.
{\bf 106}, 304 (1984).

\bibitem{kozak2} P. H. Lee and J. J. Kozak, {\it Lattice theory of reaction
efficiency in compartmentalized systems. II. Reduction of dimensionality}, 
J. Chem. Phys. {\bf 80}, 705 (1984).

\bibitem{szabo} D. Shoup, G. Lipari  and A. Szabo, {\em Diffusion-controlled
bimolecular reaction rates. The effect of rotational diffusion and orientation
constraints}, Biophys. J. {\bf 36}, 697 (1981).

\bibitem{zwan} R. Zwanzig and A. Szabo, {\it Time dependent rate of diffusion
influenced ligand binding to receptors on cell surfaces}, Biophys. J. {\bf 60},
671 (1991).

\bibitem{axelrod} D. Axelrod and M. D. Wang, {\it Reduction-of-Dimensionality
Kinetics at Reaction-Limited Cell Surface Receptors}, Biophys. J. {\bf 66},
588 (1994).

\bibitem{schmick} M. Schmick and P. I. H. Bastiaens, {\it The Interdependence
of Membrane Shape and Cellular Signal Processing}, Cell {\bf 156}, 1132 (2014).

\bibitem{mac} M. A. McCloskey and M-m Poo, {\it Rates of Membrane-associated
Reactions: Reduction of Dimensionality Revisited}, J. Cell Biology {\bf 102},
88 (1986).

\bibitem{berg2} O. G. Berg, R. B. Winter and P. H. von Hippel, {\it 
Diffusion-driven mechanisms of protein translocation on nucleic acids.
1. Models and theory}, Biochemistry {\bf 20}, 6929 (1981).


\bibitem{berg3} P. H. von Hippel and O. G. Berg, {\it Facilitated target
location in biological systems}, J. Biol. Chem. {\bf 264}, 675 (1989).

\bibitem{coppey} M. Coppey, O. B\'enichou, R. Voituriez and M. Moreau, 
{\it Kinetics of Target Site Localization of a Protein on DNA: A Stochastic
Approach}, Biophys. J. {\bf 87}, 1640 (2004). 

\bibitem{marco} S. E. Halford and J. F. Marko, {\it How do site-specific
DNA-binding proteins find their targets?}, Nucleic Acids Res. {\bf 32},
3040 (2004).

\bibitem{hu} T. Hu, A. Y. Grosberg and B. I. Shklovskii, {\it How Proteins
Search for Their Specific Sites on DNA: The Role of DNA Conformation}, 
Biophys. J. {\bf 90}, 2731 (2006).

\bibitem{mirny} L. Mirny, {\it Cell commuters avoid delays}, Nature Phys.
{\bf 4}, 93 (2008).

\bibitem{mirny2} L. Mirny, M. Slutsky, Z. Wunderlich, A. Tafvizi, J. Leith,
and A. Kosmrlj, {\it How a protein searches for its site on DNA: the mechanism
of facilitated diffusion}, J. Phys. A: Math. Theor. {\bf 42}, 434013 (2009).

\bibitem{lomholt} M. A. Lomholt, B. van den Broek, S-M. J. Kalisc, G. J. L.
Wuite and R. Metzler, {\it Facilitated diffusion with DNA coiling},
Proc. Natl. Acad. Sci. USA {\bf 106}, 8204 (2009).

\bibitem{lomholt1} B. van den Broek, M. A. Lomholt, S.-M. J. Kalisch, R.
Metzler, and G. J. L. Wuite, \emph{How DNA coiling enhances target
localization by proteins}, Proc. Natl. Acad. Sci. USA \textbf{105}, 15738
(2008).

\bibitem{klenin} K. V. Klenin, H. Merlitz, J. Langowski, and C.-X. Wu,
\emph{Facilitated Diffusion of DNA-Binding Proteins}, Phys. Rev. Lett.
\textbf{96}, 018104 (2006).

\bibitem{kolesov} G. Kolesov, Z. Wunderlich, O. N. Laikova, M. S. Gelfand,
and L. A. Mirny, \emph{How gene order is influenced by the biophysics
of transcription regulation}, Proc. Natl. Acad. Sci. USA \textbf{104},
13948 (2007).

\bibitem{otto} O. Pulkkinen and R. Metzler, \emph{Distance matters: the
impact of gene proximity in bacterial gene regulation}, Phys. Rev. Lett.
\textbf{110}, 198101 (2013).

\bibitem{slutsky} M. Slutsky and L. A. Mirny, \emph{Kinetics of Protein-DNA
Interaction: Facilitated Target Location in Sequence-Dependent Potential},
Biophys. J. \textbf{87}, 4021 (2004).

\bibitem{max} M. Bauer and R. Metzler, \emph{In vivo facilitated diffusion
model}, PLoS ONE \textbf{8}, e53956 (2013).

\bibitem{sheinman} M. Sheinman, O. B\'enichou, Y. Kafri and R. Voituriez, 
{\it Classes of fast and specific search mechanisms for proteins on DNA}, 
Rep. Prog. Phys. {\bf 75}, 026601 (2012).

\bibitem{peters} R. Peters, {\it Translocation through the nuclear pore
complex: Selectivity and speed by reduction-of-dimensionality}, Traffic
{\bf 6}, 421 (2005)\\
{\it Translocation through the nuclear pore: Kaps pave the way}, BioEssays
{\bf  31}, 466 (2009). 

\bibitem{benichou1} O. B\'enichou, M. Coppey, M. Moreau, P-H. Suet, and R.
Voituriez, {\em Optimal Search Strategies for Hidden Targets}, Phys. Rev.
Lett.{\bf  94}, 198101 (2005).



\bibitem{Benichou11a} O. B\'enichou, C. Loverdo, M. Moreau, and R. Voituriez,
{\it Intermittent search strategies}, Rev. Mod. Phys. {\bf 83}, 81 (2011).

\bibitem{Benichou10} O. B\'enichou, D. S. Grebenkov, P. Levitz, C. Loverdo,
and R. Voituriez, {\it Optimal Reaction Time for Surface-Mediated Diffusion},
Phys. Rev. Lett. {\bf 105}, 150606 (2010).

\bibitem{Benichou11} O. B\'enichou, D. S. Grebenkov, P. Levitz, C. Loverdo,
and R. Voituriez, {\it Mean First-Passage Time of Surface-Mediated Diffusion
in Spherical Domains}, J. Stat. Phys. {\bf 142}, 657-685 (2011).

\bibitem{Rupprecht12a} J.-F. Rupprecht, O. B\'enichou, D. S. Grebenkov, and
R. Voituriez, {\it Kinetics of Active Surface-Mediated Diffusion in
Spherically Symmetric Domains}, J. Stat. Phys. {\bf 147}, 891-918 (2012).

\bibitem{Rupprecht12b} J.-F. Rupprecht, O. B\'enichou, D. S. Grebenkov, and
R. Voituriez, {\it Exact mean exit time for surface-mediated diffusion},
Phys. Rev. E {\bf 86}, 041135 (2012).

\bibitem{Rojo11} F. Rojo and C. E. Budde, {\it Enhanced diffusion through
surface excursion: A master-equation approach to the narrow-escape-time
problem}, Phys. Rev. E {\bf 84}, 021117 (2011).

\bibitem{Rojo13} F. Rojo, C. E. Budde Jr., H. S. Wio, and C. E. Budde,
{\it Enhanced transport through desorption-mediated diffusion},
Phys. Rev. E {\bf 87}, 012115 (2013).

\bibitem{Benichou15} O. B\'enichou, D. S. Grebenkov, L. Hillairet, L. Phun,
R. Voituriez, and M. Zinsmeister, {\it Mean exit time for surface-mediated
diffusion: spectral analysis and asymptotic behavior}, Anal. Math. Phys.
(2015).



\bibitem{lomholt2} M. A. Lomholt, T. Koren, R. Metzler, and J. Klafter,
\emph{L{\'e}vy strategies in intermittent search processes are advantageous},
Proc. Natl. Acad. Sci. USA \textbf{105}, 11055 (2008).

\bibitem{tamm} G. Oshanin, M. Tamm and O. Vasilyev, {\em Narrow-escape times for diffusion in microdomains with a particle-surface affinity: mean-field results}, J. Chem. Phys. {\bf 132}, 06B607 (2010).

\bibitem{katja1} G. Oshanin, H. S.  Wio, K Lindenberg, and S. F. Burlatsky,
{\em Intermittent random walks for an optimal search strategy: one-dimensional
case}, J. Phys.: Condens. Matt. {\bf 19}, 065142 (2007).

\bibitem{katja2}  G. Oshanin, K. Lindenberg, H. S.  Wio, and S. Burlatsky,
{\em Efficient search by optimized intermittent random walks}, J. Phys. A:
Math. and Theor. {\bf 42}, 434008 (2009).                   

\bibitem{katja3} F. Rojo, J.  Revelli, C. E.  Budde, H. S. Wio, G. Oshanin,
and K. Lindenberg, {\em Intermittent search strategies revisited: effect of
the jump length and biased motion}, J. Phys. A: Math. and Theor. {\bf 43},
345001 (2010).	

\bibitem{vlad} V. V. Palyulin, A. V. Chechkin, and R. Metzler, \emph{L{\'e}vy
flights do not always optimize random blind search for sparse targets},
Proc. Natl. Acad. Sci. USA \textbf{111}, 2931 (2014).

\bibitem{vlad2} V. Palyulin, A. V. Chechkin, R. Klages, and R. Metzler, \emph{Search
reliability and search efficiency of combined L\'evy-Brownian motion: long
relocations mingled with thorough local exploration}, J. Phys. A \textbf{49},
394002 (2016).

\bibitem{olivier} O. B\'enichou and R. Voituriez, {\em From first-passage
times of random walks in confinement to geometry-controlled kinetics}, 
Phys. Rep. {\bf 539}, 225 (2014).

\bibitem{150} S. Redner, {\em A Guide to First Passage Processes}
(Cambridge: Cambridge University Press, 2001).

\bibitem{151} R. Metzler, G. Oshanin and S. Redner, eds., {\em First-passage
phenomena and their applications} (World Scientific Publishers: Singapore,
2014).

\bibitem{9} D. S. Grebenkov, R. Metzler, and G. Oshanin, {\em Distribution of
first-reaction times with target sites on boundaries of shell-like regions}, 
New J. Phys. {\bf 23}, 123049 (2021). 


\bibitem{McGuffee10}  S. R. McGuffee and A. H. Elcock, {\em Diffusion, Crowding and Protein Stability in a Dynamic Molecular Model of the Bacterial Cytoplasm},
PLoS ComputBiol {\bf 6}, e1000694 (2010).

\bibitem{Ghost16}  S. K. Ghosh, A. G. Cherstvy, D. S. Grebenkov, and R. Metzler, {\em Anomalous, non-Gaussian tracer diffusion in heterogeneously crowded environments},
New J. Phys. {\bf 18}, 013027 (2016).

\bibitem{ma} J. Ma, M. Do, M. A. Le Gros, C. S. Peskin, C. A. Larabell, Y. Mori,
and S. A. Isaacson, {\em Strong intracellular signal inactivation produces
sharper and more robust signaling from cell membrane to nucleus}, 
PLoS Comput. Biol. {\bf 16}, e1008356 (2020).
		
\bibitem{100} E. R. Weikum, X. Liu, and E. A. Ortlund, \emph{The nuclear
receptor superfamily: A structural perspective}, Protein Sci. \textbf{27},
1876 (2018).

\bibitem{ehr} G. Antczak and G. Ehrlich, {\em Jump processes in surface
diffusion}, Surface Science Rep. {\bf 62}, 39 (2007).

\bibitem{pap} M. D. Huber and L. Gerace, {\em The size-wise nucleus: nuclear
volume control in eukaryotes}, J. Cell Biol. {\bf 179}, 583 (2007).

\bibitem{rag} A. Rizzotto and E. C. Schirmer, {\em Breaking the scale: how
disrupting the karyoplasmic ratio gives cancer cells an advantage for
metastatic invasion}, Biochem. Soc. Trans. {\bf 45}, 1333 (2017).

\bibitem{mag} M. E. Malerba and D. J. Marshall, {\em Larger cells have
relatively smaller nuclei across the Tree of Life}, Evol. Lett. {\bf 5},
306 (2021).

\bibitem{plant} E. W. Sinnott and V. V. Trombetta, {\em The cytonuclear
ratio in plant cells}, Am. J. Botany {\bf  23}, 602 (1936).

\bibitem{carlos1} C. Mej\'{i}a-Monasterio, G. Oshanin and G. Schehr, 
{\em First passages for a search by a swarm of independent random searchers}, 
J. Stat. Mech.  P06022 (2011).

\bibitem{carlos2} T. G. Mattos, C. Mej\'{i}a-Monasterio, R. Metzler, and G.
Oshanin, {\em First passages in bounded domains: When is the mean first
passage time meaningful?} Phys. Rev. E {\bf 86} 031143 (2012).

\bibitem{alj1} A. Godec and R. Metzler, {\em Optimization and universality
of Brownian search in quenched heterogeneous media}, Phys. Rev. E {\bf 91},
052134 (2015).

\bibitem{alj2} A. Godec and R. Metzler, {\em Universal proximity effect in
target search kinetics in the few encounter limit}, Phys. Rev. X {\bf 6},
041037 (2016).


\bibitem{dist1} D. S. Grebenkov, R. Metzler  and G. Oshanin, {\em Strong
defocusing of molecular reaction times results from an interplay of geometry
and reaction control}, Comm. Chem. {\bf 1}, 96 (2018).
 
\bibitem{dist4} D. S. Grebenkov, R. Metzler and G. Oshanin, {\em Towards a
full quantitative description of single-molecule reaction kinetics in
biological cells}, Phys. Chem. Chem. Phys. {\bf 20}, 16393 (2018).

\bibitem{dist2} D. S. Grebenkov, R. Metzler and G. Oshanin, {\em Full
distribution of first exit times in the narrow escape problem}, 
New J. Phys. {\bf 21}, 122001 (2019).


\bibitem{dg1}  D. S. Grebenkov, {\em Paradigm Shift in Diffusion-Mediated Surface Phenomena}, Phys. Rev. Lett. {\bf 125}, 078102 (2020).

\bibitem{dg2}  D. S. Grebenkov, {\em Joint distribution of multiple boundary local times and related first-passage time problems with multiple targets}, 
J. Stat. Mech. 103205 (2020)

\bibitem{dg3}  D. S. Grebenkov, {\em An encounter-based approach for restricted diffusion with a gradient drift}, J. Phys. A: Math. Theor. {\bf 55}, 045203 (2022).




\bibitem{q2} D. S. Grebenkov and G. Oshanin, {\em Diffusive escape through
a narrow opening: new insights into a classic problem}, Phys. Chem. Chem.
Phys. {\bf 19}, 2723 (2017).

\bibitem{q3} G. Oshanin, M. N. Popescu and S. Dietrich, {\em Active colloids
in the context of chemical kinetics}, J. Phys. A: Math. Theor. {\bf 50},
134001 (2017).

\bibitem{q4} D. S. Grebenkov, R. Metzler, and G. Oshanin, {\em Effects of
the target aspect ratio and intrinsic reactivity onto diffusive search in
bounded domains}, New J. Phys. {\bf 19}, 103025 (2017).

\bibitem{purves} D. Purves et al, eds., {\em Neuroscience}, 2nd ed.
(Sunderland (MA): Sinauer Associates, 2001).

\bibitem{alberts} B. Alberts {\em et al}, {\em Molecular Biology of the
Cell}, $6$th ed. (New York: Garland, 2015).

\bibitem{10} D. S. Grebenkov, R. Metzler, and G. Oshanin, {\em A molecular
relay race: sequential first-passage events to the terminal reaction centre
in a cascade of diffusion controlled processes}, New J. Phys. {\bf 23},
093004 (2021).


\bibitem{Grebenkov20c} D. S. Grebenkov, {\em Surface Hopping Propagator:
An Alternative Approach to Diffusion-Influenced Reactions}, Phys. Rev. E
{\bf 102}, 032125 (2020).


\bibitem{Chao81} N. Chao, S. H. Young  and M. Poo, {\em Localization of
cell membrane components by surface diffusion into a trap}, Biophys. J.
{\bf 36}, 139 (1981).

\bibitem{Sano81} H. Sano and M. Tachiya, {\em Theory of diffusion-controlled
reactions on spherical surfaces and its application to reactions on micellar
surfaces}, J. Chem. Phys. {\bf 75}, 2870 (1981).

\bibitem{Prustel13} T. Pr\"ustel and M. Tachiya, {\em Reversible
diffusion-influenced reactions of an isolated pair on some two dimensional
surfaces}, J. Chem. Phys. {\bf 139}, 194103 (2013).

\bibitem{Grebenkov19sphere} D. S. Grebenkov, {\em Reversible reactions
controlled by surface diffusion on a sphere}, J. Chem. Phys. {\bf 151},
154103 (2019).


\bibitem{baruch} S. N. Majumdar and B. Meerson, {\em  Statistics of
first-passage Brownian functionals}, J. Stat. Mech. (2020) 023202;
Corrigendum: J. Stat. Mech. (2021) 039801.

\bibitem{smit} B. Meerson and N. R. Smith, {\em Geometrical optics of
constrained Brownian motion: three short stories}, J. Phys. A: Math.
Theor. {\bf 52}, 415001 (2019).



\bibitem{Antoine22}  C. Antoine and J. Talbot, {\em Enhancing search efficiency through diffusive echo}
(submitted; E-print: arXiv:2205.09942v1).





\bibitem{Weiss83} G. H. Weiss, K. E. Shuler, and K. Lindenberg, {\em Order
Statistics for First Passage Times in Diffusion Processes}, J. Stat. Phys.
{\bf 31}, 255 (1983).

\bibitem{Basnayake18} K. Basnayake, A. Hubl, Z. Schuss, and D. Holcman,
{\em Extreme narrow escape: Shortest paths for the first particles among
n to reach a target window}, Phys. Lett. A {\bf 382}, 3449 (2018).

\bibitem{Basnayake19} K. Basnayake, Z. Schuss, and D. Holcman,
{\em Asymptotic formulas for extreme statistics of escape times in 1, 2
and 3-dimensions}, J. Nonlinear Sci. {\bf 29}, 461-499 (2019).

\bibitem{Schuss19} Z. Schuss, K. Basnayake, and D. Holcman, {\em Redundancy
principle and the role of extreme statistics in molecular and cellular
biology}, Phys. Life Rev. {\bf 28}, 52-79 (2019).

\bibitem{Lawley20a} D. S. Lawley and J. B. Madrid, {\em A Probabilistic
Approach to Extreme Statistics of Brownian Escape Times in Dimensions 1,
2, and 3,} J. Nonlinear Sci. {\bf 30}, 1207-1227 (2020).

\bibitem{Lawley20b} D. S. Lawley, {\em Distribution of extreme first passage
times of diffusion}, J. Math. Biol. {\bf 80}, 2301 (2020).

\bibitem{Grebenkov20} D. S. Grebenkov, R. Metzler, and G. Oshanin, {\em From
single-particle stochastic kinetics to macroscopic reaction rates: fastest
first-passage time of N random walkers}, New J. Phys. {\bf 22}, 103004 (2020).

\bibitem{Majumdar20} S. N. Majumdar, A. Pal, and G. Schehr, {\em Extreme
value statistics of correlated random variables: a pedagogical review},
Phys. Rep. {\bf 840}, 1 (2020).

\bibitem{Grebenkov22}  D. S. Grebenkov, {\em Depletion of Resources by a Population of Diffusing Species}, Phys. Rev. E {\bf 105}, 054402 (2022).



\bibitem{christine} J. F. Reverey, J.-H. Jeon, H. Bao, M. Leippe, R. Metzler,
and C. Selhuber-Unkel, \emph{Superdiffusion dominates intracellular particle
motion in the supercrowded space of pathogenic Acanthamoeba castellanii},
Sci. Rep. \textbf{5}, 11690 (2015).

\bibitem{christine1} S. Thapa, N. Lukat, C. Selhuber-Unkel, A. Cherstvy,
and R. Metzler, \emph{Transient superdiffusion of polydisperse vacuoles
in highly-motile amoeboid cells}, J. Chem. Phys. \textbf{150}, 144901 (2019).

\bibitem{Witzel19} P. Witzel, M. G\"otz, Y. Lanoisel\'ee, T. Franosch, D. S. Grebenkov, and D. Heinrich, 
\emph{Heterogeneities Shape Passive Intracellular Transport}, Biophys. J. \textbf{117}, 203-213 (2019).


\bibitem{Chechkin17} 	A. V. Chechkin, F. Seno, R. Metzler, and I. M. Sokolov,
{\em Brownian yet Non-Gaussian Diffusion: From Superstatistics to Subordination of Diffusing Diffusivities},
Phys. Rev. X {\bf 7}, 021002 (2017).

\bibitem{Lanoiselee18}  Y. Lanoisel\'ee, N. Moutal, and D. S. Grebenkov, 
{\em Diffusion-limited reactions in dynamic heterogeneous media}, Nature Commun. {\bf 9}, 4398 (2018).


\bibitem{Grebenkov21_review} D. S. Grebenkov, {\em A physicist's guide to
explicit summation formulas involving zeros of Bessel functions and related
spectral sums}, Rev. Math. Phys. {\bf 33}, 2130002 (2021).

\bibitem{Erdelyi} A. Erd\'elyi, {\em Higher transcendental functions}, vol.
1. (Robert E. Krieger Publishing, Malabar FL, 1953).



\bibitem{Szmytkowski} R. Szmytkowski, {\em The parameter derivatives
$[\partial^2 P_\nu(z)/\partial\nu^2]_{\nu=0}$ and  $[\partial^3 P_\nu(z)/
\partial\nu^3]_{\nu=0}$, where $P_\nu(z)$ is the Legendre function of the
first kind,} E-print arXiv:1301.6586 (2013).

\bibitem{Laurenzi} B. J. Laurenzi, {\em Derivatives with respect to the order
of the Legendre Polynomials,} E-print arXiv:1502.06507v1 (2015).




\end{thebibliography}
\end{document}